\documentclass[twocolumn]{aastex631}
\usepackage{amsmath}
\usepackage{graphicx}
\usepackage[caption=false]{subfig}
\usepackage{float}
\usepackage{multirow,array,booktabs}

\DeclareUnicodeCharacter{2212}{-}

\begin{document}
\title{The Orbital Eccentricities of Directly Imaged Companions Using Observable-Based Priors: Implications for Population-level Distributions}
\author[0000-0001-5173-2947]{Clarissa R. Do Ó}
\affiliation{Center for Astrophysics and Space Sciences, University of California, San Diego, La Jolla, CA 92093, USA}
\author[0000-0003-2400-7322]{Kelly K. O'Neil}
 \affiliation{Department of Physics and Astronomy, University of California, Los Angeles, Los Angeles, CA 90095, USA}
\author[0000-0002-9936-6285]{Quinn M. Konopacky}
\affiliation{Center for Astrophysics and Space Sciences, University of California, San Diego, La Jolla, CA 92093, USA}
\author[0000-0001-9554-6062]{Tuan Do}
 \affiliation{Department of Physics and Astronomy, University of California, Los Angeles, Los Angeles, CA 90095, USA}
\author{Gregory D. Martinez}
\affiliation{Department of Physics and Astronomy, University of California, Los Angeles, Los Angeles, CA 90095, USA}
\author[0000-0003-2233-4821]{Jean-Baptiste Ruffio}
\affiliation{Center for Astrophysics and Space Sciences, University of California, San Diego, La Jolla, CA 92093, USA}
\author[0000-0003-3230-5055]{Andrea M. Ghez}
\affiliation{Department of Physics and Astronomy, University of California, Los Angeles, Los Angeles, CA 90095, USA}

\begin{abstract}
The eccentricity of a sub-stellar companion is an important tracer of its formation history. Directly imaged companions often present poorly constrained eccentricities. A recently developed prior framework for orbit fitting called ``observable-based priors" has the advantage of improving biases in derived orbit parameters for objects with minimal phase coverage, which is the case for the majority of directly imaged companions. We use observable-based priors to fit the orbits of 21 exoplanets and brown dwarfs in an effort to obtain the eccentricity distributions with minimized biases. We present the objects’ individual posteriors compared to their previously derived distributions, showing in many cases a shift toward lower eccentricities. We analyze the companions’ eccentricity distribution at a population level, and compare this to the distributions obtained with the traditional uniform priors. We fit a Beta distribution to our posteriors using observable-based priors, obtaining shape parameters $\alpha$ = $1.09\substack{+0.30 \\ -0.22}$\ and $\beta$ = $1.42\substack{+0.33 \\ -0.25}$. This represents an approximately flat distribution of eccentricities. The derived $\alpha$ and $\beta$ parameters are consistent with the values obtained using uniform priors, though uniform priors lead to a tail at high eccentricities. We find that separating the population into high and low mass companions yields different distributions depending on the classification of intermediate mass objects. We also determine via simulation that the minimal orbit coverage needed to give meaningful posteriors under the assumptions made for directly imaged planets is $\approx$ 15\% of the inferred period of the orbit. 
\end{abstract}

\section{Introduction} \label{intro}
The field of exoplanet direct imaging has advanced significantly in recent years, with the detection capabilities of ground based telescopes improving due to technological developments in adaptive optics (AO). Direct imaging also allows for the measurement of the relative astrometry of planets over time, which can in turn allow astronomers to better constrain the orbital parameters of these objects and the overall orbital architecture of their systems. By better understanding the architecture of these extrasolar systems, we may obtain an understanding of formation pathways of planetary systems in our Galaxy.  Thus, orbit fitting of directly imaged exoplanets has the potential to shed light on the early formation history and dynamical evolution of planetary systems with widely-separated, gas giant planets. \par
In particular, the eccentricity of a planet can tell us much about how it was formed. In the theory of protoplanetary disk formation, disk fluid elements can initially be in an eccentric orbit, but soon lose energy through collisions and settle into the minimum energy orbit, which is a circular orbit (\citealt{L08}). Therefore, planets forming in a protoplanetary disk via core or pebble accretion should have lower eccentricities, but can develop higher eccentricities through processes such as planet-planet scattering or disk migration (\citealt{Bowler_2020}, \citealt{D16}, \citealt{P06}, \citealt{Chatterjee_2008}, \citealt{2017AREPS..45..359J}). Various protoplanetary disk processes can also either damp eccentricities (e.g. \citealt{Kley_Nelson_2012}; \citealt{Bitsch_Kley_2010}) or excite eccentricities (e.g. \citealt{Moorhead_Adams_2008}; \citealt{Goldreich_Sari_2003}). Formation of planets via gravitational instability can potentially form planets with higher eccentricities. Generally, gravitational instability planet formation occurs further from the host star, causing eccentricity damping timescales to be longer at such large separations (e.g. \citealt{Mayer_2004}). These various processes may lead to a variety of eccentricities for individual planetary systems, so it is difficult to tell which mechanism is operating for individual systems. However, the eccentricity distribution of many planetary systems (i.e., the parent distribution) can give us strong constraints on which formation mechanisms are at work. Planets that slowly form from protoplanetary disks in unperturbed orbits should have low eccentricities, while planets that undergo planetary migration or outward scattering can have a range of varying eccentricities. \par
Efforts to obtain an eccentricity distribution at a population-level for sub-stellar companions have already been made for a number of exoplanet and brown dwarf populations.
\citealt{Hog2010} presents the method of hierarchical Bayesian modeling to obtain parameters for the underlying parent eccentricity distribution from the posteriors of exoplanet parameters given from orbital fits. The planet population simulated by \citealt{Hog2010} was inferred from radial velocity measurements of hot Jupiters. \citealt{2013MNRAS.434L..51K} has constrained eccentricity distributions of short-period exoplanets using data obtained from 396 companions using the radial velocity (RV) method. \citealt{Bowler_2020} used data from 27 long period, directly imaged companions to obtain an eccentricity distribution of these objects. They find that extrasolar companions present different eccentricity distributions if they are low mass or high mass. They conclude that ``exoplanets'', or lower mass companions, have lower eccentricities (so they likely formed from a disk) and ``brown dwarfs", or higher mass companions, have higher eccentricities (consistent with binary star formation). This conclusion, if true, is groundbreaking - it suggests that these companions can be observationally distinguished via their eccentricities. In this work, we revisit this analysis using a different set of priors and modified sample of directly imaged substellar companions.  \par
In order to determine the orbital eccentricities of exoplanets, the standard procedure is to fit astrometric data of the companion relative to the star at different points in time using Bayesian statistics. Within the Bayesian framework, prior probability distributions (priors) for each orbit parameter must be assumed to ultimately infer posterior probability distributions (posteriors) for each parameter. The most common priors assumed for orbital parameters are model-based priors (i.e., they assume a prior distribution in the 6 model parameters of a Keplerian orbit). They generally assume distributions that are uniform (for the eccentricity, argument of periastron and period of ascending nodes), log-uniform (semi-major axis) and sine (inclination). It is also common practice to narrow down the parameter space with physically motivated priors, such as stability constraints, coplanarity and stellar rotation rates (e.g. \citealt{PearceWyatt2015}; \citealt{Wang_2018}; \citealt{Wang_2021_pds70}; \citealt{Thompson_2023}; \citealt{Zhang_2023}). However, for orbits where less than 40\% of the orbital arc is covered, which is the majority of directly imaged companions, this standard method of model-based priors (which herein we will refer to as ``uniform" priors, as is done by \citealt{ON19}) has been shown to introduce biases in the resulting posterior distributions, generating inaccurate parameters and confidence intervals (\citealt{Greg2017}, \citealt{Lucy2014}). The eccentricity parameter is generally affected by this issue, with many objects presenting a bias towards high eccentricities when fit with uniform priors. \par
\citealt{ON19} presented a new approach to priors for the orbit fitting of long period resolved companions, which is based on uniformity in the observable parameters rather than in the model parameters. This approach is called ``observable-based" priors. This method has been shown to reduce this bias in orbital parameters where the orbital arc of the object is less than 40\% covered. In this work, we aim to obtain the eccentricity distribution of extrasolar companions at a population level, using observable-based priors to fit the orbits of 21 directly imaged companions. Given the majority of our sub-stellar companion sample is under-sampled (i.e., their current orbital coverage spans a small amount of their orbital arc),  we also perform a series of simulations to assess the minimum orbital coverage needed in order to get meaningful eccentricity posteriors for directly imaged companions. \par
Our analysis is outlined as follows. We introduce our sample and present the astrometric and radial velocity data used for our orbit fits in sections \ref{astrom} and \ref{rv}. In section \ref{of}, we present the concept of observable-based priors.  We present our results in section \ref{results}, describing the results of our orbit fitting of the 21 companions (section \ref{datares}) and simulations for the minimal orbital coverage needed (section \ref{simres}). We discuss the implications of our work in \ref{disc} and our conclusions in \ref{conc}.

\section{Data} \label{data}
\subsection{Sample Selection and Astrometric Data} \label{astrom}

The companions chosen are a sub-sample of objects from \citealt{Bowler_2020}. The criteria for choosing these objects is outlined in detail in \citealt{Bowler_2020}; we summarize them here: the objects must be at a projected separation of 5-100 AU from the host star at the time of discovery, and the hosts must be stars (M $>$ 75 $M_{Jup}$), such that any companions that are clearly part of binary systems are excluded from the sample. For the systems with more than one detected planet, such as HR 8799, PDS 70, HD 206893 and $\beta$ Pictoris, we chose only one planet as the ``representative" of the system's eccentricity in this analysis. We made this choice so that multiple planets from the same system would not bias our resulting eccentricity distributions, since it has been shown that planets with multiple systems present lower eccentricities due to stability requirements (\citealt{W09}), and therefore the eccentricities within a single system are correlated. We also do not include any companions that have less than 4 epochs of observation, as our software requires this value as a minimum amount of astrometry for orbit fitting given the number of free parameters. With these requirements, 21 companions are in the sample used in this work. The sample is presented on Table \ref{tbl:masses}.\par 
\begin{deluxetable*}{ccccccc}
\tablecaption{Companions in the Sample} \label{tbl:masses}
\tablewidth{20pt}
\tablecolumns{7}
\tabletypesize{\scriptsize}
\tablehead{\colhead{Object} & \colhead{Inferred Mass ($M_J$)} & \colhead{Mass Classification} & \colhead{Primary Spectral Type} & \colhead{Age (Myr)} & \colhead{Average Separation (as; AU)} & \colhead{References} }
\startdata
51 Eri b & 2.6 $\pm$ 0.3 &  Giant Planet & F0IV & $20\substack{+6 \\ -6}$ & 0.45; 13.2 & 1; 2; 3\\
GJ 504 b & $4.0\substack{+4.5 \\ -1.0}$\  &  Giant Planet & G0 & $160\substack{+350\\ -60}$ & 2.5; 43.5 & 4\\
HD 95086 b & 2.6 $\pm$ 0.4  &  Giant Planet & A8 & $17\substack{+4 \\ -4}$& 4.5; 56 & 1; 5\\
PDS 70 c & 4.4 $\pm$ 1.1  &  Giant Planet & K7 & $5.4\substack{+1.0 \\ -1.0}$\ & 0.2; 30.2& 6; 7\\
HR 8799 c & 8.3 $\pm$ 0.6  &  Giant Planet & A5V & $30\substack{+20 \\ -10}$ & 0.7; 38 & 1; 8\\
HIP 65426 b & 7.1 $\pm$ 1.1  &  Giant Planet & A2V & $14\substack{+4 \\ -4}$ & 0.82; 86 & 9; 10\\
$\kappa$ And b & $13\substack{+12 \\ -2}$\  & Boundary & B9V & $47\substack{+27 \\ -40}$\ & 0.91; 76.5 & 11; 12\\
$\beta$ Pic b & $11.9\substack{+2.93 \\ -3.04}$\  &  Boundary & A6V &$22\substack{+3 \\ -3}$\ & 0.25; 9.9 & 13; 14; 15\\
HR 2562 b & $10.28\substack{+5.00\\ -5.00}$\  &  Boundary & F5V & $600\substack{+300 \\ -300}$\ & 0.618; 20.3 &  16; 17\\
HD 206893 b & $28.00\substack{+2.2 \\ -2.1}$\  &  Brown Dwarf & F5V & $1,100\substack{+1,000\\ -1,000}$\ & 0.27; 10 & 18; 19 \\
1RXS0342+1216 b & 35 $\pm$ 8 &  Brown Dwarf & M4 & 60-300 & 0.83; 19.8  & 20\\
Gl 758 b & 37.9 $\pm$ 1.5  &  Brown Dwarf & K0V & $8,200\substack{+500\\ -500}$\ & 1.6; 25 & 21; 22; 35\\
HR 3549 b & 45 $\pm$ 5  &  Brown Dwarf & A0V & 100-500 & 0.9; 83.2 & 23\\
HD 1160 b & $33\substack{+12\\ -9}$\  &  Brown Dwarf & A0V & 80-120 & 0.78; 81 & 24; 25\\
HD 19467 b & 52 $\pm$ 4.3 &  Brown Dwarf & G3V & 4,600-10,000 & 1.65; 51.1 & 26\\
HR 7672 b & 61.5 $\pm$ 6.5 &  Brown Dwarf & G1V & 1,000-3,000 & 0.79; 14 & 27\\
PZ Tel b & 64 $\pm$ 5 &  Brown Dwarf & G9IV & $23\substack{+3\\ -3}$\ & 0.50; 23.65 & 1; 15\\
Gl 229 b & 35 $\pm$ 15 &  Brown Dwarf & M1V & 7,000-10,000 & 6.03; 34.7 & 28; 29; 30\\
HD 4747 b & 66.6 $\pm$ 3.5 &  Brown Dwarf & G9V & $3,300\substack{+2,300\\ -1,900}$\ &  0.6; 11.3 & 31; 32\\
HD 984 b & 61 $\pm$ 4 &  Brown Dwarf & F7V & $80\substack{+120\\ -50}$\ & 0.22; 28 & 33\\
HD 49197 b & $63.2\substack{+12.6 \\ -26.32}$\ &  Brown Dwarf & F5V & 290-790 & 0.95; 44 & 34\\
\enddata
\tablecomments{ References: (1) \citealt{Nielsen_2019}; (2) \citealt{Bell2015}; (3) \citealt{Macintosh_2015}; (4) \citealt{Kuzuhara_2013}; (5) \citealt{Meshkat_2013}; (6) \citealt{Mesa_2019}; (7) \citealt{Kep2018}; (8) \citealt{Marois_2010}; (9) \citealt{Carter_2022}; (10) \citealt{Chauvin_2017}; (11) \citealt{Currie_2018}; (12) \citealt{Jones_2016};  (13) \citealt{Lacour_2021}; (14) \citealt{Gray_2006}; (15) \citealt{Mamajek_2014}; (16) \citealt{Zhang_2023}; (17) \citealt{K16_2}; (18) \citealt{Hinkley_2022}; (19) \citealt{Mil2016}; (20) \citealt{Bowler_2014}; (21) \citealt{Brandt2019}; (22) \citealt{Mamajek_2008}; (23) \citealt{Mesa_2016}; (24) \citealt{Nielsen_2012}; (25) \citealt{Garcia_2017}; (26) \citealt{JensenClem_2016}; (27) \citealt{Liu_2002}; (28) \citealt{Nakajima_1995}; (29) \citealt{Byrne_1985}; (30) \citealt{Brandt_2020}; (31) \citealt{Xuan_2020}; (32) \citealt{Crepp_2016}; (33) \citealt{Franson_22}; (34) \citealt{Metchev_2005}; (35) \citealt{Bowler_2018} }
\end{deluxetable*}

 The astrometry used, as well as a fixed total system mass and distance, include points up to 2018 from \citealt{Bowler_2020}, as well as updated data from the literature for 8 of the companions. The new astrometry points used are presented on Table \ref{tbl:1}. The astrometry from \citealt{Bowler_2020} includes points from the literature for HD 984 b (\citealt{M15}; \citealt{J17}), HD 1160 b (\citealt{N12}; \citealt{M16}; \citealt{C18}), HD 19467 B (\citealt{Cr14}; \citealt{Cr15}), 1RXS0342+1216 b (\citealt{B15b}; \citealt{J12}; \citealt{Janson_2014}; \citealt{B_14}), 51 Eri b (\citealt{M19}), HD 49197 b (\citealt{metchev2004lowmass}; \citealt{S09}; \citealt{B16}), HR 2562 b (\citealt{K16}; \citealt{M2018}), HR 3549 b (\citealt{M2015}; \citealt{M2016}), HD 95086 b (\citealt{R16}; \citealt{chauvin2018investigating}), GJ 504 b (\citealt{Ku13}; \citealt{Bon18}), HIP 65426 b (\citealt{Ch2017}; \citealt{Ch2019}), \citealt{Mul2018}; \citealt{Wag2018}), PZ Tel b (\citealt{Mug2012}; \citealt{Bil2010}; \citealt{Beu2016}; \citealt{Gin2014}; \citealt{Mai2016}), HD 206893 b (\citealt{Mil2016}; \citealt{Del2017}; \citealt{Gra2019}), $\kappa$ And B (\citealt{Car2013}; \citealt{C18}), \citealt{Bry2016}), HD 4747 b (\citealt{Brandt2019}), Gl 229 b (\citealt{Brandt2019}), HR 7672 b (\citealt{Brandt2019}), Gl 758 b (\citealt{Brandt2019}) and HR 8799 c (\citealt{K16_2}, \citealt{Wang_2018}).

\startlongtable
\begin{deluxetable*}{ccccc}
\tablecaption{Updated Astrometry Points\label{tbl:1}}
\tablewidth{20pt}
\tablecolumns{5}
\tabletypesize{\scriptsize}
\tablehead{\colhead{Object} & \colhead{Epoch} & \colhead{Sep (mas)} & \colhead{PA(\textdegree)} & \colhead{Reference}}

\startdata
HD 19467 b & 2018.793 & 1631.4±1.6 & 238.88±0.12 & \citealt{2020} \\
HIP 65426 b & 2018.432 &827.26 ± 8.26 &149.56 ± 0.55  & \citealt{2020_2} \\
HD 206893 B  & 2018.434 &  269.53 ± 12.15 &  62.76 ± 2.16 & \citealt{2020_2} \\
HD 206893 B  & 2018.679 &  248.6 ± 4.9 & 41.8 ± 0.1 & \citealt{refId0} \\ 
HD 206893 B  & 2018.685 &  248.6 ± 4.9 &  41.8 ± 0.1 & \citealt{refId0} \\
HD 206893 B  & 2018.812 &  239.12 ± 17.55 &   42.53 ± 2.17 & \citealt{2020_2} \\
51 Eri b  & 2014.961 &   454.24 ± 1.88 &   171.22 ± 0.23 & \citealt{Rosa_2019} \\
51 Eri b  & 2015.079 &  451.81 ± 2.06 &   170.01 ± 0.26 &\citealt{Rosa_2019} \\
51 Eri b  & 2015.082 &   456.80 ± 2.57 &   170.19 ± 0.30 & \citealt{Rosa_2019} \\
51 Eri b  & 2015.085 &  461.5 ± 23.9 &  170.4 ± 3.0 & \citealt{Rosa_2019} \\
51 Eri b  & 2015.665 &  455.10 ± 2.23 &  167.30 ± 0.26 & \citealt{Rosa_2019} \\
51 Eri b  & 2015.847 &  452.88 ± 5.41 &  166.12 ± 0.57 & \citealt{Rosa_2019} \\
51 Eri b  & 2015.961 &  455.91 ± 6.23 &   165.66 ± 0.57 & \citealt{Rosa_2019} \\
51 Eri b  & 2015.966 & 455.01 ± 3.03 &  165.69 ± 0.43 & \citealt{Rosa_2019} \\
51 Eri b  & 2016.0726 &  454.46 ± 6.03 &   165.94 ± 0.51 & \citealt{Rosa_2019} \\
51 Eri b  & 2016.714 &  454.81 ± 2.02 &  161.80 ± 0.26 & \citealt{Rosa_2019} \\
51 Eri b  & 2016.722 &  451.43 ± 2.67 &   161.73 ± 0.31 & \citealt{Rosa_2019} \\
51 Eri b  & 2016.959 &  449.39 ± 2.15 &  160.06 ± 0.27 & \citealt{Rosa_2019} \\
51 Eri b  & 2017.861 &  447.54 ± 3.02 &   155.23 ± 0.39 & \citealt{Rosa_2019} \\
51 Eri b  & 2018.884 & 434.22 ± 2.01 &  149.64 ± 0.23 & \citealt{Rosa_2019} \\
PZ Tel b  & 2018.434 & 564.22 ± 1.57 &   58.76 ± 0.42 & \citealt{2020_2}  \\
PDS 70 c & 2016.3675 &  215.1 ± 7.0 &  285.0±1.5 & \citealt{Benisty_2021} \\
PDS 70 c & 2016.4167 & 254.1 ± 10.0 &  283.3±2.0 & \citealt{Benisty_2021} \\
PDS 70 c & 2018.1493 & 209.0 ± 13.0 &  281.2±0.5 & \citealt{Benisty_2021} \\
PDS 70 c & 2018.4671 &  235.5 ± 25.0 & 277.0±6.5 & \citealt{Benisty_2021}\\
PDS 70 c  & 2019.1767 &  225.0 ± 8.0 &  279.9±0.5 & \citealt{Benisty_2021}\\
PDS 70 c & 2019.432&  140.9 ± 2.2 &280.4 ± 2.0 & \citealt{2020AJ....159..263W} \\
$\beta$ Pic b  & 2003.859 & 413 ± 22 &  34 ± 4 & \citealt{N19}  \\
$\beta$ Pic b  & 2008.862 &  210 ± 27 & 211.49 ± 1.9 & \citealt{N19} \\
$\beta$ Pic b  & 2009.8151 & 299 ± 14 &   211 ± 3 & \citealt{N19}  \\
$\beta$ Pic b  & 2009.922 & 339 ± 10 & 209.3 ± 1.8 & \citealt{N19} \\
$\beta$ Pic b  & 2009.922 &  323 ± 10 &   209.2 ± 1.7 & \citealt{N19}  \\
$\beta$ Pic b  & 2009.9932 &  306 ± 9 &  212.1 ± 1.7 & \citealt{N19}  \\
$\beta$ Pic b  & 2010.2726 &   346 ± 7 & 209.9 ± 1.2 & \citealt{N19} \\
$\beta$ Pic b  & 2010.7411 &  383 ± 11 &  210.3 ± 1.7 & \citealt{N19}  \\
$\beta$ Pic b  & 2010.875 &  387 ± 8 & 212.4 ± 1.4 & \citealt{N19} \\
$\beta$ Pic b  & 2010.878 &  390 ± 13 & 212 ± 2 & \citealt{N19}  \\
$\beta$ Pic b  & 2010.982 &   407 ± 5 & 212.8 ± 1.4 & \citealt{N19} \\
$\beta$ Pic b  & 2011.0863 &  408 ± 9 & 211.1 ± 1.5 & \citealt{N19} \\
$\beta$ Pic b  & 2011.2315 & 426 ± 13 & 210.1 ± 1.8 & \citealt{N19}  \\
$\beta$ Pic b  & 2011.801 &  452 ± 3 & 211.6 ± 0.4  & \citealt{N19} \\
$\beta$ Pic b  & 2011.801 &  455 ± 5 & 211.9 ± 0.6 & \citealt{N19} \\
$\beta$ Pic b  & 2012.242 &  447 ± 3 & 210.8 ± 0.4 & \citealt{N19}  \\
$\beta$ Pic b  & 2012.242 &   448 ± 5 & 211.8 ± 0.6 & \citealt{N19} \\
$\beta$ Pic b  & 2012.919 &  461 ± 14 & 211.9 ± 1.2 & \citealt{N19}  \\
$\beta$ Pic b  & 2012.925 &  470 ± 10 & 212.0 ± 1.2 & \citealt{N19} \\
$\beta$ Pic b  & 2013.875 & 430.8 ± 1.5 & 212.43 ± 0.17 & \citealt{N19} \\
$\beta$ Pic b  & 2013.875 &  429.1 ± 1.0 & 212.58 ± 0.15 & \citealt{N19}  \\
$\beta$ Pic b  & 2013.8808 &  430.2 ± 1.0 & 212.46 ± 0.15& \citealt{N19}  \\
$\beta$ Pic b  & 2013.941 & 425.5 ± 1.0 & 212.51 ± 0.15 & \citealt{N19} \\
$\beta$ Pic b  & 2013.941 & 424.4 ± 1.0 & 212.85 ± 0.15 & \citealt{N19}  \\
$\beta$ Pic b  & 2013.944 & 425.3 ± 1.0 & 212.47 ± 0.16 & \citealt{N19} \\
$\beta$ Pic b  & 2014.853 & 356.2 ± 1.0 & 213.02 ± 0.19 & \citealt{N19} \\
$\beta$ Pic b  & 2014.936 &  350.51 ± 3.20 & 212.60 ± 0.66 & \citealt{N19} \\
$\beta$ Pic b  & 2015.0644 & 335.5 ± 0.9 & 212.88 ± 0.20 & \citealt{N19}  \\
$\beta$ Pic b  & 2015.2507 & 317.3 ± 0.9 & 213.13 ± 0.20 & \citealt{N19}  \\
$\beta$ Pic b  & 2015.341 & 332.42 ± 1.70 & 212.58 ± 0.35 & \citealt{N19} \\
$\beta$ Pic b  & 2015.749 & 262.02 ± 1.78 & 213.02 ± 0.48 & \citealt{N19}  \\
$\beta$ Pic b  & 2015.848 &  250.5 ± 1.5 & 214.14 ± 0.34 & \citealt{N19} \\
$\beta$ Pic b  & 2015.914 &  242.05 ± 2.51 & 213.30 ± 0.74 & \citealt{N19} \\
$\beta$ Pic b  & 2015.927 & 240.2 ± 1.1 & 213.58 ± 0.34 & \citealt{N19} \\
$\beta$ Pic b  & 2015.974 &  234.5 ± 1.0 & 213.81 ± 0.30 & \citealt{N19} \\
$\beta$ Pic b  & 2015.985 & 234.84 ± 1.80 & 213.79 ± 0.51 & \citealt{N19} \\
$\beta$ Pic b  & 2016.0533 & 227.23 ± 1.55 & 213.15 ± 0.46 & \citealt{N19} \\
$\beta$ Pic b  & 2016.056 & 222.6 ± 2.1 & 214.84 ± 0.44 & \citealt{N19}  \\
$\beta$ Pic b  & 2016.234 & 203.66 ± 1.42 & 213.90 ± 0.46 & \citealt{N19} \\
$\beta$ Pic b  & 2016.291 & 197.49 ± 2.36 & 213.88 ± 0.83& \citealt{N19} \\
$\beta$ Pic b  & 2016.709 & 142.36 ± 2.34 & 214.62 ± 1.10& \citealt{N19} \\
$\beta$ Pic b  & 2016.7855 & 134.50 ± 2.46 & 215.50 ± 1.22 & \citealt{N19} \\
$\beta$ Pic b  & 2016.8811 & 127.12 ± 6.44 & 215.80 ± 3.37 & \citealt{N19} \\
$\beta$ Pic b  & 2018.711& 140.46 ± 03.12 & 29.71 ± 1.67 & \citealt{N19} \\
$\beta$ Pic b  & 2018.7219 & 141.9 ± 5.3 & 28.16 ± 1.82 & \citealt{N19}  \\
$\beta$ Pic b  & 2018.881 & 164.5 ± 1.8 & 28.64 ± 0.70 & \citealt{N19} \\
HD 984 b  & 2019.514 & 233.8 ± 1.8 & 57.64 ± 0.29 & \citealt{Franson_22} \\
HD 984 b & 2020.578 & 242.9 ± 1.7 & 51.61 ± 0.26 & \citealt{Franson_22} \\
HR 2562 b & 2018.0836 &	669.44 ± 1.24 & 298.55 ± 0.2 & \citealt{Zhang_2023} \\
HR 2562 b & 2018.2452 & 670.84 ± 2.83 & 298.74 ± 0.26 & \citealt{Zhang_2023} \\
HR 2562 b & 2018.8836 & 685.76 ± 1.25 & 298.89 ± 0.13 & \citealt{Zhang_2023} \\
\enddata
\tablecomments{The data from \citealt{Benisty_2021} includes astrometry points from \citealt{2019}, \citealt{2019_2} and \citealt{2020AJ....159..263W}. The data from \citealt{N19} includes astrometry points from \citealt{Currie_2011}, \citealt{refId2}, \citealt{Nielsen_2014} and \citealt{2019_3}. }
\end{deluxetable*}

 \begin{figure*}
  \begin{center}
    \centering
    \centering{{\includegraphics[width= 7.7cm]{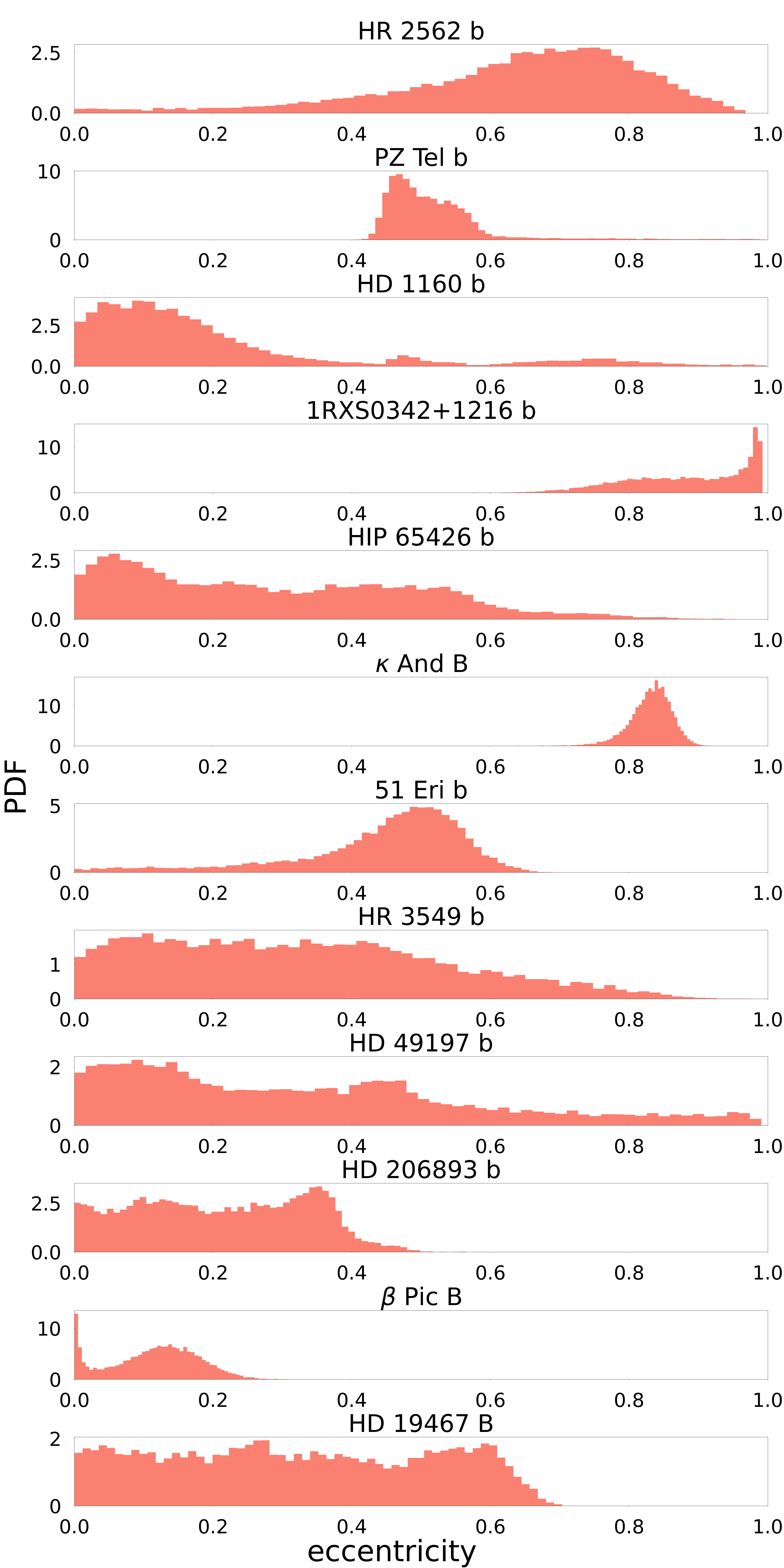} }}%
    \qquad
    \centering{{\includegraphics[width=7.7cm]{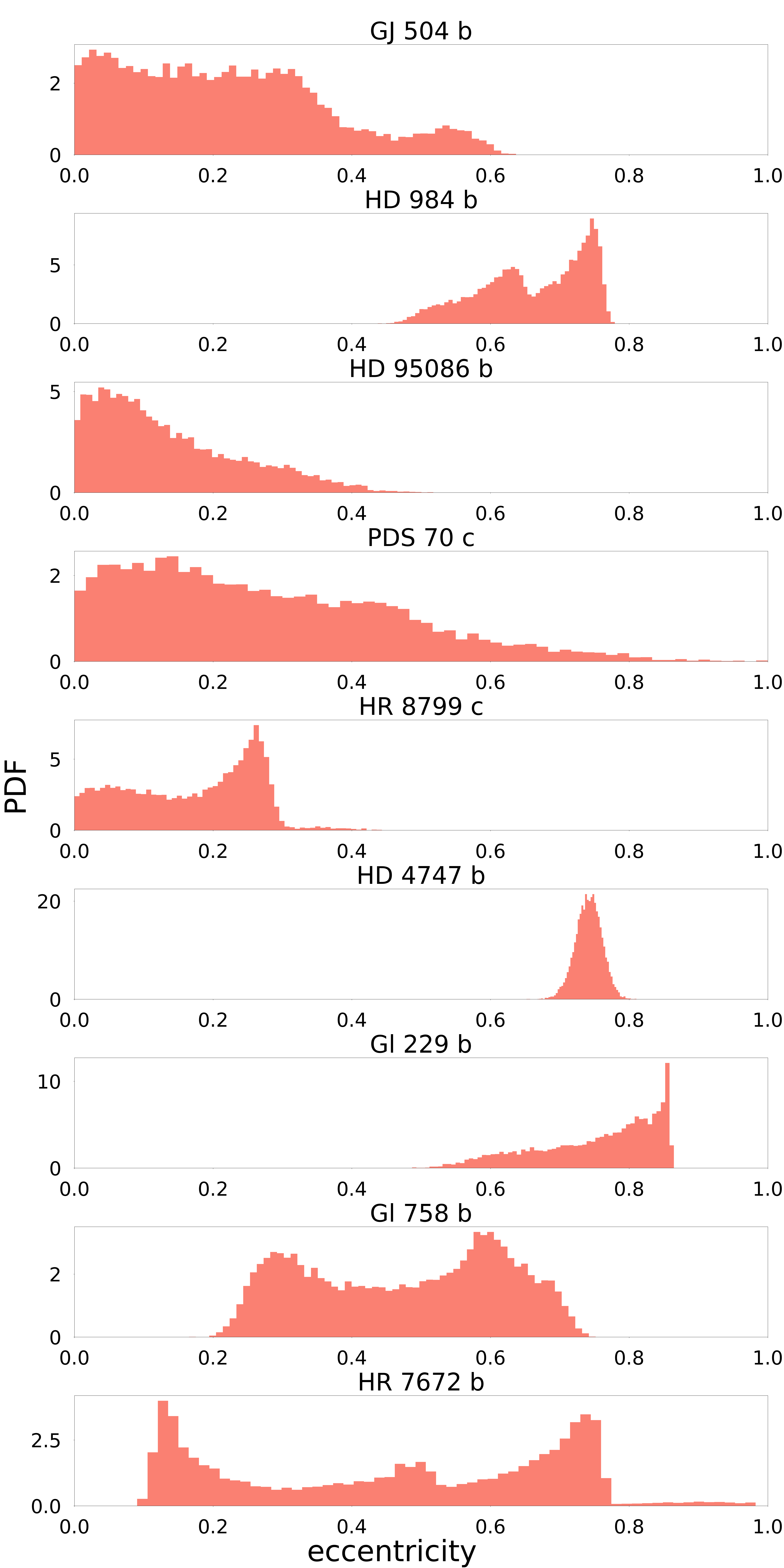} }}%
\caption{The weighted eccentricity distribution derived via our orbital fits for the sample of 21 companions using observable-based priors. The x-axis represents the eccentricity values and the y-axis represents the normalized incidence of eccentricities. Here, the fits include both astrometric and radial velocity data points}.
\label{fig:eccind}
  \end{center}
\end{figure*}

\par

\subsection{Radial Velocity Data} \label{rv}
Relative radial velocities (RV), when available, contribute a great deal of information to visual orbit fits.  Since RV measurements are challenging to obtain for directly imaged companions, only six objects in this study have radial velocities available in the literature. The RVs used are shown in Table \ref{tbl:4}. \par
Using RVs for these six objects allows us to eliminate the degeneracy in $\Omega$ and therefore better constrain the orbital plane of the object. Additionally, using RV data for HD 1160 b changed the eccentricity distribution of the companion when using uniform priors (see Section \ref{shifts}).

\begin{deluxetable*}{cccc}
\tablecaption{Radial Velocity Data Included \label{tbl:4}}
\tablewidth{20pt}
\tablecolumns{4}
\tabletypesize{\scriptsize}
\tablehead{\colhead{Object} & \colhead{Epoch} & \colhead{Relative RV (km/s)} &  \colhead{Reference}}
\startdata
$\kappa$ And b & 2016.311 & −1.4 ± 0.9 & \citealt{kw20} \\
$\kappa$ And b & 2017.308 & −1.4 ± 0.9 & \citealt{kw20} \\
HD 1160 b & 2017.843 &  7.343 ± 2.046 & J. Beas Gonzalez; in prep \\
HD 1160 b  & 2018.558 & 6.811 ± 2.670 & J. Beas Gonzalez; in prep \\
$\beta$ Pic b  & 2013.9603 &  -15.4 ± 1.7 &  \citealt{Snellen_2014}\\
HR 8799 c & 2010.5356 &	-0.3 ±  2.1 &  \citealt{Ruffio_2019}\\
HR 8799 c & 2010.8425 &	0.5 ± 2.1 &  \citealt{Ruffio_2019}\\ 
HR 8799 c & 2011.5575 &	1.5 ± 2.4 &  \citealt{Ruffio_2019} \\ 
HR 8799 c & 2011.5603 &	0.7 ± 2.9 & \citealt{Ruffio_2019} \\
HR 8799 c & 2011.563 &	0.5 ± 3.0 & \citealt{Ruffio_2019}\\
HR 8799 c & 2013.5658 &	-4.3 ± 4.0 & \citealt{Ruffio_2019}\\
HR 8799 c & 2017.8397 &	1 ±2&  \citealt{Ruffio_2019}\\
HR 8799 c & 2020.5751 &	0.9 ± 2.4  & \citealt{Wang_2021}\\
HR 7672 b & 2020.4358 & -6.074 ± 0.646 & \citealt{Ruffio2023}\\
HR 7672 b & 2020.4385 & -5.51 ± 0.58  &  \citealt{Ruffio2023}\\
HR 7672 b & 2020.7418 & -6.899 ± 0.701 & \citealt{Ruffio2023}\\
HR 7672 b & 2021.2644 & -6.366 ± 0.368 & \citealt{Ruffio2023}\\
HD 4747 b & 2020.7418 & -5.107 ± 0.101 & \citealt{2022ApJ...937...54X} \\
\enddata
\end{deluxetable*}

\section{Methods} 

\subsection{Orbit Fitting} \label{of}
We use observable-based priors for fitting the relative astrometry to orbits for all 21 directly imaged companions. These priors are implemented in the orbit fitting software Efit5 (\citealt{Mey2012}). Efit5 uses MULTINEST (\citealt{Fer2008}; \citealt{Fer2009}), a multimodal (or nested) sampling algorithm, to perform a Bayesian analysis on the data. For all of these fits, we use 3,000 live points in the nested sampling algorithm. We also include relative radial velocities (RVs) when available. In order to compare these results to the uniform prior approach, we also fit for the orbits of these companions with uniform priors. For our orbit fitting with uniform priors, we use both the Efit5 package and the orbitize! package (\citealt{Blunt_2020}) with the Markov-Chain Monte Carlo (MCMC) approach. We verify that both fitting methods with uniform priors give consistent results. For the uniform priors with orbitize, we fit for 50 million possible orbits, using 1,000 walkers, 20 temperatures, 20 threads, 10,000 burn steps and a thin factor of 10, as is done in \citealt{Bowler_2020}.
\par
After fitting the orbits of the 21 companions using observable-based priors, we use their posteriors to fit for the eccentricity distribution of our entire population of 21 companions. We use a Beta distribution as the model for our parent eccentricity distribution, as is done in other works (e.g. \citealt{Bowler_2020}; \citealt{2013MNRAS.434L..51K}). In order to recover the parent distribution from the sample's posteriors, we use Maximum Likelihood Estimate fitter in a bootstrapping manner. The details of our recovery technique are presented in Sections \ref{hbm} and \ref{bf}. We then split the sample into subsamples to assess possible distinctions between the population distributions.
\par
The purpose of observable-based priors is to improve orbital estimates for orbits where the data covers a low  percentage of the orbital arc ($<$ 40\%). Here, we briefly summarize the formulation of observable-based priors. A detailed formulation is outlined in \citealt{ON19}. \par
In its general form, observable-priors assume that all regions of observable parameter space that can be observed are equally likely, thus emphasizing uniformity in observables rather than in model parameters. In the case of orbit fitting, our fit starts with measured observables D from the astrometry:
\begin{equation}
    D = \{x(t), y(t)\}
\end{equation}
or, for astrometry and radial velocities:
\begin{equation}
    D = \{x(t), y(t), v_z(t)\}
\end{equation}
where x and y are the object's positions ( right ascension (R.A.) and declination (Dec)) in the plane of the sky relative to the position of the primary ($x_o$ and $y_o$) and $v_z$ is the velocity relative to the star. These measured observables are linearly related to the orbital observables (which describe the position and motion in the orbital plane) by the Thiele-Innes constants (e.g
\citealt{Hartkopf1989}, \citealt{WrightHoward2009}). Due to this linear relationship, a uniform distribution in the measured observables would imply a uniform distribution in the orbital observables. 
The orbital observables, denoted here as X, Y, $V_x$ and $V_y$, are also connected to the model parameters according to the following equations (e.g. \citealt{hilditch_2001};\citealt{Ghe2003}):
\begin{equation}
   X = a(cos E - e)
\end{equation}
\begin{equation}
   Y = a(\sqrt{1 - e^2}sin E)
\end{equation}
\begin{equation}
   V_x = - \frac{sin E}{1 - e cos E}\sqrt{\frac{GM}{a}}
\end{equation}
\begin{equation}
   V_y = \frac{\sqrt{1-e^2} cos E}{1 - e cos E}\sqrt{\frac{GM}{a}}
\end{equation}
where G is the gravitational constant, E is the eccentric anomaly, e is the eccentricity, a is the semimajor axis and M is the mass of the system.
By transforming between measured observables and orbital observables, and then between orbital observables and model parameters, we can transform between measured observables and model parameters. This allows us to express a distribution that is uniform in the measured observables in terms of model parameters. 
\par
This fitting method reduces biases in orbital parameters by a factor of two in orbits with low phase coverage (\citealt{ON19}). For the eccentricity parameter, observable-based priors reduce the bias towards artificially high eccentricities usually found in fits with flat priors. Flat priors lead to a biased region in periastron passage ($T_o$) parameter space, where the $T_o$ tends to artificially coincide with the observation epochs (\citealt{K16_2}). This bias is mitigated in observable-based priors because they suppress this biased region of the parameter space when sampling it. Thus, observable-based priors present a suitable fitting method for the orbital parameters of our object sample, which all have phase coverage  $<$40\%, and provide a different view on the eccentricity distribution of the companions when compared to the standard method for orbit fitting. 

\subsection{Recovering the Population Eccentricity Distribution} 
\label{hbm}

  The individual eccentricity posterior distributions obtained with the observable-based prior method and the data described in the previous sections are presented on Figure \ref{fig:eccind}. From these individual distributions, we now seek to determine the eccentricity distribution of the population.  We adopt a Beta distribution for our underlying parent distribution for exoplanet eccentricity, similar to  \citealt{Bowler_2020}, \citealt{Hog2010}, \citealt{2013MNRAS.434L..51K} and \citealt{VE2019}. The Beta distribution is a convenient choice for this study because its values, like the eccentricities, range from 0 to 1. It also only requires two positive parameters, $\alpha$ and $\beta$, which allow it to present a variety of shapes, such as linear, Gaussian and uniform. The Beta distribution is represented by:

\begin{equation}
    f(e, \alpha, \beta) = \frac{\Gamma(\alpha + \beta)e^{\alpha - 1}(1 - e)^{\beta - 1}}{\Gamma(\alpha)\Gamma(\beta)}
\end{equation}
Here, $\Gamma$ is the the Gamma function. We use the posteriors in our sample and the beta function fitter present on the SciPy package (\citealt{SciPy20}), which uses the Maximum Likelihood Estimate (MLE) method, for fitting our posteriors to the distribution. We perform our fitting using a bootstrapping method by repeating the following steps: (1) sampling an eccentricity point from each posterior chain and then (2) fitting a Beta distribution to those points using the SciPy Beta fitting functionality. \par

\subsection{Beta Distribution Recovery Simulation}\label{bf}

\begin{figure*}
    \centering
    \centering{{\includegraphics[width= 8.5cm]{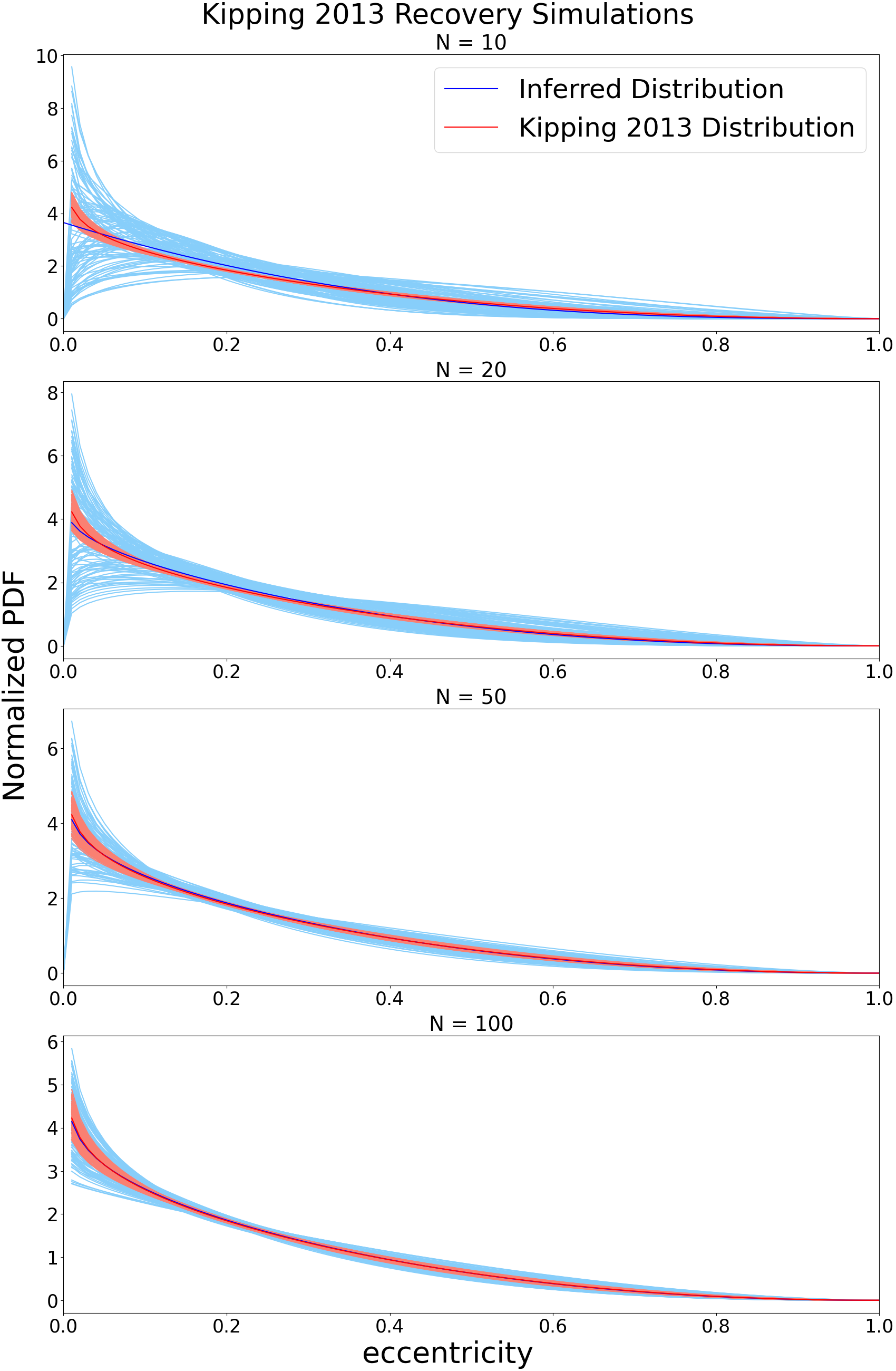} }}%
    \qquad
    \centering {{\includegraphics[width=8.5cm]{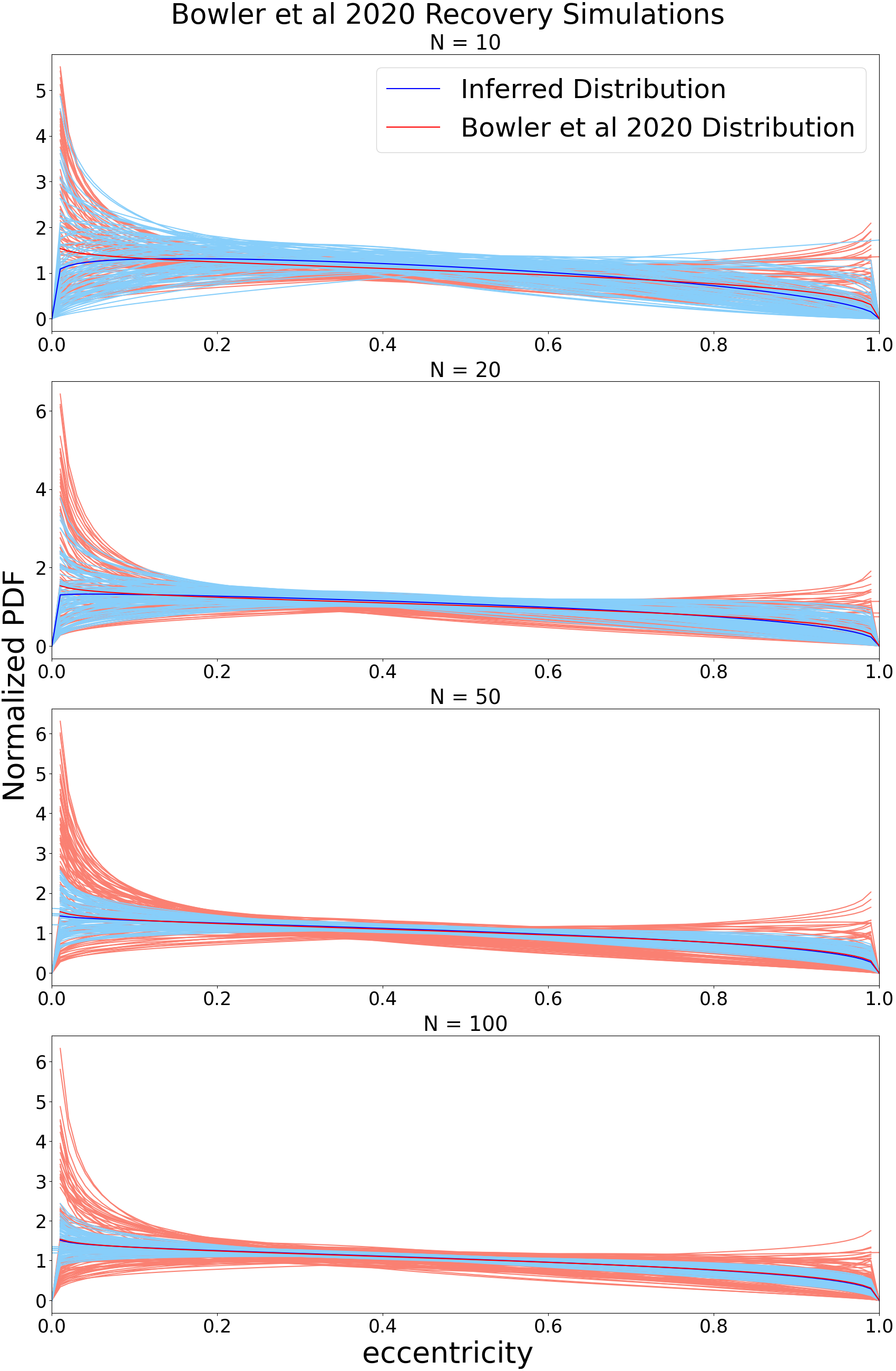} }}
    \caption{The resulting distributions of our recovery simulations plotted on top of the distributions from where the synthetic data is drawn for \citealt{2013MNRAS.434L..51K} (a) and \citealt{Bowler_2020} (b). The original distributions are represented in red, and the obtained recovered distributions are presented in blue. The different columns are labeled with the distributions obtained for differing N values (i.e. number of planets in the sample, or number of points drawn from the true input distribution). The shaded regions represented in lighter blue and red represent the possible distributions encompassed by our confidence intervals. The purpose of this exercise is to validate our bootstrapping technique and check that we get matching results from the original input (or true) distributions. }%
    \label{fig:recoverdists}
\end{figure*}

In order to validate the choice of using the Maximum Likelihood Estimate fitter for obtaining our estimated $\alpha$ and $\beta$ distributions, we develop a procedure similar to the recovery analysis presented in \citealt{Bowler_2020}. The purpose of this exercise is to test our bootstrapping method. We obtain points from the $\alpha$ = 0.867, $\beta$ = 3.03 distribution presented by \citealt{2013MNRAS.434L..51K} for warm Jupiters and points from the $\alpha$ = 0.95, $\beta$ = 1.30 distribution presented by \citealt{Bowler_2020}. We vary the number of points taken, taking N = 10, 20, 50 and 100. N represents the number of companions that contribute to our fitting sample.  Our resulting recovered distributions are presented in Tables \ref{tbl:2} (Kipping) and \ref{tbl:3} (Bowler). In the plots with Beta distribution PDFs (Figure \ref{fig:recoverdists}), we plot the ``true" input distribution in a darker blue/red shade along with 100 randomly sampled distributions from within the confidence intervals in a lighter blue/red shade. \par
\startlongtable
\begin{deluxetable}{ccc}
\tablecaption{Beta Distribution Shape Parameters (Kipping) \label{tbl:2}}
\tablewidth{20pt}
\tablecolumns{3}
\tabletypesize{\scriptsize}
\tablehead{\colhead{N} & \colhead{$\alpha$} & \colhead{$\beta$} }
\startdata
10 & $1.00\substack{+0.60 \\ -0.32}$  & $3.65\substack{+2.70 \\ -1.37}$ \\
20 & $0.93\substack{+0.34 \\ -0.22}$  & $3.31\substack{+1.46 \\ -0.93}$ \\
50  & $0.89\substack{+0.18 \\ -0.14}$  & $3.13\substack{+0.79 \\ -0.58}$  \\
100  & $0.88\substack{+0.12 \\ -0.10}$  & $3.08\substack{+0.51 \\ -0.42}$ \\
\enddata
\tablecomments{The original input distribution parameters are $\alpha$ = 0.867, $\beta$ = 3.03.}

\end{deluxetable}

\startlongtable
\begin{deluxetable}{ccc}
\tablecaption{Beta Distribution Shape Parameters (Bowler) \label{tbl:3}}
\tablewidth{20pt}
\tablecolumns{3}
\tabletypesize{\scriptsize}
\tablehead{\colhead{N} & \colhead{$\alpha$} & \colhead{$\beta$} }
\startdata
10 & $1.10\substack{+0.69 \\ -0.37}$  & $1.53\substack{+1.02 \\ -0.54}$ \\
20 & $1.02\substack{+0.38 \\ -0.25}$  & $1.40\substack{+0.56 \\ -0.36}$\\
50 & $0.98\substack{+0.21 \\ -0.16}$  & $1.34\substack{+0.30 \\ -0.23}$ \\
100 & $0.96\substack{+0.14 \\ -0.11}$  & $1.32\substack{+0.20 \\ -0.16}$ \\
\enddata
\tablecomments{The original input distribution parameters are $\alpha$ = 0.95, $\beta$ = 1.30.}
\end{deluxetable}

\section{Results} \label{results}

In this section, we show our results for the eccentricity distributions of individual objects and different populations of directly imaged companions. We find that some objects presented significant eccentricity shifts when using observable-based priors instead of uniform priors (\ref{shifts}). For our full sample of 21 objects, we obtain shape parameters $\alpha$ = $1.09\substack{+0.30 \\ -0.22}$\ and $\beta$ = $1.42\substack{+0.33 \\ -0.25}$ (\ref{wholesample}). We test whether leaving one object out of the sample changes our results (\ref{leaveout}), and find that our sample does not have any true outliers. We also test splitting the sample by companion mass (low and high), and find that the consistency between the distributions changes depending on where intermediate mass objects are placed (\ref{planetsvsbd}). Finally, we simulate how much orbital coverage is needed to obtain reliable posteriors in our distributions, both for uniform and observable priors (\ref{simres}). We obtain 15\% of the period of the orbit as a result.

\subsection{Eccentricity Distributions} \label{datares}
\par

\subsubsection{Population Eccentricity Distribution} \label{wholesample}
We use the posteriors from 21 companions to investigate changes in the inferred eccentricity of individual systems due to observable-based priors, as well as to obtain an inferred eccentricity parent distribution. Following the procedure from section \ref{bf}, we randomly sample the posteriors to obtain a possible range of $\alpha$ and $\beta$ shape parameters of the parent Beta distribution.\par
\begin{figure}
    \centering
    \centering{{\includegraphics[width= 7.5cm]{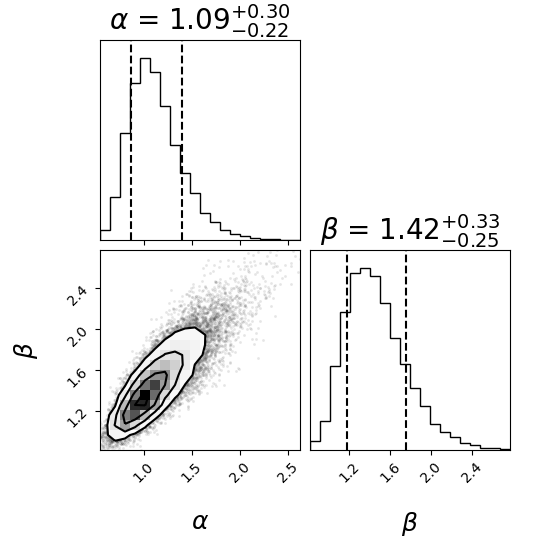} }}%
    \qquad
    \centering{{\includegraphics[width=9.2cm]{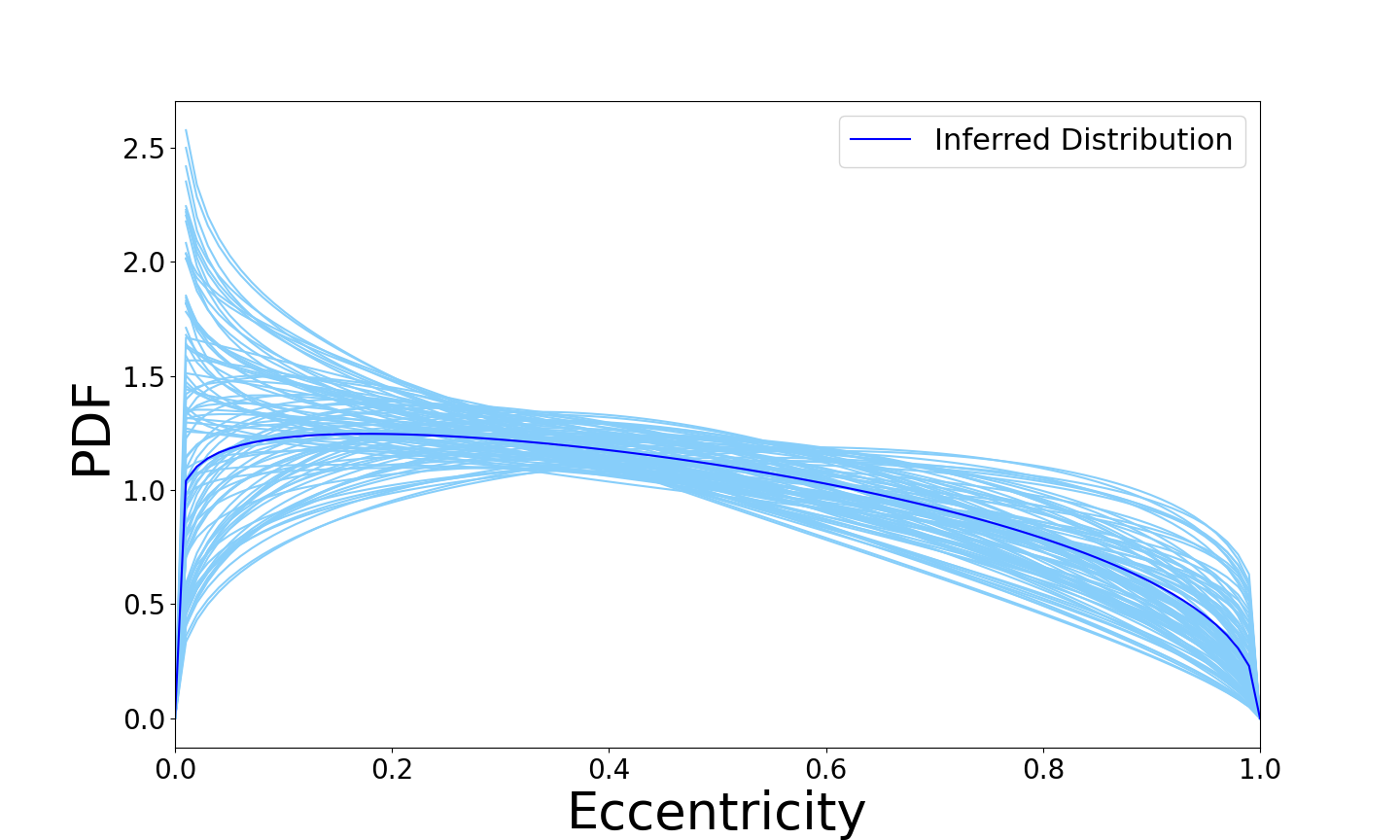} }}%
    \caption{Inferred distribution for the sample of directly imaged companions. The beta fit results (68th percentile) from sampling the posteriors are presented in (a). The distribution (dark blue) is then plotted with 100 distributions (light blue) that encompass our uncertainties. The lighter curves indicate the possible ranges of $\alpha$ and $\beta$ from the fitting distribution.}%
    \label{fig:wholesample}
\end{figure}

Our result for the eccentricity distribution of the entire sample is presented in Figure \ref{fig:wholesample}. We obtain a beta distribution with $\alpha$ = $1.09\substack{+0.30 \\ -0.22}$\ and $\beta$ = $1.42\substack{+0.33 \\ -0.25}$. This shape presents a near-uniform distribution for the eccentricities of substellar companions, with a slight tendency towards lower eccentricities. \par
In order to determine if two derived population distributions are consistent with each other, we estimate the consistency of their respective parameters $\alpha$, and $\beta$. The procedure defined here is a general method, which we use to determine consistency of two Beta probability distributions throughout this work. We define two random variables as the differences $\delta\alpha=\alpha_2-\alpha_1$ and $\delta\beta=\beta_2-\beta_1$ where $\alpha_1$, $\beta_1$ defines the first population, and $\alpha_2$, $\beta_2$ defines the second population. If the two populations are consistent, the distributions of both $\delta\alpha$ and $\delta\beta$ should be consistent with zero. The distributions of $\delta\alpha$ and $\delta\beta$ are calculated by sampling the respective distributions of $\alpha_1$, $\alpha_2$, $\beta_1$, and $\beta_2$ in a bootstrapping manner from the resulting fits (i.e., sampling the $\alpha_1$, $\beta_1$ and $\alpha_2$, $\beta_2$ pairs from the fit chains), and calculating their differences at every iteration. 
 We then calculate the p-value of the sample:
 \begin{equation}
    \mathcal{P} = p_{value}
 \end{equation}
 by integrating the 2D distribution of ($\delta\alpha$,$\delta\beta$) inside the contour of constant density that goes through the origin. The integration is performed using a Gaussian kernel density estimator. An example of this process along with more details on how $\mathcal{P}$ is calculated is shown in Appendix \ref{appB}.
 \par
 We alternatively calculate $\mathcal{P}$ using the Kolmogorov–Smirnov Test (KS Test) using in the package ndtest in Python (github.com/syrte/ndtest). For this test, the discrete values of x and y for the PDF must be used, so we sample 100 pairs of $\alpha$ and $\beta$ from the chains of two distributions, obtain 1 value from each of the pairs and compare the two using the KS Test. We obtain fully consistent results between the kernel density estimate and the KS test, and thus report $\mathcal{P}$ from the kernel density herein. \par
\begin{figure}[htb!]
  \begin{center}
\centerline{\includegraphics[width=4in]{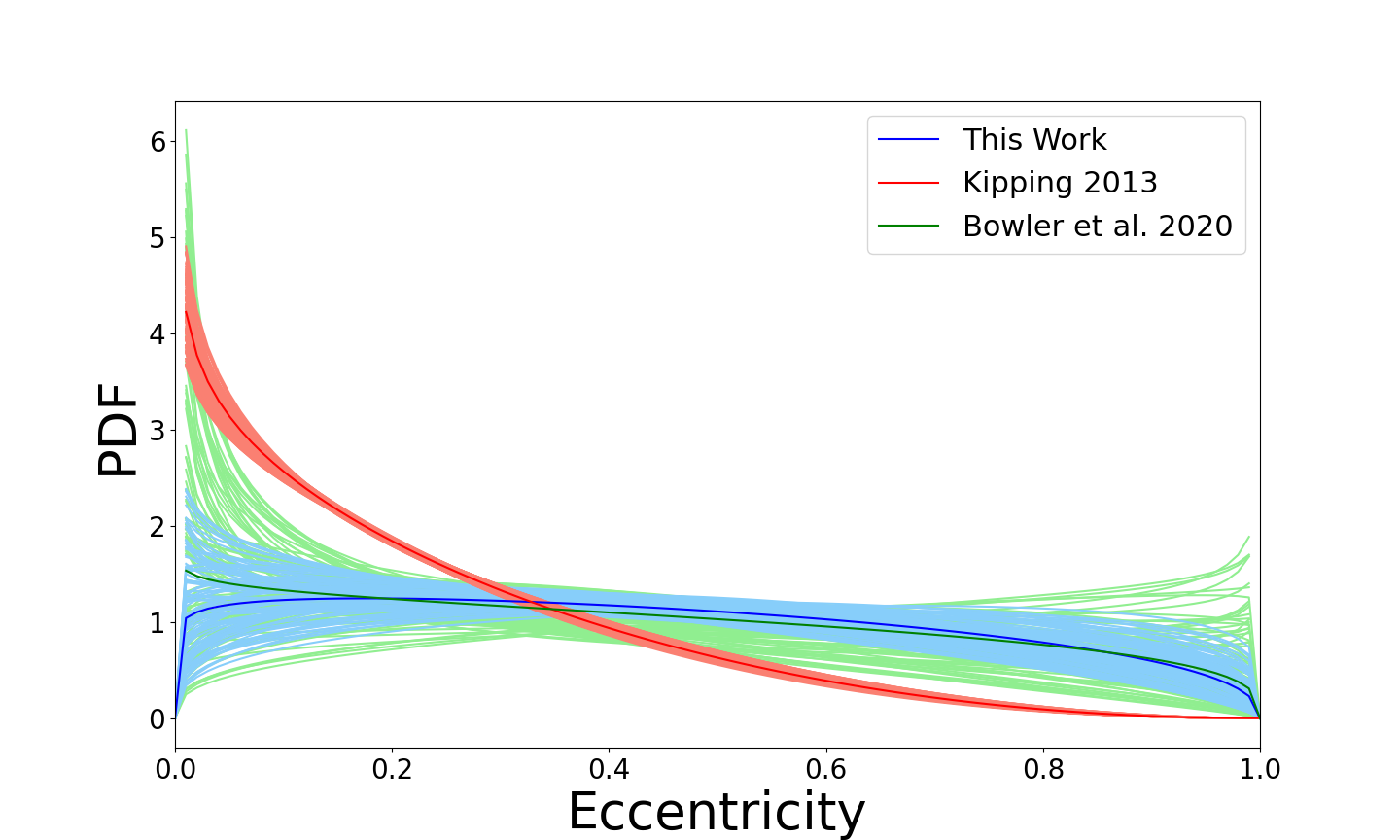}}
\caption{Comparison of the obtained beta distribution for the entire sample with the distributions of \citealt{Bowler_2020} and \citealt{2013MNRAS.434L..51K}. We note that the distribution obtained by \citealt{2013MNRAS.434L..51K} is for short period warm Jupiters, whereas \citealt{Bowler_2020}'s distribution is for long period extrasolar planets and brown dwarfs. Our sample closely resembles the one used by the latter's work, but with a lower high eccentricity incidence - likely due to the change from uniform to observable priors.}
\label{fig:comparison.png}
  \end{center}

\end{figure}

 For the distributions presented in \citealt{Bowler_2020} and \citealt{2013MNRAS.434L..51K}, we do not have the posterior chains of $\alpha$ and $\beta$ pairs so we randomly sample from uniform distributions with the $\alpha$ and $\beta$ ranges (including the uncertainties) given by these works.  When comparing our Beta distribution to the distributions found by \citealt{Bowler_2020} and \citealt{2013MNRAS.434L..51K}, we find that our distribution is consistent with \citealt{Bowler_2020}'s ($\mathcal{P}$ of 0.77) but not with \citealt{2013MNRAS.434L..51K}'s  ($\mathcal{P}$ of 0.000) (Figure \ref{fig:comparison.png}).\par
The reasons for the inconsistency with \citealt{2013MNRAS.434L..51K} could involve the fact that \citealt{2013MNRAS.434L..51K}'s distribution is for short-period exoplanets (therefore not including the long period objects that populate our sample), the lower N in our sample (21 vs. 396 objects) and the difference in detection method for the objects. \citealt{2013MNRAS.434L..51K}'s sample used RVs as the primary detection method of the companions, while we used direct imaging. \citealt{2013MNRAS.434L..51K} also has much smaller confidence intervals than both our shape parameters and \citealt{Bowler_2020}'s shape parameters. This, again, is likely due to a small confidence interval in the individual eccentricities as well as the limited sample size ($<$ 10\%) of the direct imaging sample when compared to the RV sample.\par

As mentioned in section \ref{of}, the traditional prior framework often presents a bias towards higher eccentricities. This is why, when comparing our distribution to the one obtained by \citealt{Bowler_2020}, we note that our parent distribution presents a reduced curved peak near high eccentricities, favoring slightly lower eccentricities. This could indicate that the bias towards higher eccentricities in individual orbits can also provide a bias towards higher eccentricities in the underlying parent distribution, particularly when the sample size is small. The observable-based prior aims to avoid such a bias in individual orbit fits; however, it is not completely capable of eliminating it. We also note that our low-eccentricity end presents a smaller peak (near e = 0), likely due to the fact that we did not include more than one planet from multi-planet systems (e.g. HR 8799 b,d, and e are not included) into our sample. Including those objects could cause an artificial increase in the PDF at low eccentricities since their eccentricities are correlated by being in the same system and being stable.

\subsubsection{Individual Objects with Significant Eccentricity Shifts} \label{shifts}

Some objects present a significant shift in their eccentricity distributions when comparing orbit fits from uniform priors and observable-based priors. The orbits with significant eccentricity shifts in our sample are HD 49197 b, HR 2562 b, HIP 65426 b, HD 1160 b and PZ Tel b. These 5 objects compose more than 20\% of our companion sample. 
\par
HD 49197 b had its eccentricity posterior change from $0.62\substack{+0.33 \\ -0.43}$ to $0.28\substack{+0.29 \\ -0.20}$, (weighted median of the posterior distribution (50th percentile)), with the uncertainties encompassing the 68th percentile values, favoring lower eccentricity solutions when fit with observable-based priors. HR 2562 b had its eccentricity posterior change from $0.49\substack{+0.44 \\ -0.36}$ to $0.66\substack{+0.16 \\ -0.33}$, changing from a bimodal distribution with eccentricity modes at $\approx$  0.2 and 0.9 to a fit that spans a large amount of eccentricity values but disfavors eccentricities $>$ 0.9. HIP 65426 b had its posteriors change from $0.57\substack{+0.29 \\ -0.38}$ to $0.26\substack{+0.41 \\ -0.20}$, favoring lower eccentricity values with the observable-prior fit. \par
One object that is of particular interest is HD 1160 b, one of two known companions orbiting HD 1160.  When fitting the orbit of HD 1160 b using uniform priors, there is a strong preference for higher eccentricities ($\geq$ 0.9; see Figure \ref{fig:hd1160comp}). With observable-based priors, however, the eccentricity distribution strongly shifts towards lower values. We validate the lower-eccentricity solution in two ways: (1) by checking consistency with an updated orbit fit that incorporates the two new RV measurements reported in Section \ref{rv}, and (2) by assessing the stability of the system, given its age. As discussed below, both approaches validate the lower-eccentricity solution, indicating that the observable-based prior, in this case, is doing exactly what it was designed to do: minimize the biases in the posteriors when in the prior-dominated regime.

\begin{figure}[htb!]
  \begin{center}
\centerline{\includegraphics[width=4in]{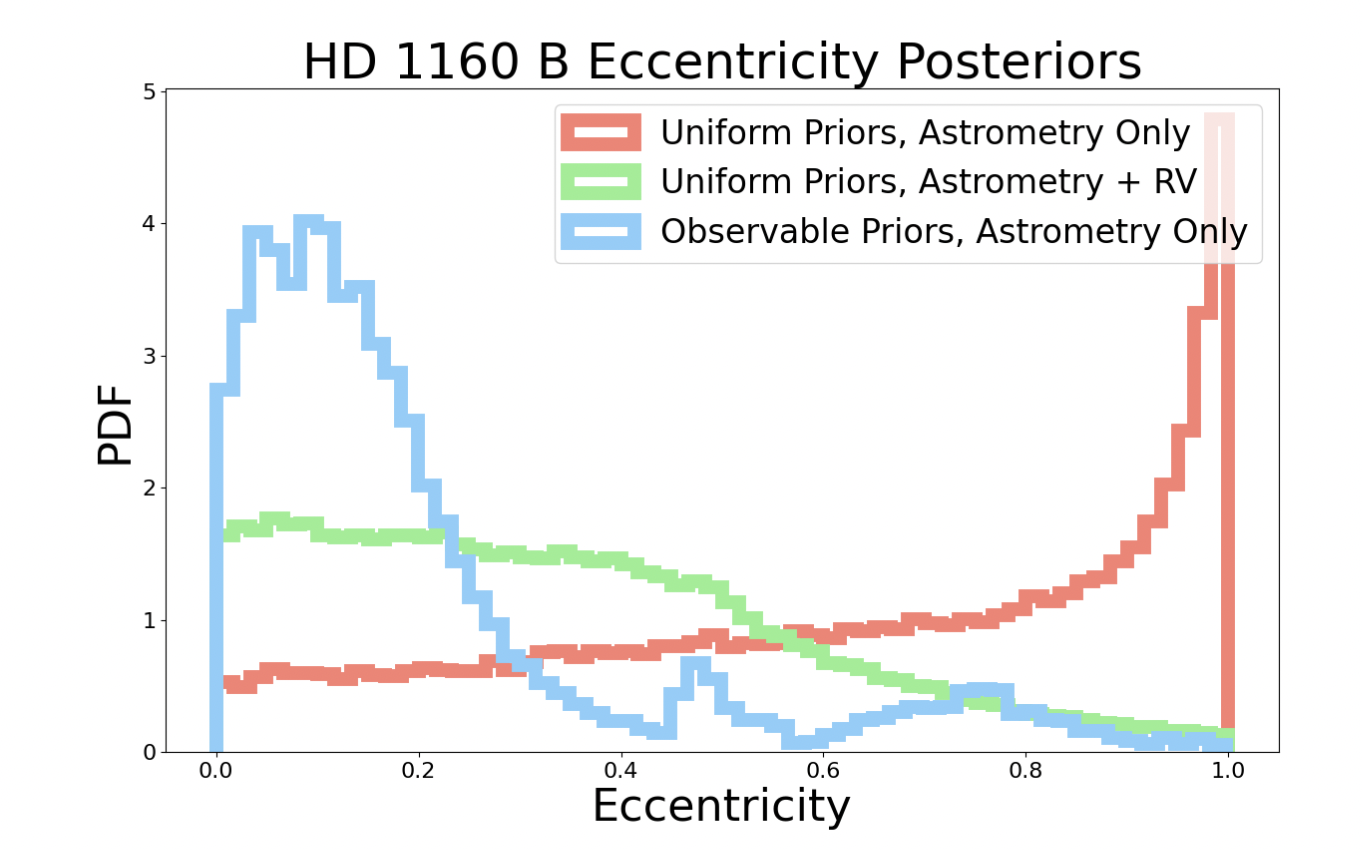}}
\caption{Comparison of the eccentricity posteriors for 3 different prior + data configurations of HD 1160 b.}
\label{fig:priorcomparisonhd1160b}
  \end{center}

\end{figure}

\begin{figure}[htb!]
  \begin{center}
\centerline{\includegraphics[width=3.5in]{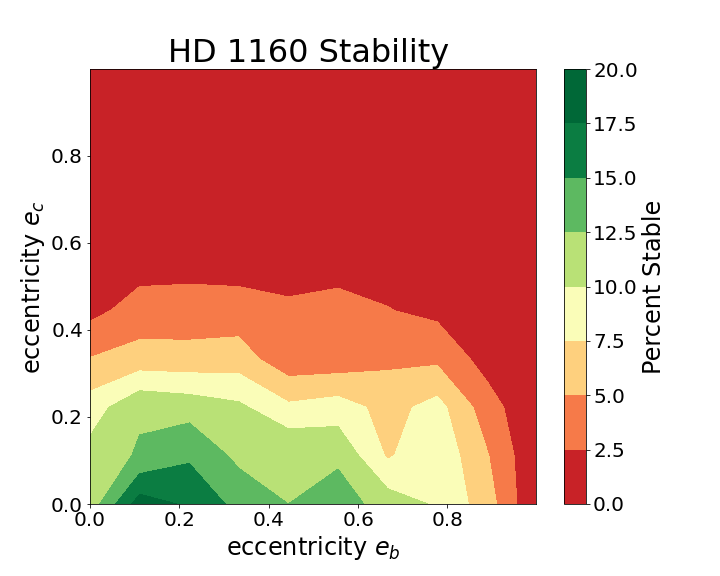}}
\caption{Stability contour for the HD 1160 system obtained by binning the companions' posteriors in eccentricity bins of 0.1 width. The colors represent percent of the 1,000 random draws for each bin that remain stable for the age of the system. HD 1160 b and c orbits are shown to disfavor very eccentric orbits by the stability constraint.} 
\label{fig:hd1160contour}
  \end{center}

\end{figure}

\begin{enumerate}
    \item 
 We fit HD 1160 b's orbit using uniform priors with orbitize!. The main difference from the previous fit is that we include two radial velociy (RV) data points, presented in Section \ref{rv}. The addition of two RV data points was sufficient to shift the eccentricity distribution, yielding the distribution presented in Figure \ref{fig:rvhd1160}. This distribution resembles the observable-based priors distribution (even without RVs) much more closely (see Figure \ref{fig:priorcomparisonhd1160b}). 

\item
To analyze whether the system would allow for such a high eccentricity for one of the companions, we ran stability simulations for the age of the system. Using the WHFast integrator on REBOUND (\citealt{Rein_2015}), we integrate the system for 120 Myr, which is the oldest estimate for the system's age given by \citealt{Garcia_2017}. We assess stability using the Mean Exponential Growth of Nearby Orbits (MEGNO) (\citealt{2003PhyD..182..151C}). Orbits that present a MEGNO $\leq$ 2 over a period of time are considered stable. We keep their masses fixed to their best fit values of 0.1 $M_\odot$ (Beas-Gonzalez et al, in prep) for HD 1160 b and 0.22 $M_\odot$ (\citealt{Nielsen_2012}) for HD 1160 c. We use the orbital posteriors for both HD 1160 b and HD 1160 c from orbitize! to assess the stability of the system. We bin the b and c posteriors into 10 bins, with each bin spanning a 0.1 spacing in eccentricity (e.g., 0.1 $<$ $e_b$ $<$ 0.2, 0.3 $<$ $e_c$ $<$ 0.4). For each bin, we draw 1,000 random orbit combinations from the posteriors and assess whether they remain stable for the age of the system. This yields 100,000 possible orbit combinations from the objects' posteriors, with 1,000 per bin (and a total of 100 bins). We then assess how many orbits in each bin remain stable using the MEGNO parameter. We do this for two different posteriors for HD 1160: with and without the RVs. \par
We present a contour plot of the 100,000 possible orbit combinations of HD 1160 b and c, with the contours representing what percent of the random draws remained stable for the age of the system. The contour is shown in Figure \ref{fig:hd1160contour}. The figure illustrates that HD 1160 b is unlikely to have eccentricity values $>$ 0.9, which was a value favored by the uniform prior fit without radial velocity measurements. When we test this simulation set but with HD 1160 b's orbital posteriors without RVs, we obtain an equivalent plot, showing that the majority of solutions for HD 1160 b and c disfavor very eccentric orbits for both objects.

\end{enumerate}

\begin{figure}
  \begin{center}
\centerline{\includegraphics[width=4in]{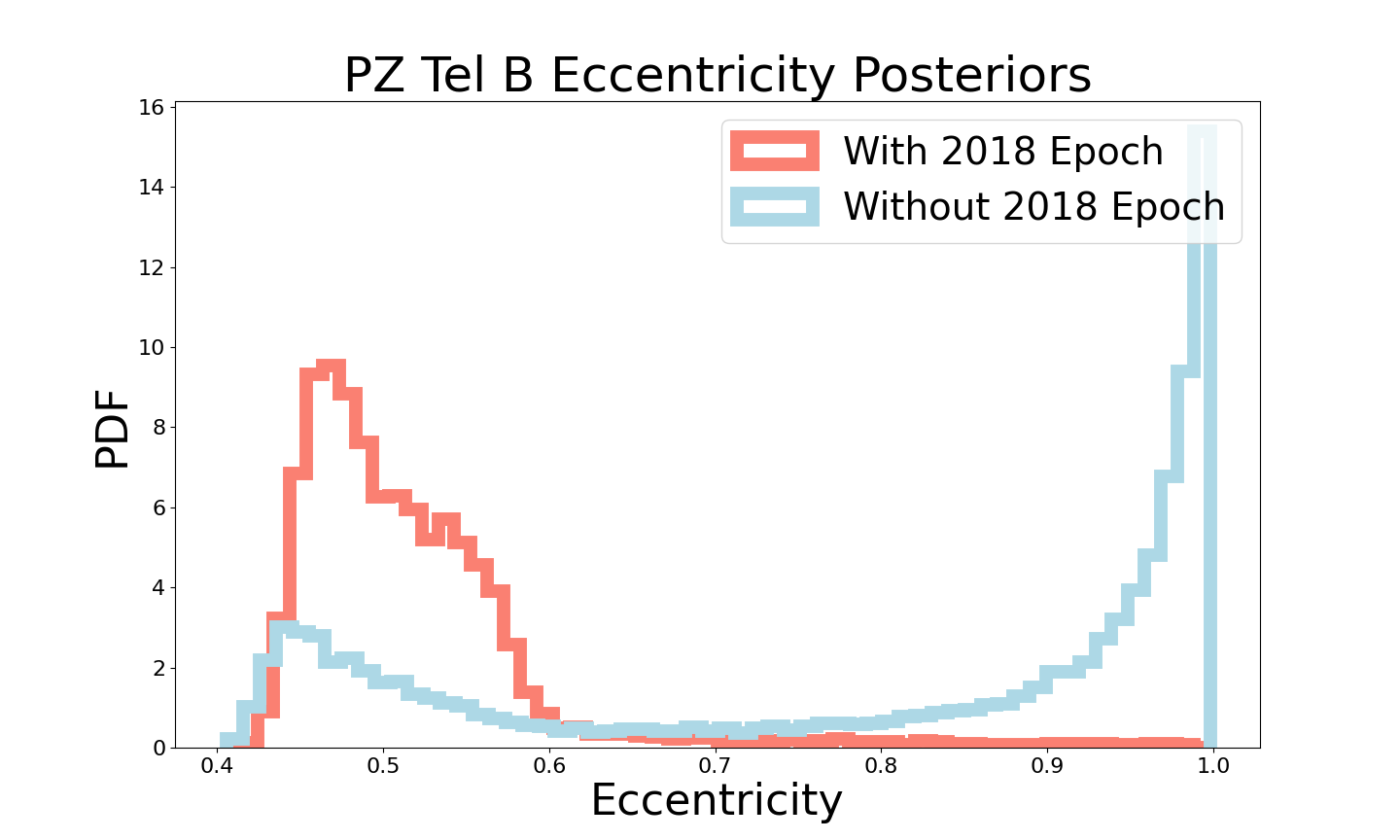}}
\caption{Eccentricity posteriors of PZ Tel b with (red) and without (blue) the 2018 epoch.}
  \label{fig:pztelhist}
  \end{center}

\end{figure}

For PZ Tel b, a new astrometry point was sufficient to also change the eccentricity distribution of the object. The 2018 epoch added to our fits is shown in Table \ref{tbl:1}.  Prior to 2018, works that presented orbit fits for this object indicated that it had a very high eccentricity (e.g.,\citealt{Gin2014}, \citealt{Beu2016}, \citealt{Mai2016}). The new astrometry point for PZ Tel b, obtained in 2018, changes the orbital period coverage of the companion from $3.9\substack{+1.9\\ -1.3}$ \% to $6.2\substack{+2.9\\ -2.1}$ \%, calculated from the weighted median of the posterior distribution (50th percentile), with the uncertainties encompassing the 68th  percentile of our orbit fits for the period and the current astrometric coverage increase from 7 to 11 years. This new point's impact on the distribution implies that there could be a minimal orbital period coverage needed in order to obtain eccentricity posteriors that are meaningful for directly imaged planets. Our posteriors obtained with observable-based priors with and without the 2018 epoch are presented in Figure \ref{fig:pztelhist}.

\subsubsection{Leave-One-Out Cross Validation} \label{leaveout}
Given the small sample size of directly imaged companions, there is a possibility that a single object with a narrow eccentricity posterior may noticeably impact the resulting population distribution.  To test for this, we analyze the impact of removing one single object from the sample for each of the 21 objects. We do that by re-running the beta distribution fit on the entire sample, but excluding one object at a time. We characterize a large impact as a change in expected $\alpha$ and $\beta$ parameters for the underlying distribution of eccentricity that falls outside of the 1-$\sigma$ confidence level for the parameter values shown in Section \ref{wholesample}. \par
We find that removing any single object from our population does not yield $\delta\alpha$ and $\delta\beta$ to be outside of the 68\% confidence interval using $\mathcal{P}$ defined in the previous Section. We therefore conclude that our sample does not have obvious outliers.
The object which produces the most different distribution is 1RXS0342+1216 b, which is a high mass companion with eccentricity of $0.94\substack{+0.04 \\ -0.11}$ (given from our orbit fit's 68th percentile, with the central value being the weighted median of the posteriors). Even in this case, we estimate that the two distributions have parameters that are consistent with a $\mathcal{P}$ of 0.50.

\subsubsection{Separating the Sample by Companion Mass} \label{planetsvsbd}

We also separate the sample in two populations based on the mass of the companion. The goal is to identify any possible differences in the eccentricity as a function of mass.  Such a distinction could be correlated with the masses expected for gas giant planets (companions with mass $\leq$ 15 $M_J$) and brown dwarf companions (companions with mass $>$ 15 $M_J$). If differences in eccentricity distributions are found between the populations, eccentricities can potentially be used to constrain formation mechanisms.  The complexity here is the use of a mass boundary for distinguishing between populations.  If the definition of a planet versus a brown dwarf is related to formation, it is unlikely that 15 $M_J$ is a meaningful boundary (e.g., \citealt{Bodenheimer_2013}; \citealt{Mordasini_2012}).  Additionally, most of the masses assumed for these companions are from evolutionary models and rely on uncertain properties such as system age (e.g., \citealt{Carson_2013}; \citealt{Hinkley_2013}).  Even with these caveats, it is still beneficial to look for differences between more and less massive companions in the context of distinction between these populations. \par
 \begin{figure*}
  \begin{center}
\centerline{\includegraphics[width=5in]{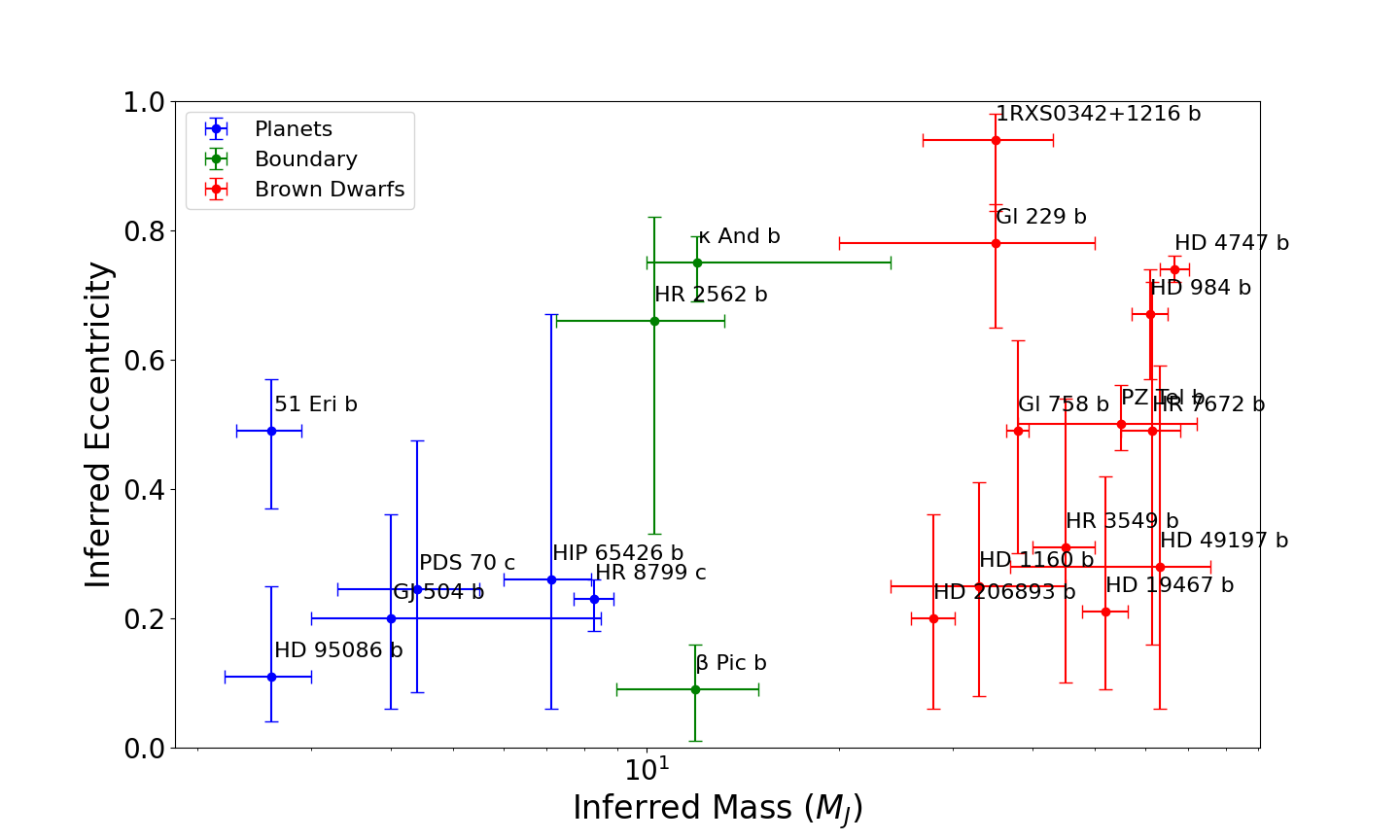}}
\caption{The mass of the companions plotted against the eccentricity of the companions. Mass estimates are from the literature, while eccentricity estimates are from this work. Error bars on the eccentricity represent the 68th percentile of the orbit fit. The blue dots represent ``planets", or objects under 15 $M_J$, while red dots represent ``brown dwarfs", or objects with masses above 15 $M_J$. Green dots are ``boundary" objects: intermediate mass objects that could be in either distribution.}
\label{fig:planetsvsbds}
  \end{center}
\end{figure*}

\begin{deluxetable*}{ccc}
\tablecaption{Consistency Between Planets vs. Brown Dwarfs Distributions  represented as $\mathcal{P}$ (P-value)} \label{tbl:confidencelevels}
\tablewidth{20pt}
\tablecolumns{2}
\tabletypesize{\scriptsize}
\tablehead{\colhead{  } & \colhead{$\kappa$ And b (Planet)} & \colhead{$\kappa$ And b (Brown Dwarf)}}
\startdata
HR 2562 b (Planet) &  0.373 &  0.001 \\
HR 2562 b (Brown Dwarf) & 0.139  & 0.000 \\
\enddata
\tablecomments{Impact of the classification of intermediate mass objects on the consistency between the planet and brown-dwarf populations. The consistency is measured as $\mathcal{P}$ measuring the similarity of the estimated parameters $\alpha$ and $\beta$ that define the Beta distribution. A $\mathcal{P}$ $>$0.68 would mean that the two distributions are consistent, while $<$0.01 means that the two distribution are likely different. For this table, $\beta$ Pic b is considered a planet, although considering it a brown dwarf produces no significant changes in $\mathcal{P}$.}
\end{deluxetable*}
When classifying these objects into the low-mass and high-mass categories, the intermediate mass objects at the boundary, namely $\beta$ Pic b, HR 2562 b and $\kappa$ And b, could have a strong impact on the final result. Thus, we re-run our ``planet" vs. ``brown dwarf" population placing them in either of the populations to analyze whether the eccentricity distribution of each population changes. The mass estimates are plotted against the eccentricity estimates (68th percentile) in Figure \ref{fig:planetsvsbds}. The mass estimates and respective references are shown in Table \ref{tbl:masses}.  \par
Including all of the boundary objects in the planet sample makes it comprised of 9 companions. When none of them are in the sample, the planet population is comprised of only 6 objects. This low sample size can cause an outlier to significantly skew the eccentricity of the entire population. \par
\begin{figure*}
    \centering
    \centering{{\includegraphics[width= 12cm]{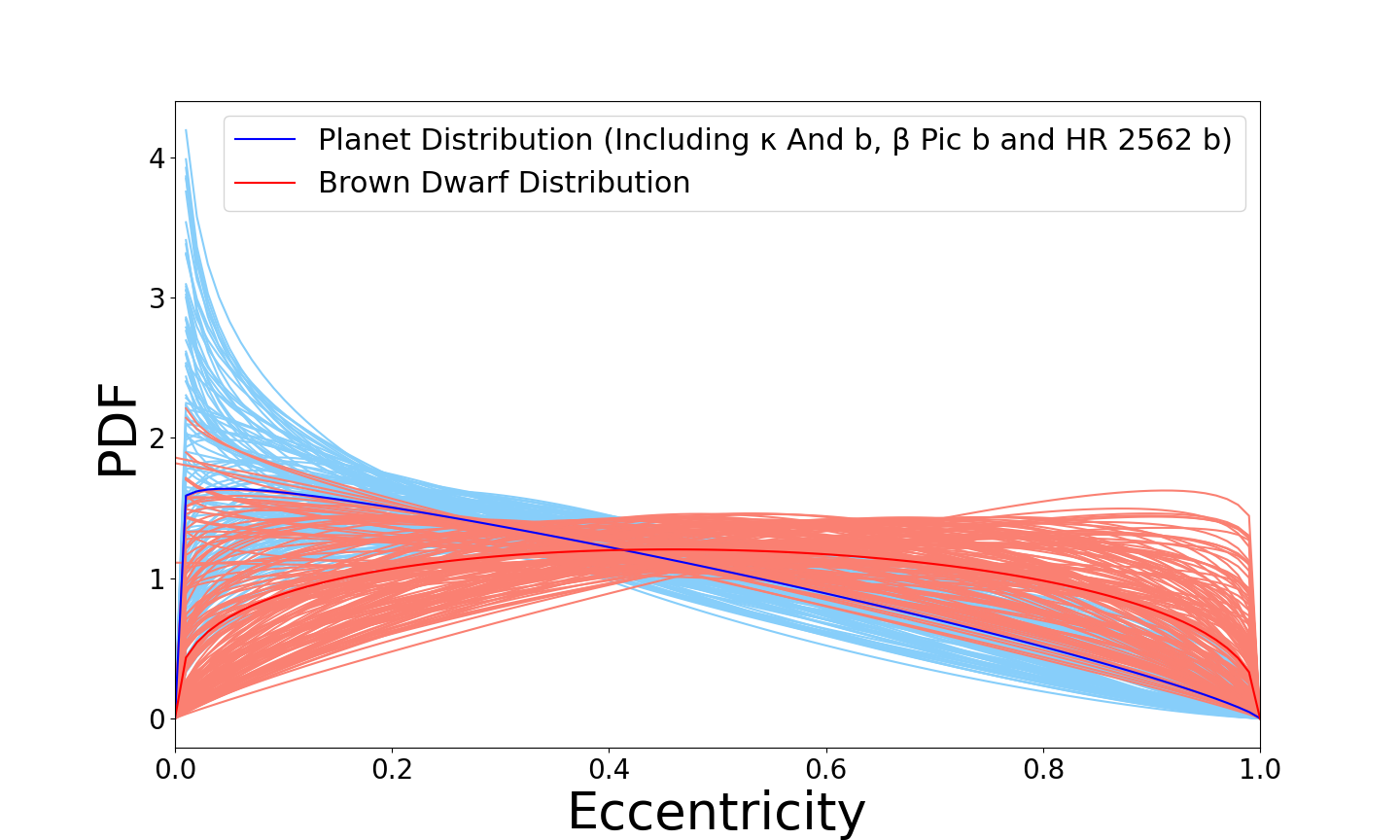} }}%
    \qquad
    \centering{{\includegraphics[width=12cm]{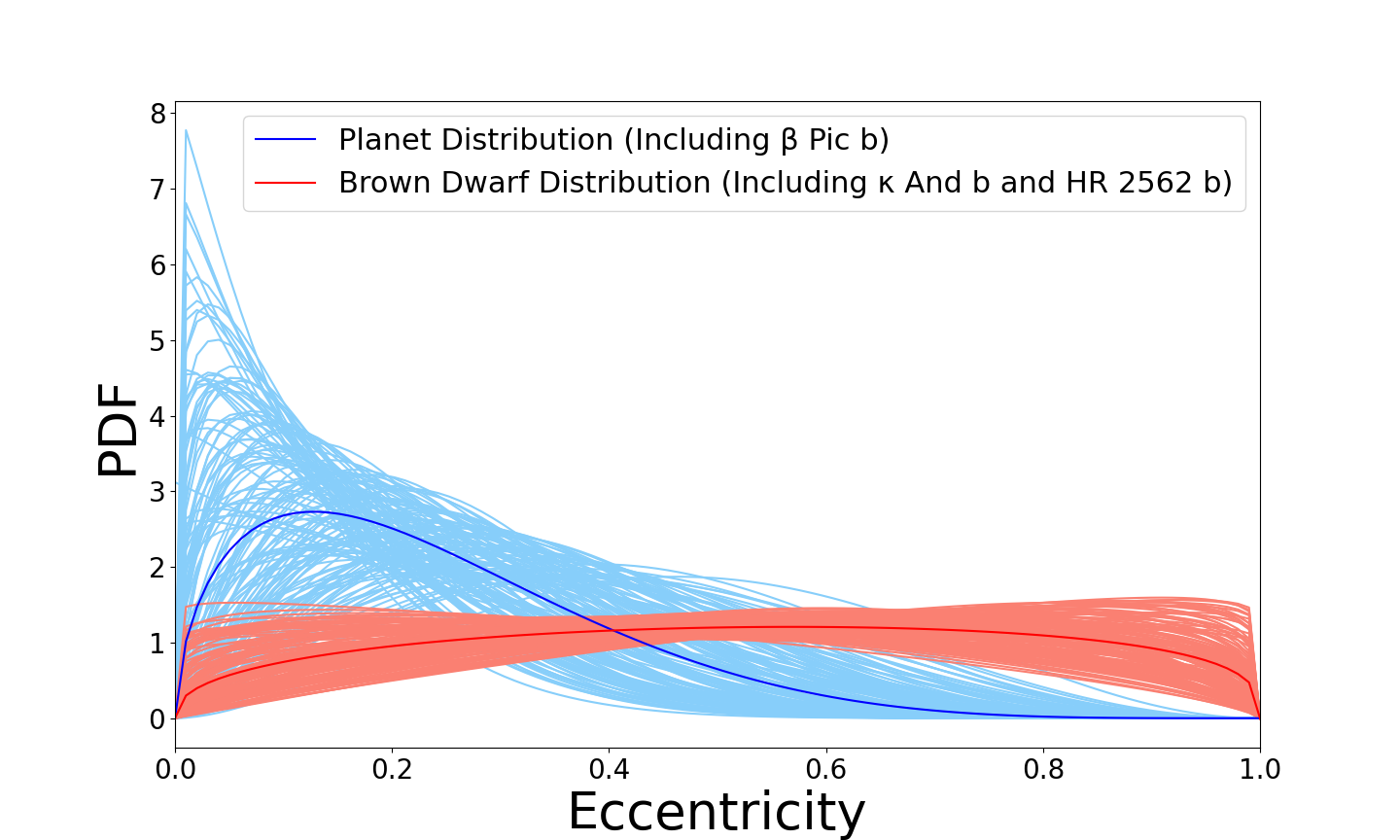} }}%
    \caption{Comparison of ``Planet" vs ``Brown Dwarf" distributions for different combinations of intermediate mass objects: for our highest $\mathcal{P}$ case (i.e., most similar distributions) and lowest $\mathcal{P}$ case (i.e., most different distributions).   }%
    \label{fig:confidencecomparison}
\end{figure*}

When switching $\beta$ Pic b from the planet to brown dwarf population, we find no significant change in our $\mathcal{P}$ values that measure the consistency between the two populations. This is not the case for $\kappa$ And b and HR 2562 b, in particular the former. Table \ref{tbl:confidencelevels} represents our Consistency Parameter (defined as the $\mathcal{P}$ value) for different combinations of these objects' classifications. $\kappa$ And b, when considered a brown dwarf, produces distributions of planets and brown dwarfs that are different from each other. We illustrate two of these extreme cases (highest $\mathcal{P}$ and lowest $\mathcal{P}$) in Figure \ref{fig:confidencecomparison}. The ``extreme" cases yield, respectively, planet population parameters of $\alpha$ = $1.04\substack{+0.41 \\ -0.28}$\ and $\beta$ = $1.82\substack{+0.58 \\ -0.42}$ vs $\alpha$ = $1.59\substack{+1.26 \\ -0.60}$\ and $\beta$ = $5.03\substack{+3.73 \\ -2.01}$. Their brown dwarf distributions are parameterized by $\alpha$ = $1.33\substack{+0.64 \\ -0.39}$\ and $\beta$ = $1.39\substack{+0.54 \\ -0.34}$ vs $\alpha$ = $1.41\substack{+0.62 \\ -0.40}$\ and $\beta$ = $1.31\substack{+0.43 \\ -0.29}$.. It is clear from Figure \ref{fig:confidencecomparison} and Table \ref{tbl:confidencelevels} that a small portion of objects can shift whether the two populations have similar distributions or not. This is likely due to the small sample size both for the entire population but especially for the sub-samples separating high and low mass objects. \par
We also test the mass separation distributions completely omitting the three boundary objects, and find planet and brown dwarf populations that are similar to Figure \ref{fig:confidencecomparison}, bottom panel. However, the low sample size of the population (only 6 planets, none of which have eccentricities above $\approx$ 0.7 in their posteriors' 68th percentile) is significantly affected by a single object with higher eccentricity, such that our population distribution derivation is prone to higher uncertainties and to being skewed by any outliers in the sample.\
Given these results and the uncertainties and caveats around masses, we conclude that our current sample and its components' individual eccentricity distributions do not provide enough constraints to affirm that different formation pathways are underway for sub-stellar companions above and below $\sim$15 M$_{Jup}$.

\subsubsection{Separating the Sample by Companion Separation} \label{sepcomps}

\begin{figure*}
  \begin{center}
\centerline{\includegraphics[width=5in]{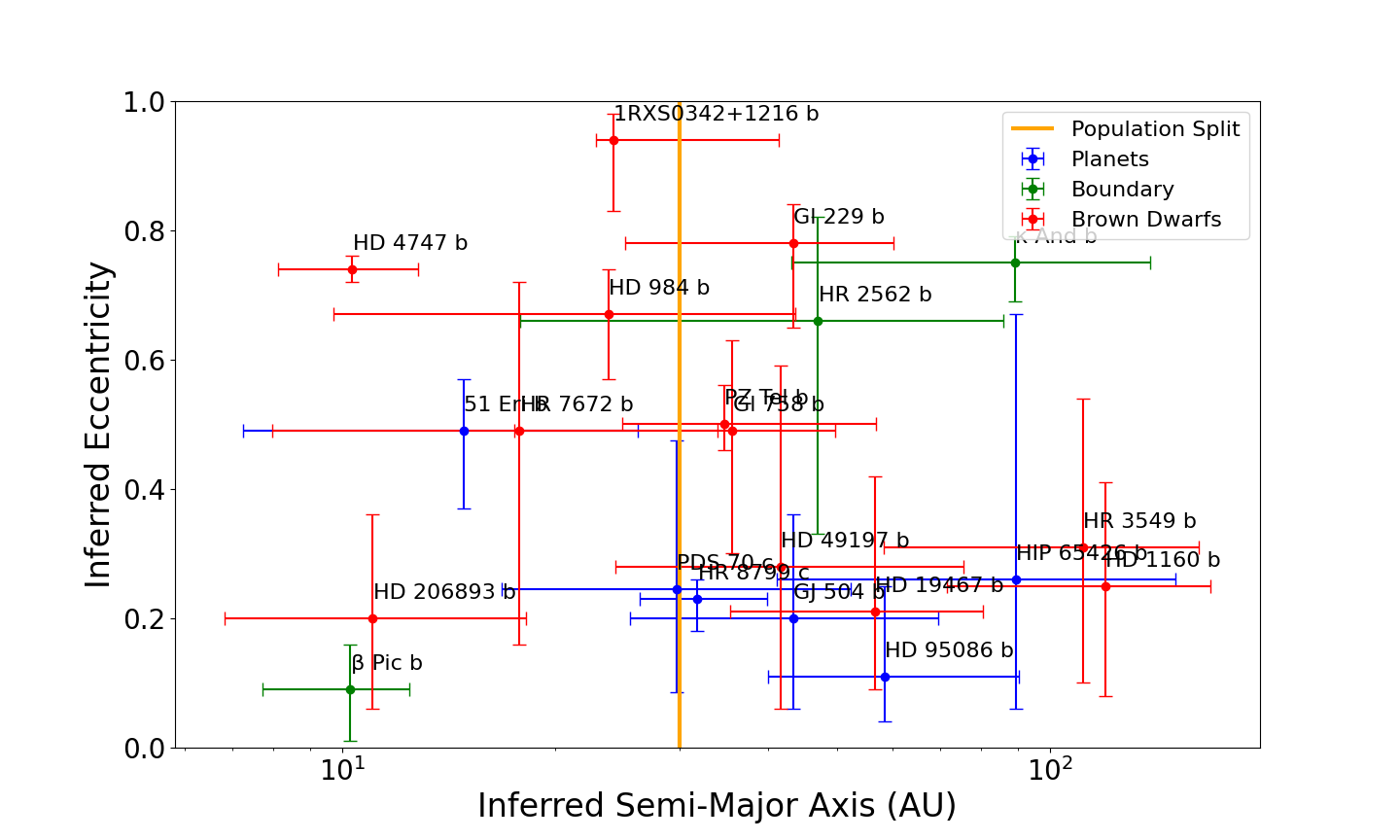}}
\caption{The semi-major axis plotted against the eccentricity of the companions. Semi-major axis values are calculated from the 68th percentile of our orbital fits. Error bars on the eccentricity represent the 68th percentile of the orbit fit. The blue dots represent ``planets", or objects under 15 $M_J$, while red dots represent ``brown dwarfs", or objects with masses above 15 $M_J$. Green dots are ``boundary" objects: intermediate mass objects that could be in either distribution. In orange, we plot the separation cutoff where we split our sample between ``Close" and ``Wide" separations.}
\label{fig:separationvseccentricity_scatter}
  \end{center}
\end{figure*}

We also test splitting our companion sample by separation from the host star. The interest in performing this comparison comes from the possible distinction between core accretion and gravitational instability companions, since gravitational instability models tend to favor formation further away from the host star  (e.g. \citealt{Mayer_2004}), while core/pebble accretion companions are most efficiently formed closer to the host star (e.g. \citealt{DodsonRobinson_2009}). The 68th percentile eccentricity posteriors are plotted against the separation of the companions, in AU, in Figure \ref{fig:separationvseccentricity_scatter}. \par

\begin{figure*}
  \begin{center}
\centerline{\includegraphics[width=5in]{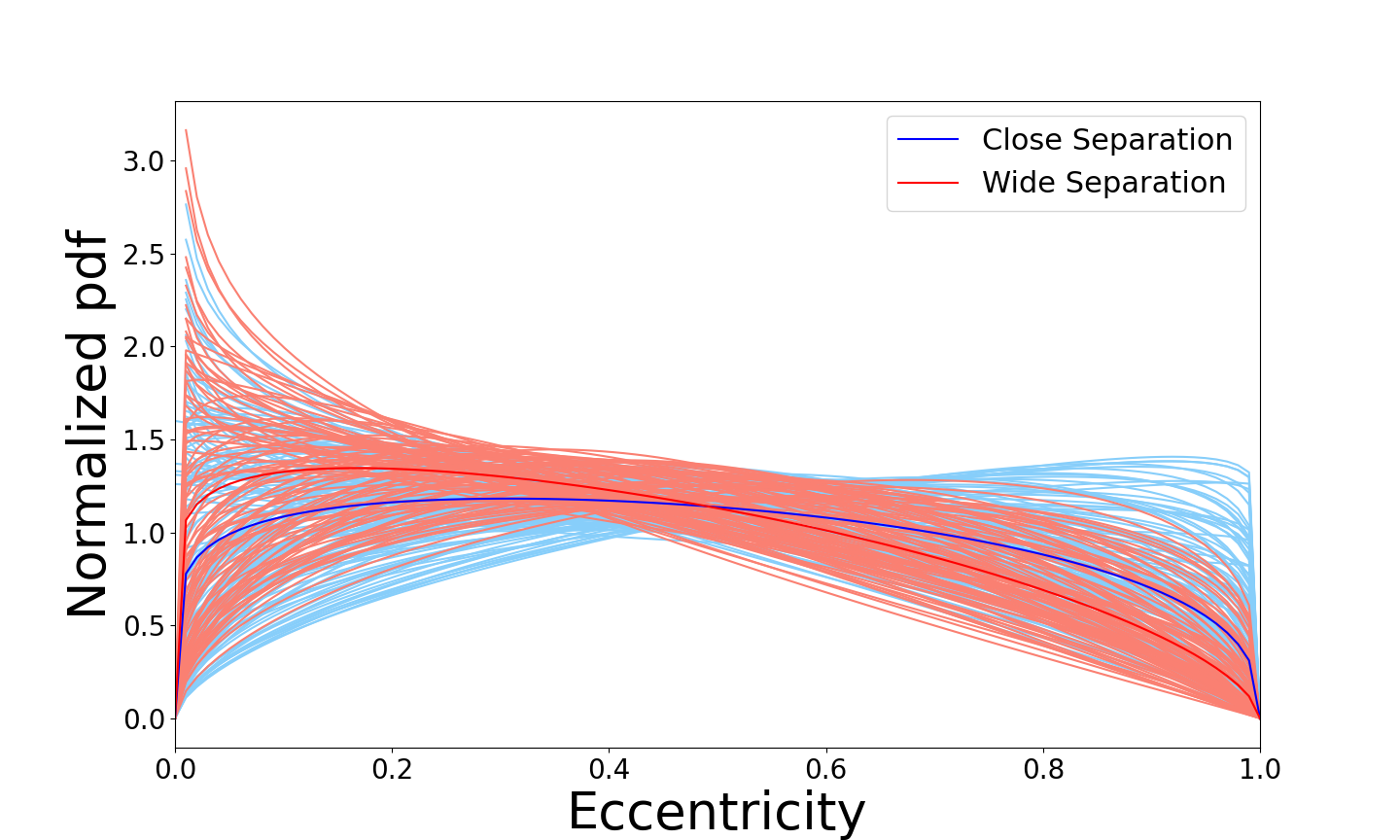}}
\caption{The ``Close" vs ``Wide" separation companion populations, with the split happening at 30 AU. The two distributions are consistent with each other, with a $\mathcal{P}$ of 0.85.} 
\label{fig:separationvseccentricity_pop}
  \end{center}
\end{figure*}

We also test splitting the population into ``Close" and ``Wide" separation companions, with the cutoff of 30 AU. This cutoff allows the companion sample to be split $\approx$ in half between the 5-30 AU and 30-100 AU bins. It is also close to the 35 AU threshold set by \citealt{DodsonRobinson_2009}, beyond which the critical mass obtained by core accretion objects cannot occur and where gravitational instability becomes a more efficient companion formation mechanism. We find that the two populations are consistent with each other, with a $\mathcal{P}$ of 0.85. Their distributions are shown in Figure \ref{fig:separationvseccentricity_pop}.

\subsection{Minimal Orbital Coverage Simulations} \label{simres}

\begin{figure}
  \begin{center}
\centerline{\includegraphics[width=4.5in]{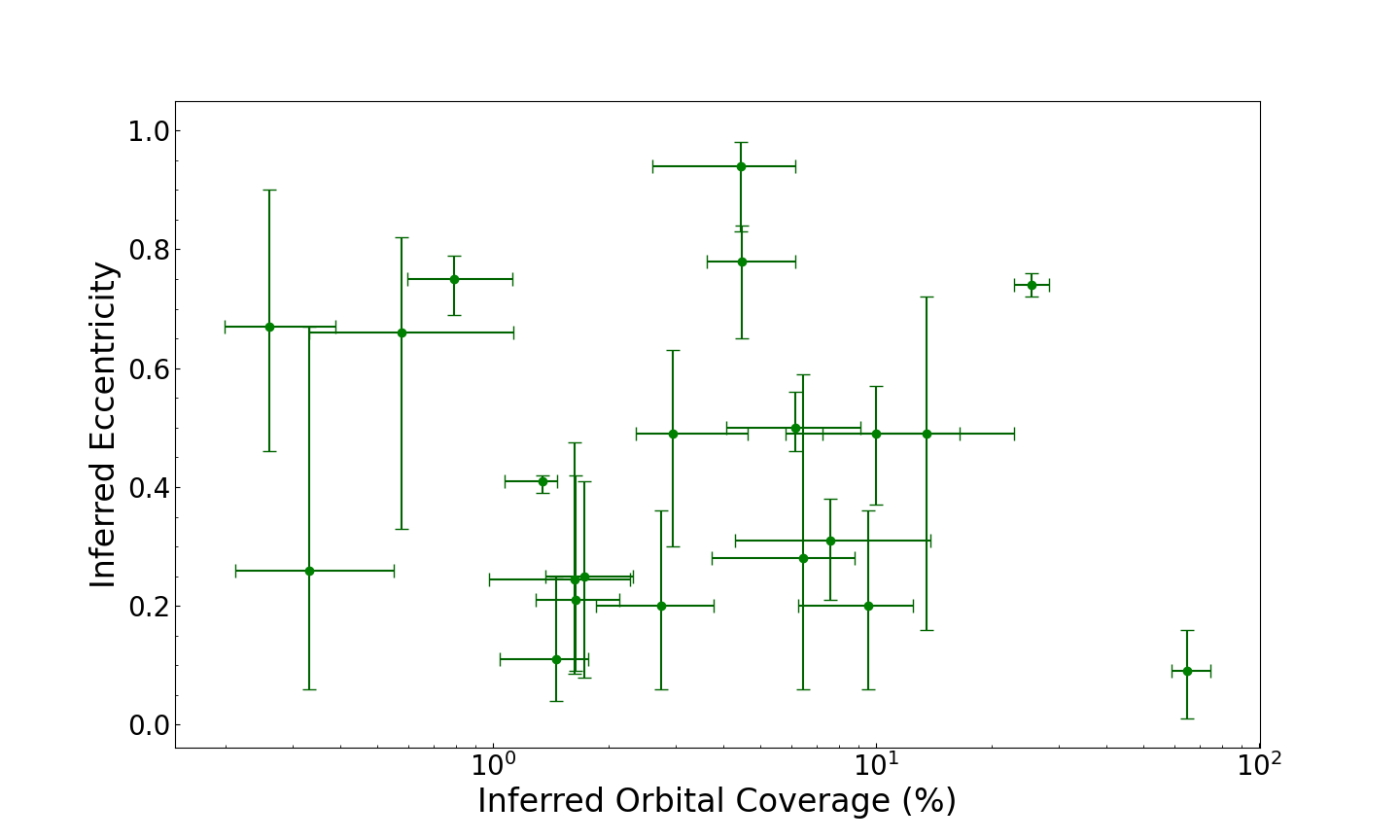}}
\caption{The inferred orbital coverage of each object in the sample, calculated as a fraction of the astrometric coverage and the period fit, plotted against the eccentricity of the object. This is from the real data used in our sample.  Most of the sources have less than 10\% orbital period coverage, with 4 objects presenting less than 1\% period coverage.}
\label{fig:datacov}
  \end{center}
\end{figure}

Given the major shift in the eccentricity distribution of PZ Tel b given the addition of new data (Section \ref{shifts}), we simulate how much orbital period coverage we need in order to obtain a meaningful posterior distribution for eccentricity. The main goal of this analysis is to stipulate whether we can obtain meaningful results on these objects with the currently available astrometry. For context given the sample we have been exploring, we examine the best estimate for the orbital coverage as a fraction of the period (in \%), plotted against the best estimate for the eccentricity of each source, with error bars encompassing the possible values presented in each individual eccentricity distribution. We obtain the orbital coverage by dividing the current astrometric coverage for each object by its inferred orbital period. The results are presented in Figure \ref{fig:datacov}.  The average orbital period coverage for the sources in our sample is 7.4\%, calculated from weighted median values of our sample's orbital period fits and the astrometric coverage, in years, for each object in the sample. \par

In order to assess the minimum required orbital phase coverage, we simulate data to indicate how an object's orbital configuration affects the amount of coverage needed to obtain reliable posteriors.   We perform this simulation in hopes of finding a trend for how much information must be obtained for each system such that a meaningful posterior can be extracted - and, in turn, a meaningful eccentricity underlying parent distribution can be obtained. \par
First, we simulated orbits with properties that are representative of the average object in our sample. We define
a period of 200 years, and inclination of 60 degrees and periastron passage occurring in 2150. The total system
 mass is fixed at 0.68 $M_\odot$. We vary the eccentricity of a given orbit to see how its value affects the posteriors for
a given orbital period coverage. We calculate astrometry points based on these orbital parameters, starting in the
year 2021. We generate astrometry points in order to simulate increasing orbital phase coverage. The
astrometry ``time step'' represents 1\% of the period of the orbit. The astrometry encompasses uncertainties on the
order of milliarcsecond precision, as is typical for high contrast imaging instruments (\citealt{Konopacky_2014}). To
simulate instrument differences and other systematic factors, the simulated astrometry points also have noise randomly
sampled from a Gaussian distribution centered at 0 and with the width of the astrometric uncertainty. This simulated
astrometry is used to run orbit fits analogous to those in the rest of our analysis. We run our observable-based prior
and uniform prior orbit fitter 100 times for each orbital eccentricity value, increasing the astrometry and hence the
phase coverage with each successive run. In order to consider a simulation ”successful”, the real input eccentricity, To,
inclination and period must be within the 68\% confidence interval given by the fit.

 \begin{figure*}
    \centering
    \centering{{\includegraphics[width= 14cm]{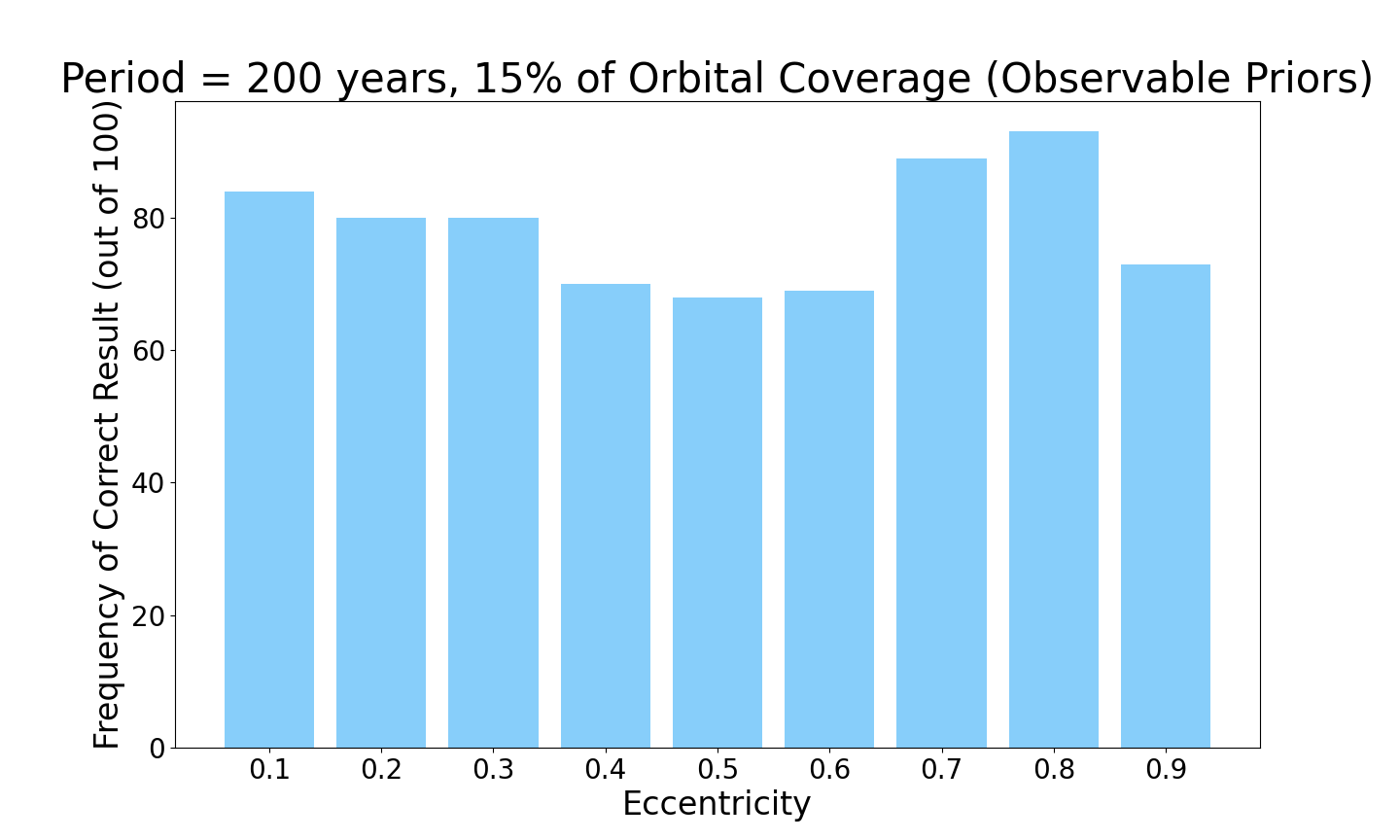} }}%
    \qquad
    \centering{{\includegraphics[width=14cm]{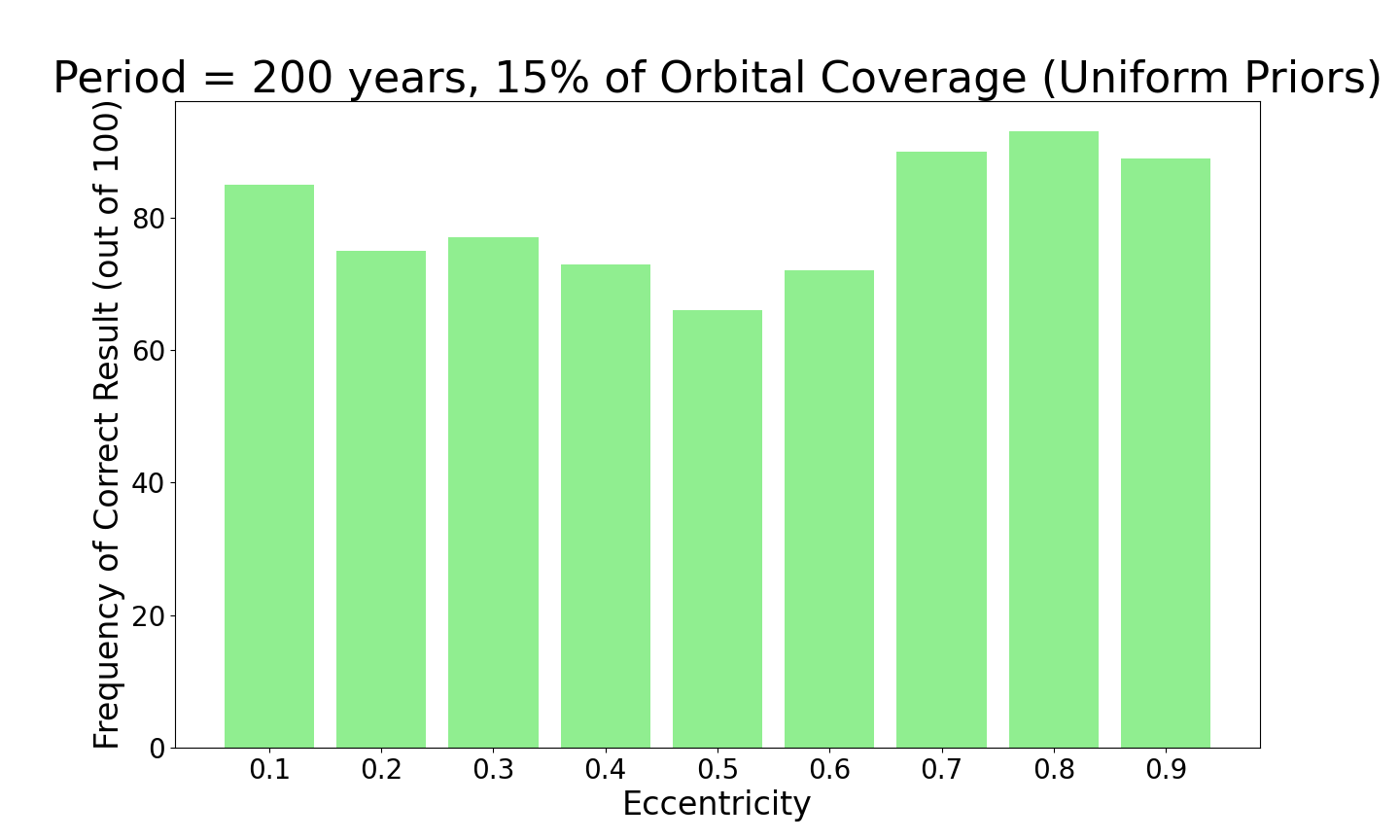} }}%
    \caption{ The bar plot of how many successful tries (i.e., true parameters are within 1 $\sigma$ of posterior distribution) we obtain as a function of eccentricity for observable and uniform priors. For 15\% of the orbital period in coverage, all of the eccentricity values from 0.1 to 0.9 have over 68 of the 100 trials considered successful.}%
    \label{fig:barplots}
\end{figure*}
We find that the orbital coverage of 15\% was the minimum value needed to obtain 68 successful posteriors (out of 100 trials) for all eccentricity value variations, both for observable and uniform priors. This encompasses 30 years of observations from 2021 to 2051 for this specific orbit.  The orbital coverage value is defined differently from the \citealt{Lucy2014} 40\% minimum orbital phase coverage, since here it is based off of the period of the orbit and not the orbital arc. Thus the 15\% period coverage is dependent on an orbital arc where the planet does not go through periastron passage during observation, which should be the case for the majority of directly imaged planets, as the planet spends the least amount of time of its orbit at periastron. The phase coverage obtained by this specific orbit spans $\approx$ 3 - 15\% of the orbit's phase (depending on eccentricity, which was varied in the interval from 0.1 to 0.9 in steps of 0.1), which is less than the value obtained by \citealt{Lucy2014}.  Another difference is that in \citealt{Lucy2014} the total system mass was a free parameter in the fit, while for our case (as is standard for most direct imaging orbit fits) the mass was kept fixed. Having the total system mass as a free parameter is likely a reason as to why \citealt{Lucy2014}’s phase coverage requirement is larger than the ones obtained with our simulations. We also test the needed orbital coverage for 100-year and 400-year period orbits and find no dependence on period as long as the orbital period coverage remains the same (at least 15\% of period). Note that this is a rule of thumb only - some orbits may require more coverage than this, while others may require less. The 15\% coverage number found here represents a minimum point at which the orbit posteriors for directly imaged planets should realistically be used to derive population distributions. Our results are presented in Figure \ref{fig:barplots}.  \par

For the median period of objects in our sample ($\approx$ 251 years), we would need 38 years of data to begin to have reliable posteriors. Looking at specific example cases, HD 19467 B has an estimated period of $426.14\substack{+115.89\\ -98.13}$ years (weighted median of the fit, with uncertainties spanning the 68th percentile of orbital fit) but our astrometry only covers 7 years (or $1.6\substack{+0.5\\ -0.4}$\%), calculated from the 68th percentile of the fit, of the estimated orbital period - which means that our parameter posteriors for this object may not be reliable. This result is also exemplified in Section \ref{shifts}, where PZ Tel b's orbital coverage increase from $3.9\substack{+1.9\\ -1.3}$ completely shifted the eccentricity distribution of the companion. However, given that we still have less than 15\% coverage for PZ Tel b, we might expect additional changes in the posteriors as more data is collected. \par

\section{Discussion} \label{disc}

In this work, we use observable-based priors to fit for the orbital parameters of 21 directly imaged substellar companions with updated astrometry and radial velocity measurements from the literature. Revisiting the analysis done by \citealt{Bowler_2020} with a different set of priors, the goal is to obtain the eccentricity distribution of the entire population and analyze if there is any difference in the eccentricity distributions of planet and brown dwarf populations. Differences in these populations' eccentricity distributions could potentially be signs of different formation mechanisms: with planets mainly forming from protoplanetary disks and thus presenting low eccentricities, and brown dwarfs forming similarly to binary star systems (for instance via disk fragmentation) and therefore spanning all values of eccentricities (\citealt{Bowler_2020}).
\par
\subsection{Choice of Priors}
The vast majority of exoplanets and brown dwarfs in our directly imaged population have long periods, due to their fairly large distance from the host star. Consequently, their orbital arcs covered by current astrometry span a small fraction of their orbital periods, yielding orbital fits that are severely undersampled. Such undersampling of data can cause all results of orbit fitting to be in a prior-dominated regime. Choosing uniform priors for all orbital parameters results in equal weighting between all possible phases of the orbit.  This is demonstrably not the case from basic orbit physics, which dictates that the planet presents the highest orbital speed during periastron passage ($T_o$).  Thus the probability of catching these long period companions right at $T_o$ is low, since it spends the least amount of time there throughout its orbit.  In the case of well-sampled data covering a large fraction of the orbit, the information contained in the data itself is sufficient to overcome the choice of prior.  However, in the case of minimally sampled data, which corresponds typically to linear motion of the planet without measureable higher order motion such as acceleration, landing in the prior-dominated regime can result in unintended effects. One such example is that astrometric systematics could be interpreted by the fitter as rapid acceleration, hence leading to a best fit that is near periastron.  This known bias caused by using uniform priors warranted a different approach to priors when it came to orbit fitting. Observable-based priors (\citealt{ON19}) provided such an approach, where the uniformity was in the orbital observables rather than in the orbital parameters. These priors decrease the bias towards high eccentricities and periastron passage during observations, effectively by down-weighting the likelihood of an orbit fit that has $T_o$ near the time of observations.  While it is the case that proper approaches to prior definition in Bayesian statistics is a matter of debate, it is important to recognize that in the prior-dominated regime, as we are here, prior choices have measurable consequences in the posteriors and mitigation of biases is desirable.
\par Differences in eccentricity posteriors of some of the objects in our sample are indeed noticed when it comes to comparing these two different priors - in particular for HR 2562 b, HD 1160 b, HD 49197 b, HIP 65426 b and PZ Tel b (for corner plots of these objects in comparison to uniform priors, see Appendix \ref{appA}). In some cases, the long tail of high eccentricities completely disappears (e.g. HD 1160 b) when using observable priors. This is particularly important given our simulation showing that the system only remains stable if HD 1160 b has an eccentricity of $<$ 0.9. Indeed, when radial velocity data is added to HD 1160 b's orbital fit with uniform priors,  this is validated - we see a significant shift in its eccentricity posteriors to lower eccentricities, more consistent with what observable priors obtained with an astrometry-only solution.  Stability arguments have been successfully used in other cases to help inform the orbital parameters of directly imaged planets, and represent a powerful means of constraint beyond astrometry alone (e.g., \citealt{Wang_2018}). \par
\subsection{Population Results}
Given the major shift in many of the posteriors using observable-based priors, we expected some change in the population distribution from the combined dataset.  While the specific values of $\alpha$ and $\beta$ are different from \citealt{Bowler_2020}, the overall result is very similar  - a nearly uniform distribution, but with a lower incidence of high eccentricity orbits. Since we did not include all planets from the same multi-planet system in the sample (but rather chose one as the system's ``representative") due to the likely preference that multiple planet systems have for lower eccentricities due to stability requirements (\citealt{WrightHoward2009}), it is possible that inclusion of these objects (such as HR 8799 b,d,e) could have yielded a population eccentricity distribution with a stronger preference for lower eccentricities.  Overall, since the distributions are statistically consistent with each other, we can conclude that current data yields an eccentricity distribution that is largely flat with eccentricity.  \par

\subsection{Planets and Brown Dwarfs}
Given the fact that the sample size is small and some of the posteriors are prior-dominated, care must be taken when interpreting parameter-based subdivisions of the sample.  This is exemplified when we separate the sample by mass into ``planets" and ``brown dwarfs" (i.e. low and high mass objects), as we have 3 objects (HR 2562 b, $\kappa$ And b and $\beta$ Pic b) that fall into a ``boundary" or ``intermediate" mass classification - objects whose mass estimates allow for placement in both classifications. In such cases, we tested placing them in either population to see if results would change. Indeed, placing $\kappa$ And b and HR 2562 b into the planet population yields completely different results from placing them into the brown dwarf population.  Given these results, we conclude the uncertainty in both model-derived masses and individual eccentricity distributions are too large to allow us to distinguish between different populations as a function of mass.  This means that at this time, we cannot use eccentricities to say that different formation pathways are underway for sub-stellar companions above and below $\sim$15 M$_{Jup}$.  We note again that the distinction between populations at this mass boundary is not particularly meaningful from a formation perspective, but rather is a useful ``breakpoint" to explore these possibilities.  \par 

\subsection{Minimal Orbital Coverage}
Since we are mostly in the prior-dominated regime and there were major changes in posteriors with only a small addition of data, it is important to have some handle on how much data is needed to begin to make meaningful population distributions for directly imaged planets.  Our approach here, to look at when we have enough orbital coverage for the eccentricity distribution to cover the true value in simulated data, is one possible way to answer that question.  Our finding of 15\% coverage is valid for both observable-based and model-based, or uniform, priors.  The results converge at that point, suggesting that ultimately the distinction between these two prior sets really only matters for interpreting results in the prior-dominated regime.  However, this is not the only possible set of simulations that should be conducted to get a firm handle on the necessary orbital phase coverage.  For instance, we did not consider the role of some of the orbital parameters, such as inclination, in defining the needed coverage.  Edge-on orbits are often more quickly constrained than face-on, suggesting that this parameter is important to consider.  Furthermore, we did not explore in depth the role that true time from periastron plays in the defining of meaningful posteriors.  Instead, we take 15\% as a useful guideline for a general orbit where the companion is not close to periastron, which should encompass the majority of real systems.\par
As noted above, the average percent coverage of our sample is 7.4\%, with most objects having astrometry spanning less than 10\% of the orbital period. Given these results, we conclude again that the  undersampling of the orbital period for each individual object in our sample coupled with our small sample size does not allow us to affirm that planets and brown dwarfs have different formation pathways. Until the orbits of these objects are well sampled enough to provide meaningful posteriors, the eccentricity of the population of directly imaged substellar companions remains essentially unconstrained. \par

\section{Conclusion} \label{conc}
The main findings of this study are:
\begin{itemize}
    \item We derive new orbital parameter posteriors for a set of 21 directly imaged substellar companions using observable-based priors.
    \item Several companions have resulting eccentricity distributions that change significantly from previous results.  The inclusion of radial velocity (RV) data points or new astrometry in the orbit fitting process shifts the resulting posteriors for both observable-based and uniform prior fits.  
    \item We derive a population-level eccentricity distribution for the 21 companions and obtain shape parameters $\alpha$ = $1.09\substack{+0.30 \\ -0.22}$\ and $\beta$ = $1.42\substack{+0.33 \\ -0.25}$. These values are consistent with \citealt{Bowler_2020}'s parameters obtained using uniform priors, but with a lower incidence of high eccentricity objects.
    \item Separating the population into ``giant planets" and ``brown dwarfs" produces different results depending on where intermediate mass objects are placed. This result implies that our current sample size and large uncertainties may not be sufficient to determine whether these objects do in fact present distinct eccentricity populations.
    \item From our orbital coverage simulations, we find that one generally needs 15\% orbital period coverage to obtain a reliable posteriors on eccentricity, period and $T_o$ posterior .

\end{itemize}

Following the conclusions of this work, the addition of new astrometry or radial velocity points for directly imaged companions can help further constrain the eccentricity posteriors of directly imaged companions to reliable intervals. This can in turn allow for a more robust eccentricity distributions at a population level. 

\section{Acknowledgements}
We thank Christopher Theissen and Sarah Blunt for helpful discussions on the analysis and results presented in this work.  We also thank an anonymous referee for their comments which helped improve this manuscript. \par
Some of the data presented herein were obtained at the W. M. Keck Observatory, which is operated as a scientific partnership among the California Institute of Technology, the University of California, and the National Aeronautics and Space Administration. The W. M. Keck Observatory was made possible by the financial support of the W. M. Keck Foundation. The authors wish to acknowledge the significant cultural role that the summit of Maunakea has always had within the indigenous Hawaiian community. The author(s) are extremely fortunate to conduct observations from this mountain. Portions of this work were conducted at the University of California, San Diego, which was built on the unceded territory of the Kumeyaay Nation, whose people continue to maintain their political sovereignty and cultural traditions as vital members of the San Diego community. \par
C.D.O. is supported by the National Science Foundation Graduate Research Fellowship under Grant No. DGE-2038238.  K.K.O. is supported by the Preston Graduate Fellowship. Further support for this work at UCLA was provided by the W. M. Keck Foundation, and NSF Grant AST-1909554. Any opinions, findings, and conclusions
or recommendations expressed in this material are those of the author(s) and do not necessarily reflect
the views of the National Science Foundation.

\clearpage
\bibliographystyle{aa_bst.bst}
\bibliography{main.bib}
\clearpage
\appendix

\section{Corner Plots with Uniform vs. Observable Based Priors} \label{appA}

We present in this section the corner plot comparisons between the orbital eccentricities of HD 1160 b, HD 49197 b, HR 2562 b, HIP 65426 b and PZ Tel b when evaluated using uniform and when using observable based priors. We choose these specific objects because, when fit with uniform priors, they present the high-eccentricity bias discussed in Sections \ref{intro} and \ref{of}. When fit with observable-based priors, however, this bias appears to be mostly suppressed.  
\begin{figure} [htb!]
    \centering
    \centering{{\includegraphics[width= 8.6cm]{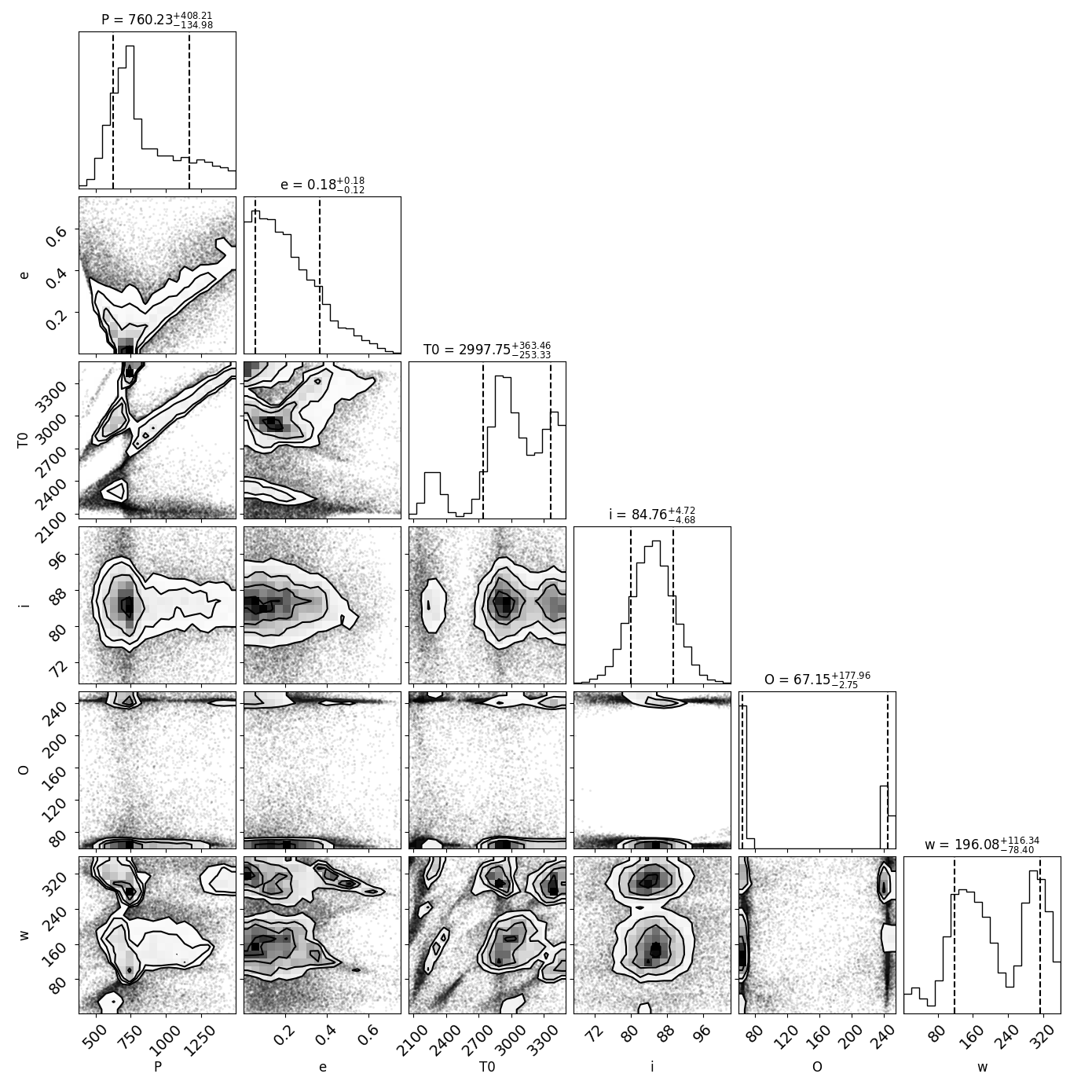} }}%
    \qquad
    \centering{{\includegraphics[width=8.5cm]{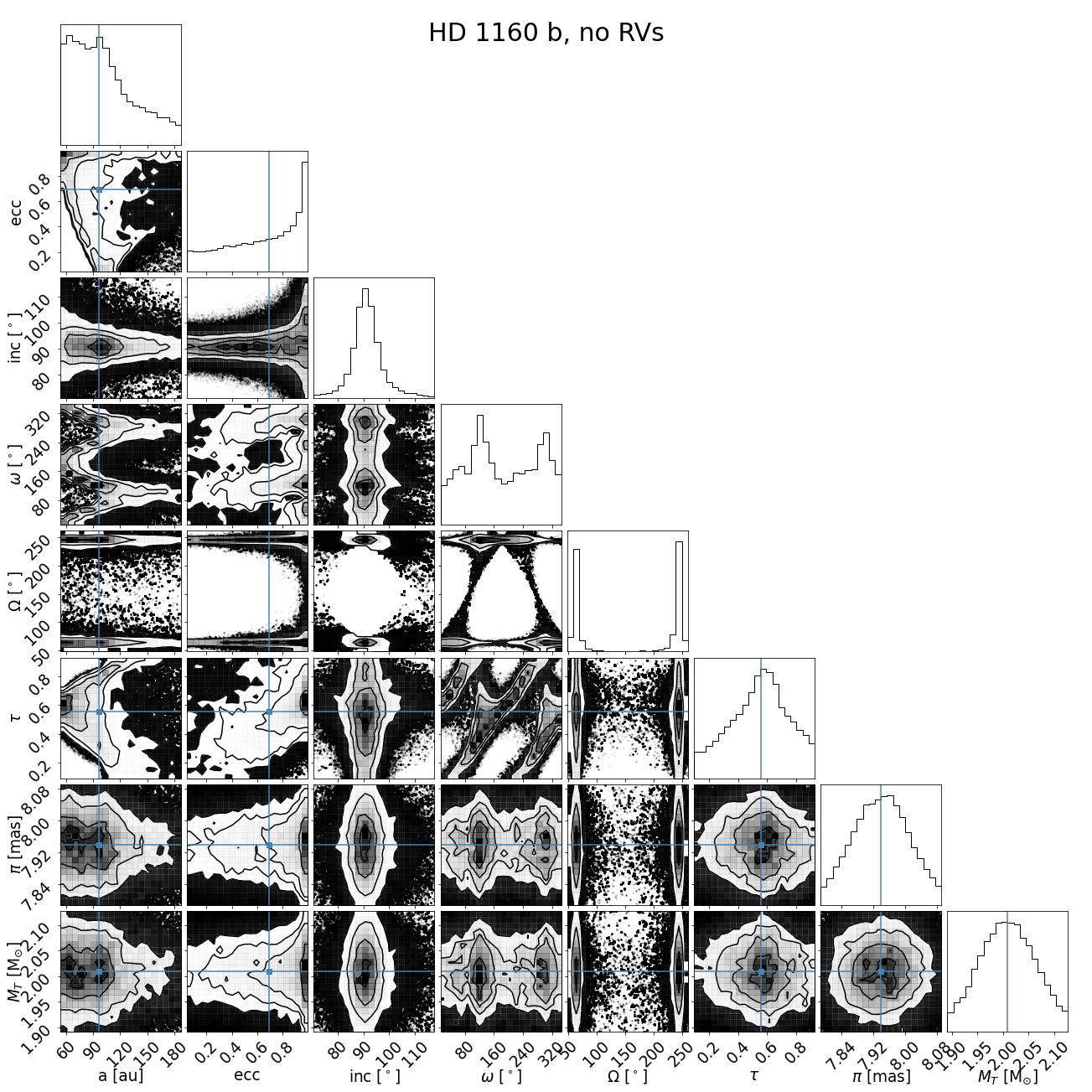} }}%
    \caption{The HD 1160 b orbital parameter corner plots using observable-based priors (left) and uniform priors (right). For this plot, both priors had only astrometry measurements. For observable priors, the columns represent, respectively, the period (P; in years), the eccentricity (e), the epoch of periastron passage ($T_0$; in years), the inclination (i; in $^{\circ}$), Longitude of ascending node ($\Omega$; in $^{\circ}$) and argument of periapsis ($\omega$; in $^{\circ}$). For uniform priors, the columns represent, respectively, the semi-major axis (a; in AU), the eccentricity (e), the inclination (i; in $^{\circ}$), the argument of periapsis ($\omega$; in $^{\circ}$), the longitude of ascending node ($\Omega$; in $^{\circ}$), the epoch of periastron passage ($\tau$), the distance of the system ($\pi$; in mas) and the system mass (M; in $M_\odot$).}%
    \label{fig:hd1160comp}
\end{figure}

 \begin{figure}[htb!]
  \begin{center}
\centerline{\includegraphics[width=3.5in]{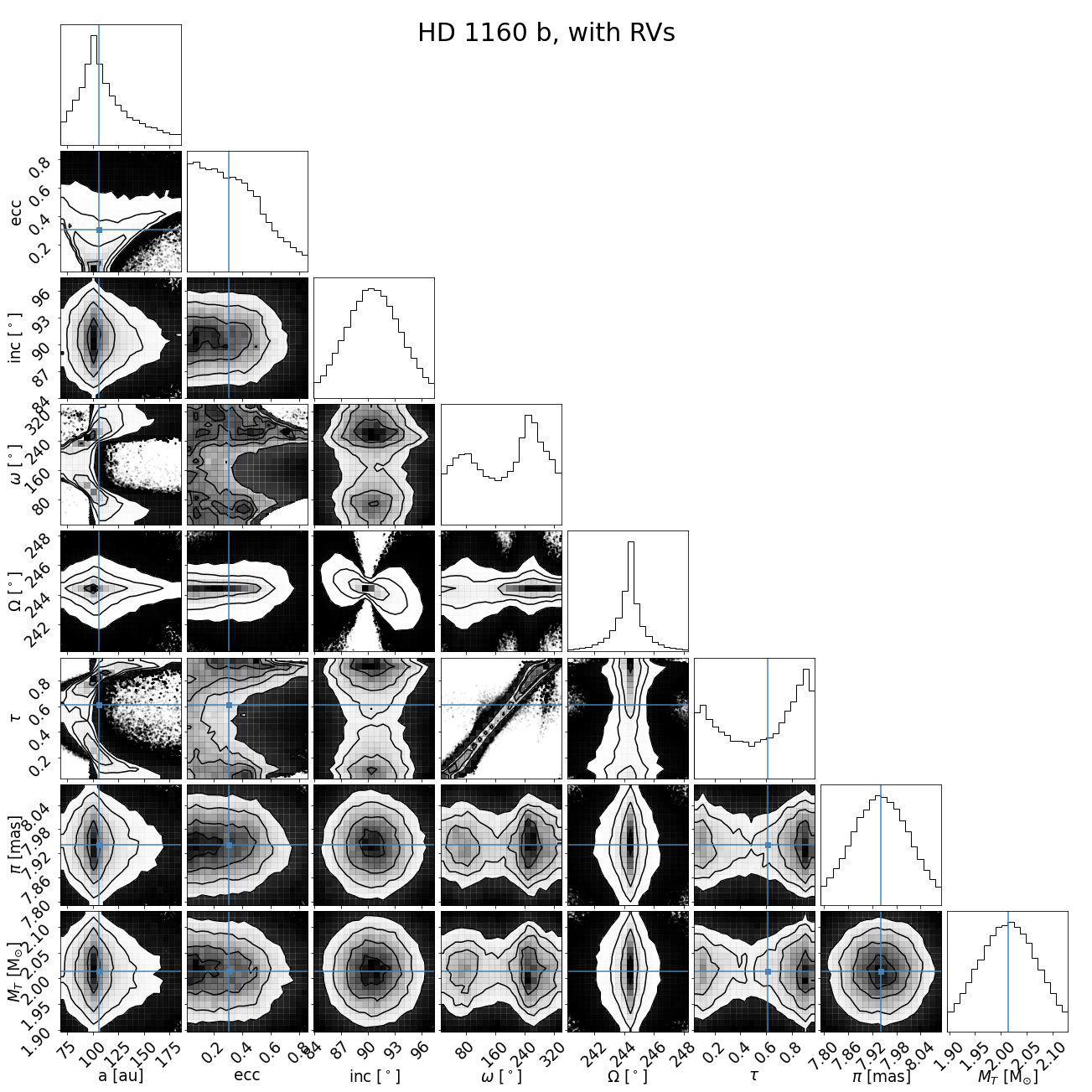}}
\caption{Uniform priors corner plot for HD 1160 b with two RV points added.}
\label{fig:rvhd1160}
  \end{center}
\end{figure}

\begin{figure}
    \centering
    \centering{{\includegraphics[width= 8.6cm]{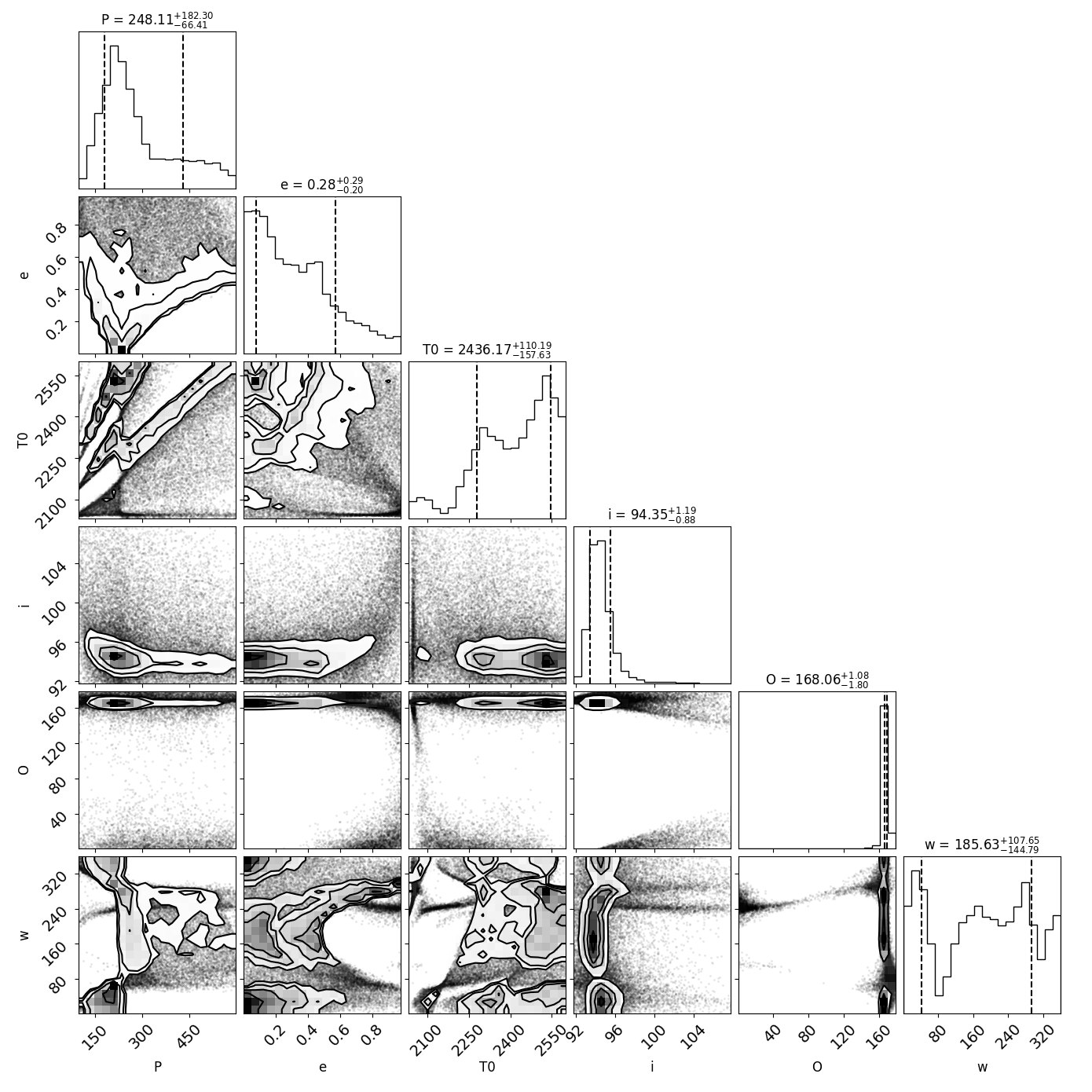} }}%
    \qquad
    \centering{{\includegraphics[width=8.5cm]{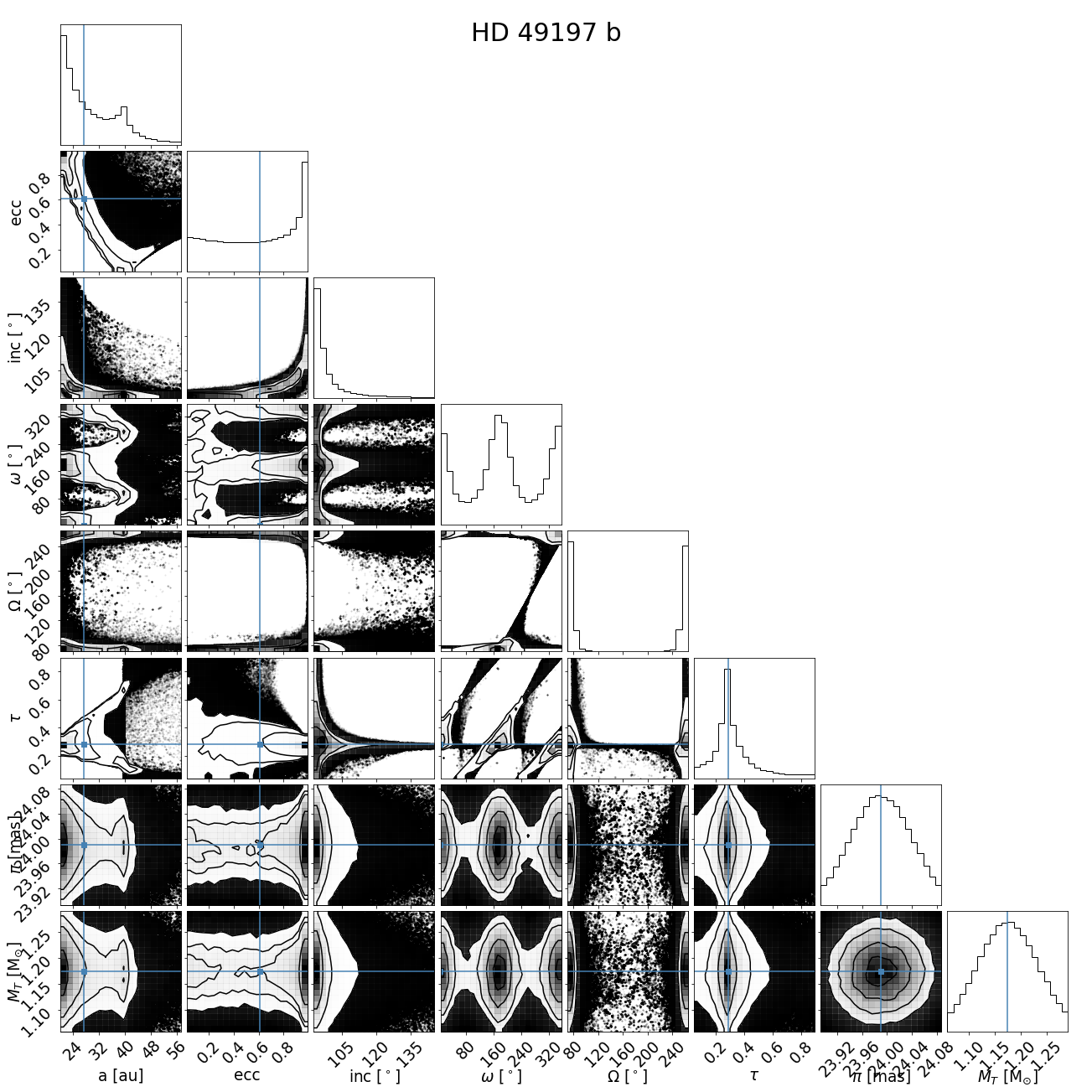} }}%
    \caption{Same as \ref{fig:hd1160comp}, but for HD 49197 B.}%
    \label{fig:hd49197b}
\end{figure}

\begin{figure}
    \centering
    \centering{{\includegraphics[width= 8.6cm]{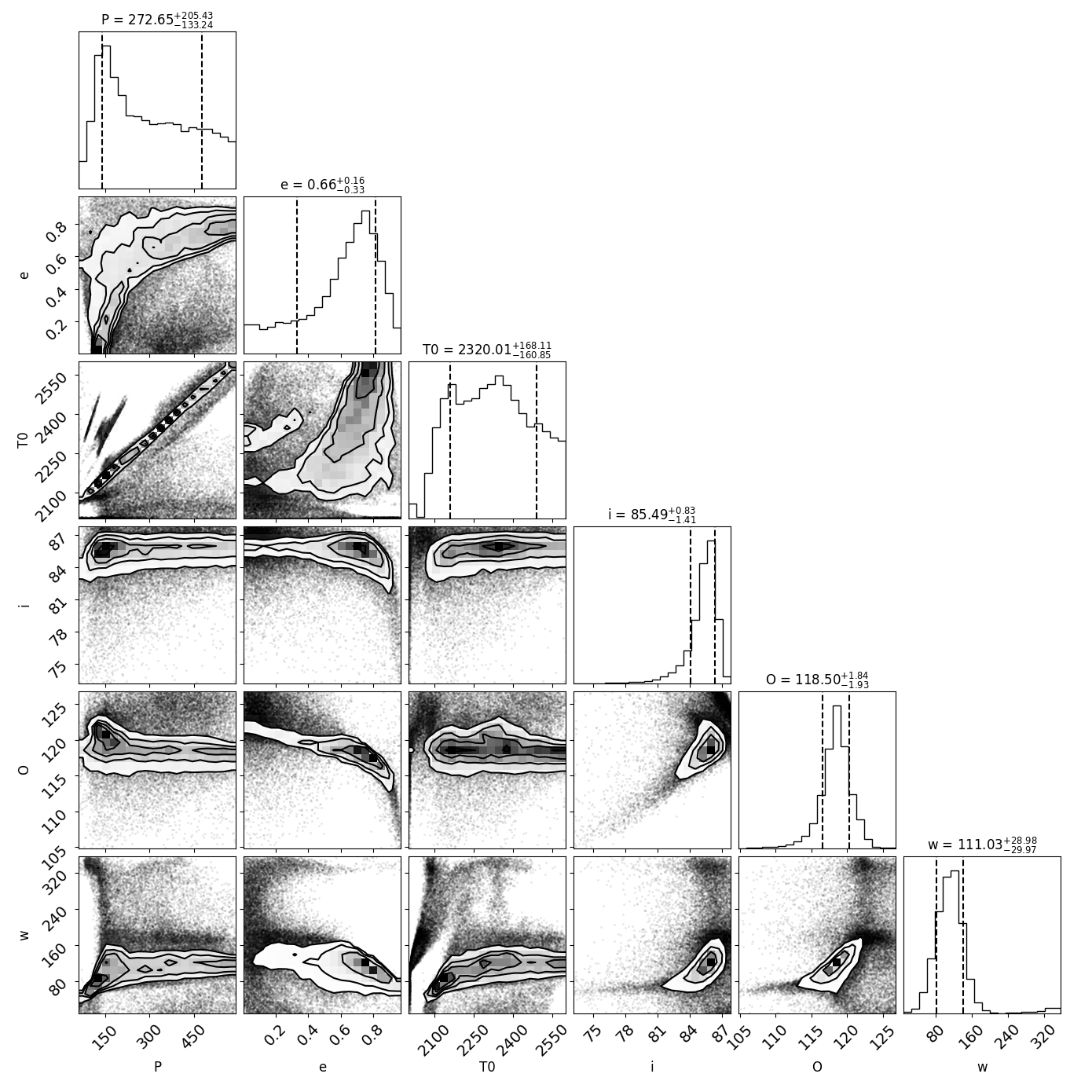} }}%
    \qquad
   \centering{{\includegraphics[width=8.5cm]{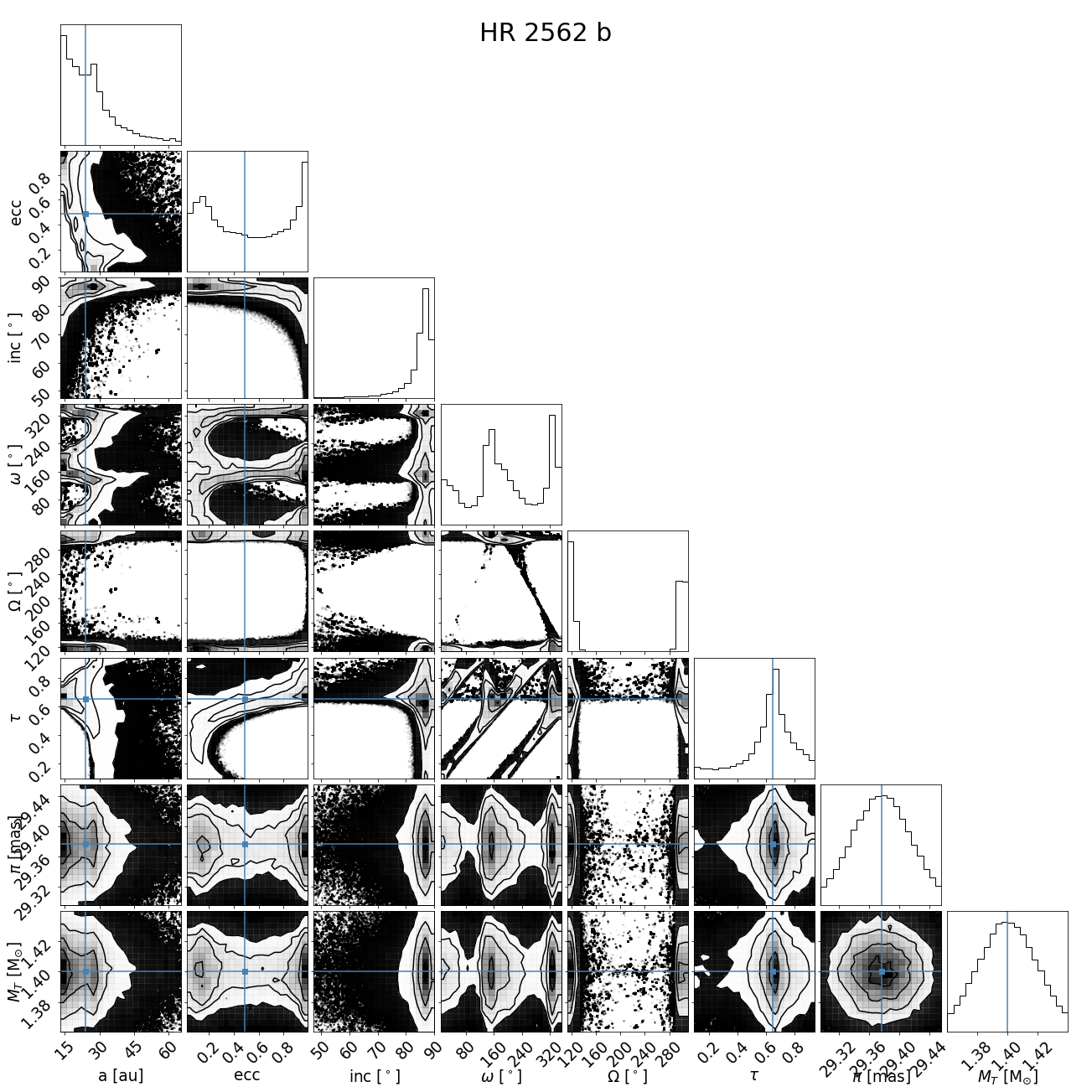} }}%
    \caption{Same as \ref{fig:hd1160comp}, but for HR 2562 B.}%
    \label{fig:hr2562b}
\end{figure}
\begin{figure}
    \centering
    \centering{{\includegraphics[width= 8.6cm]{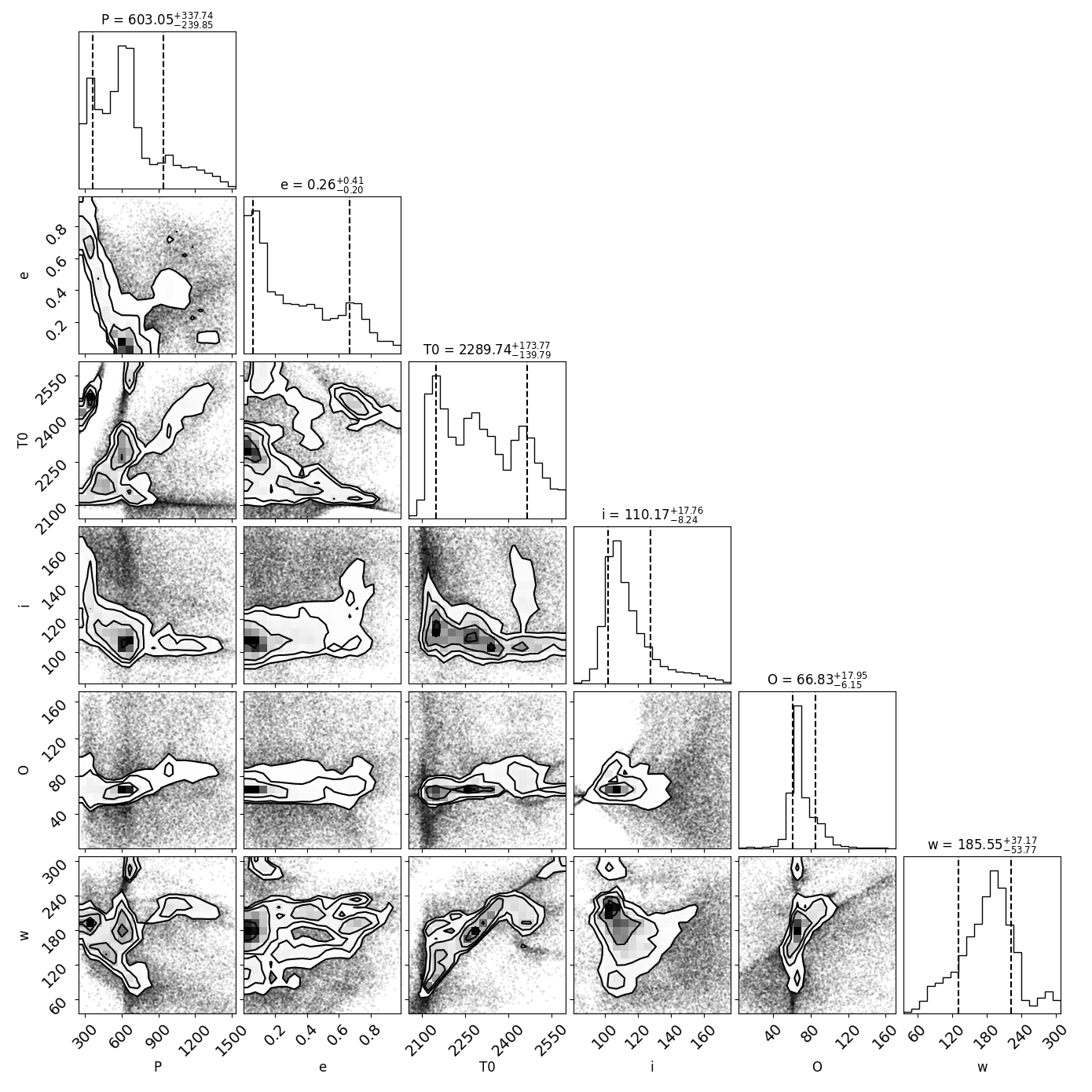} }}%
    \qquad
    \centering{{\includegraphics[width=8.5cm]{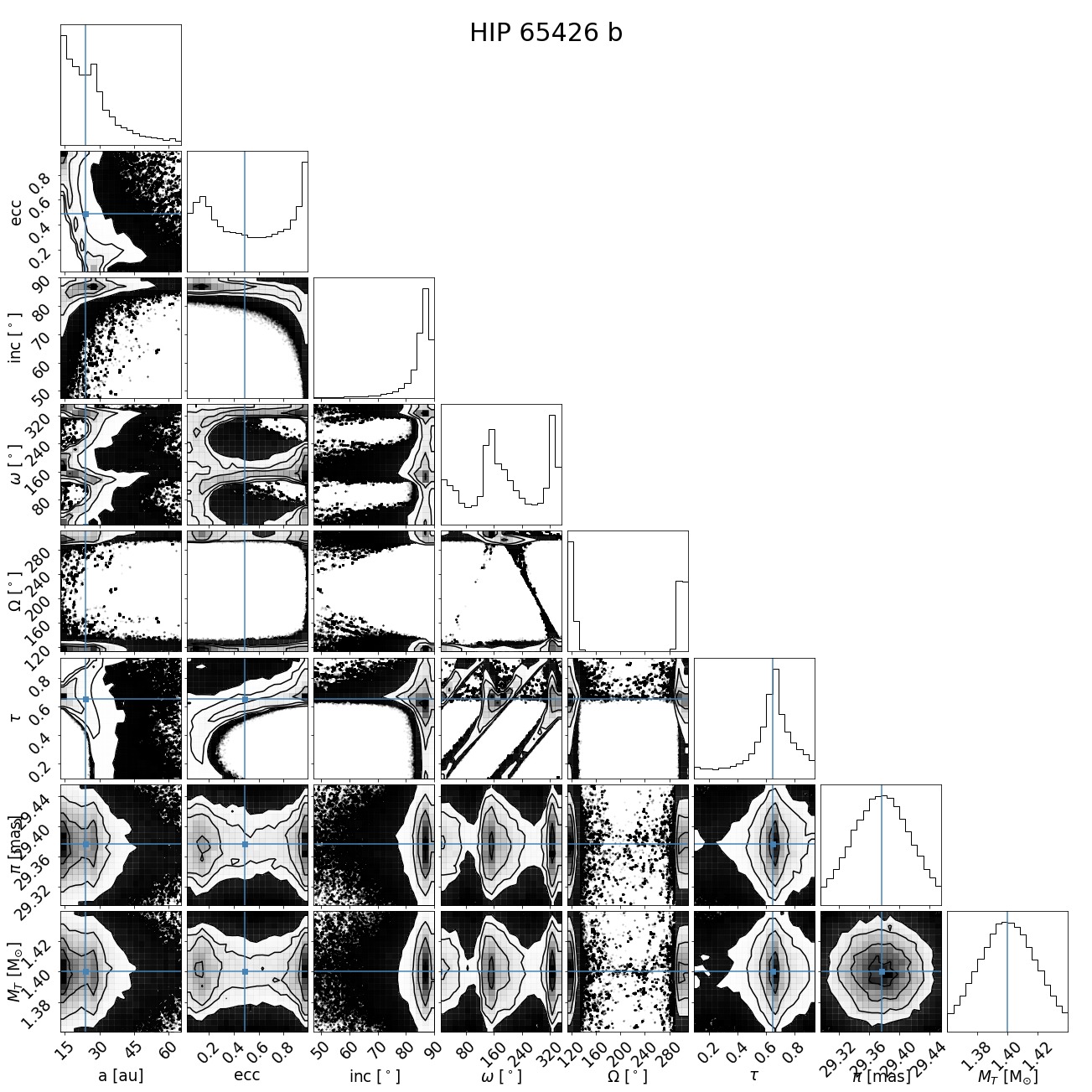} }}%
    \caption{Same as \ref{fig:hd1160comp}, but for HIP 65426 B.}%
    \label{fig:hip65426b}
\end{figure}
\begin{figure}
    \centering
    \centering{{\includegraphics[width= 8.6cm]{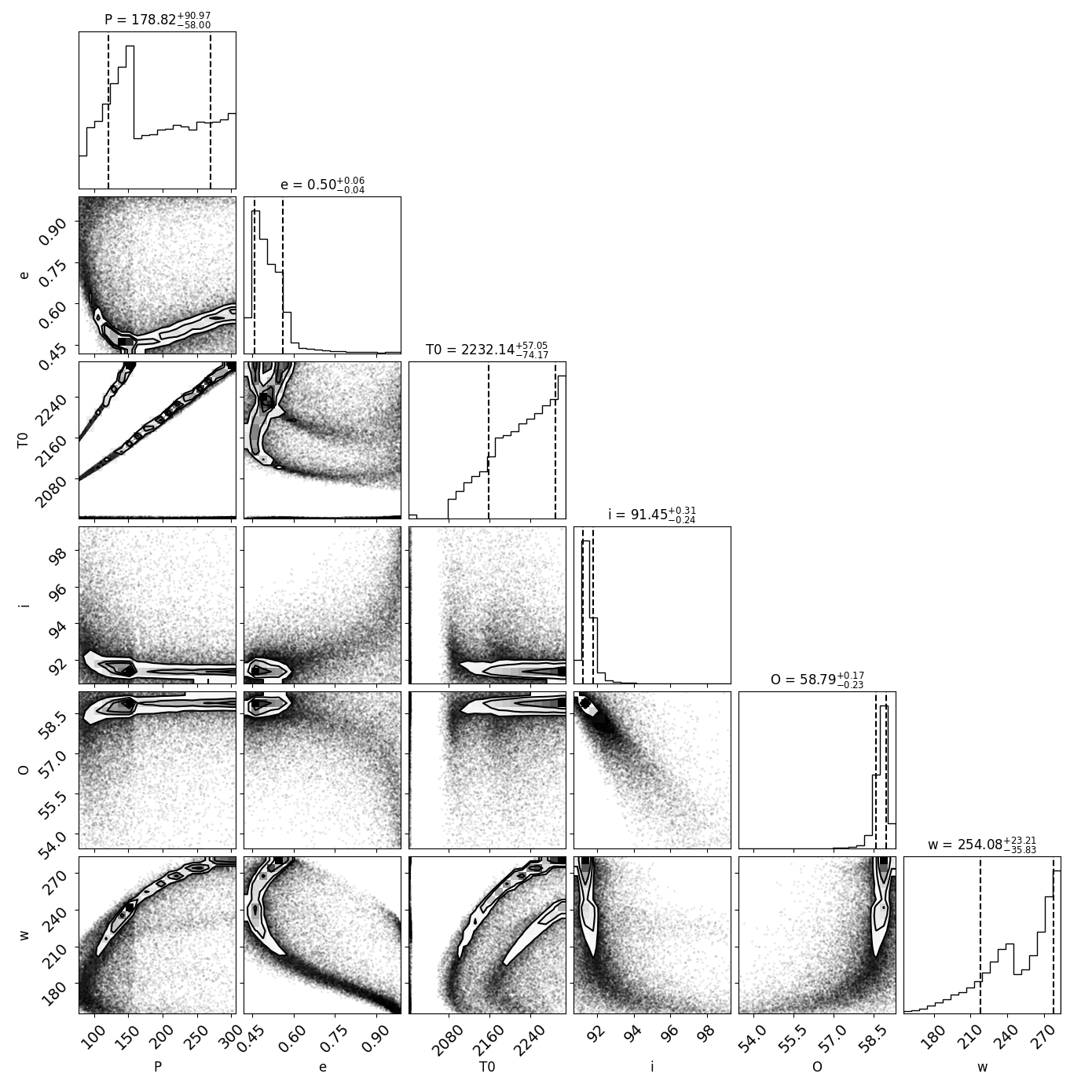} }}%
    \qquad
    \centering{{\includegraphics[width=8.5cm]{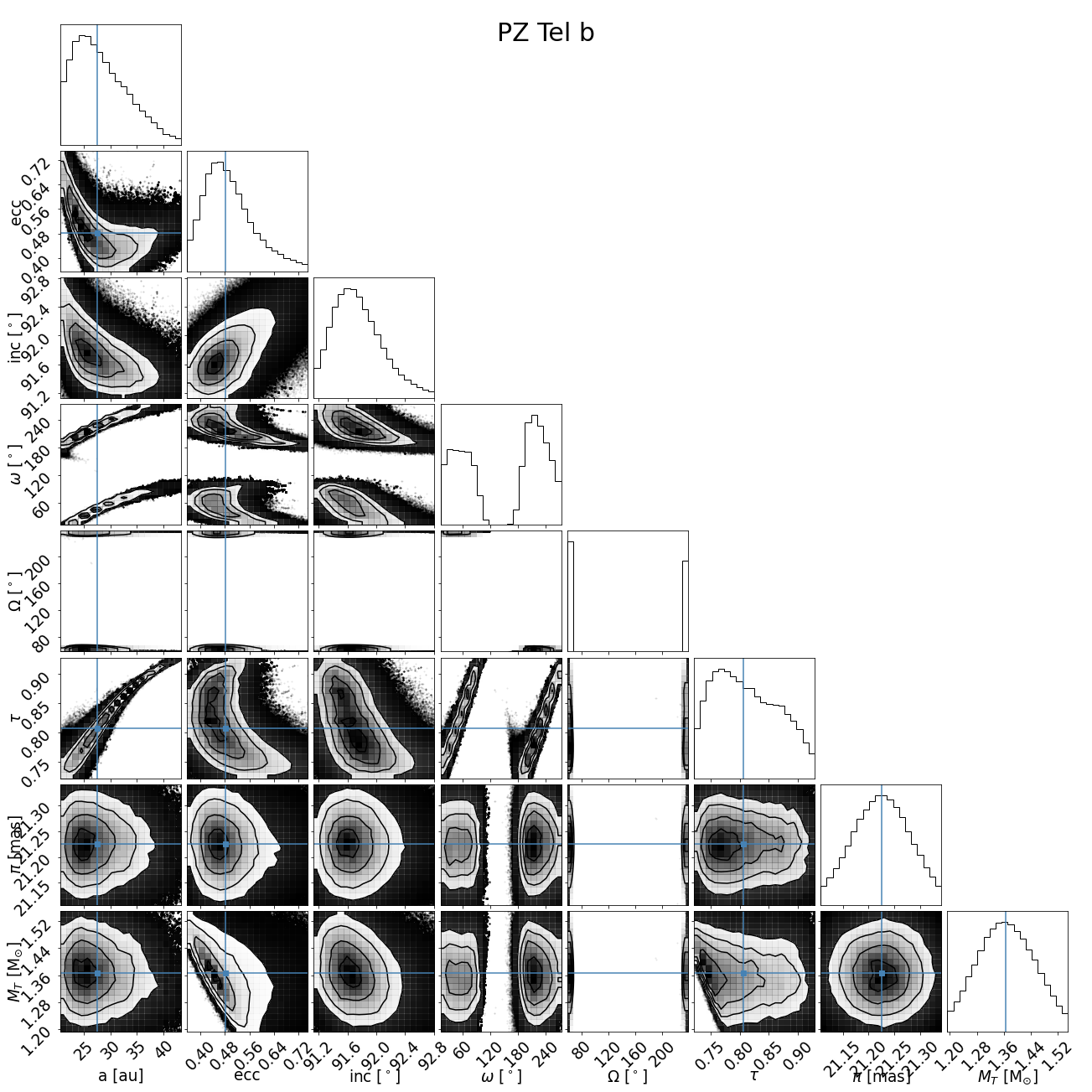} }}%
    \caption{Same as \ref{fig:hd1160comp}, but for PZ Tel b. Both fits include the 2018 epoch from Table \ref{tbl:1}. See discussion on this result in Section \ref{shifts}.}%
    \label{fig:pztelb}
\end{figure}

\section{Calculating $\mathcal{P}$} \label{appB}

\begin{figure}
    \centering
    \centering{{\includegraphics[width=8.5cm, height=8.5cm]{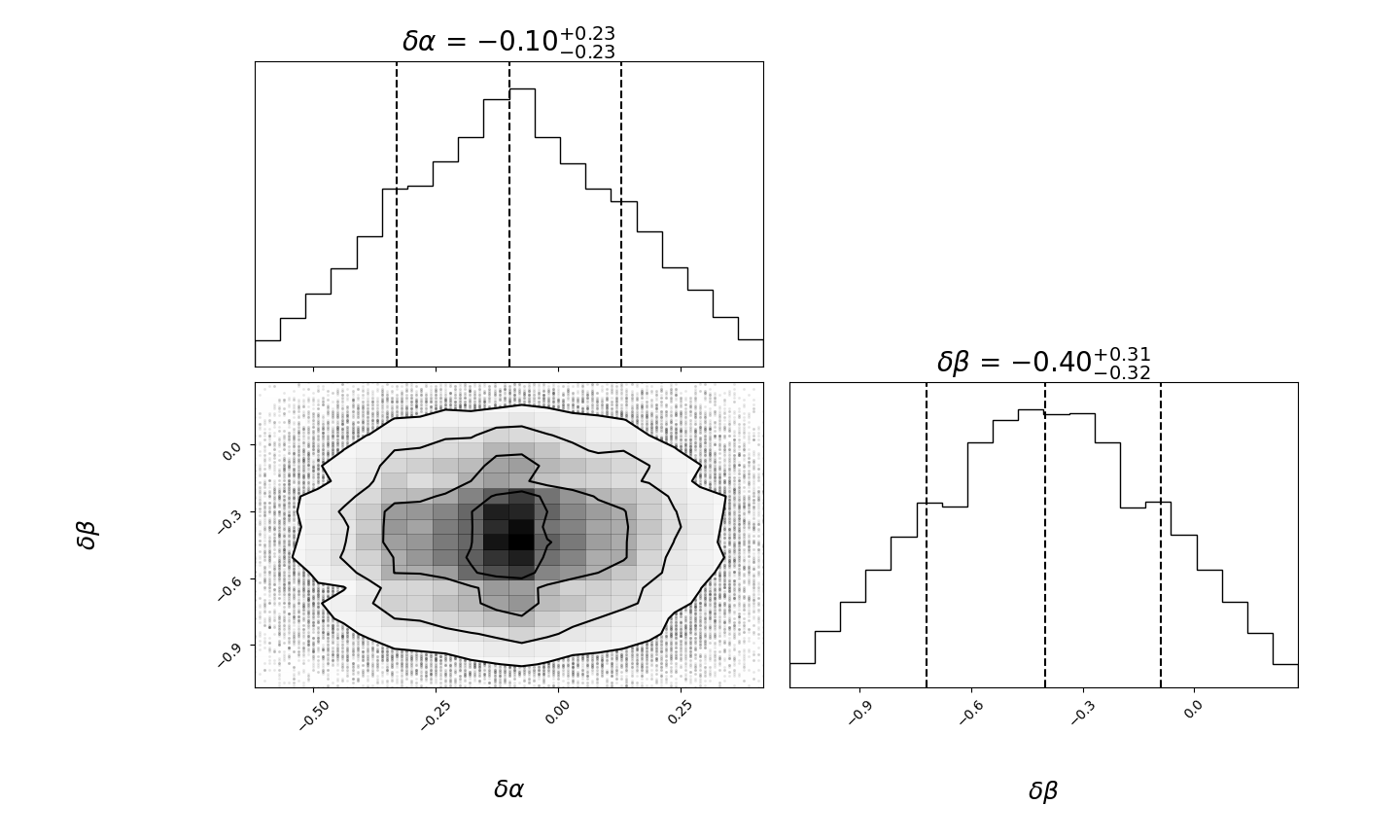} }}%
    \qquad
    \centering{{\includegraphics[width=8.5cm, height=8.5cm]{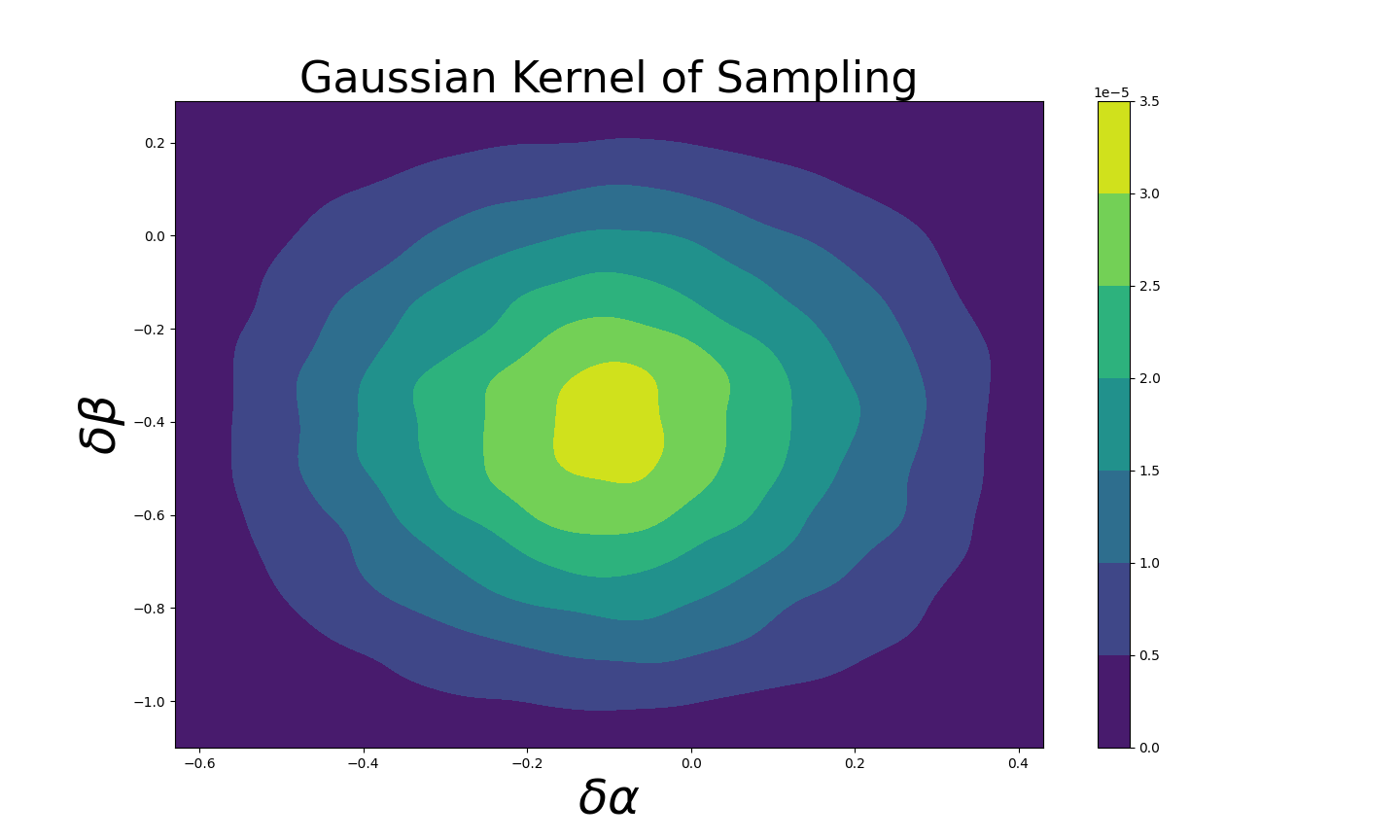} }}%
    \caption{Sample of $\delta\alpha$ and $\delta\beta$ for our distributions with the whole population compared to the distributions with the whole population not including 1RXS0342+1216 B (a). Gaussian kernel fit to the contour of our sampling (b). From this kernel, we integrate the distribution outside the contour of constant density that goes through the origin.}%
    \label{fig:gaussiancurves}
\end{figure}

Here we show an example of the calculation to obtain our ``consistency parameter"  $\mathcal{P}$, defined as the p-value, for two Beta probability distributions. We begin with two Beta distributions for different populations, each with their own pairs of  ($\alpha_1$, $\beta_1$) and ($\alpha_2$, $\beta_2$) from their fits. We then sample pairs from each distribution and compute $\delta\alpha=\alpha_2-\alpha_1$ and $\delta\beta=\beta_2-\beta_1$  for each iteration. An example of this method for our entire population distribution compared with the whole population minus the object 1RXS0342+1216 B is shown in Figure \ref{fig:gaussiancurves} (a). However, this is merely a sample from our distributions, and not a 2-D PDF that we can evaluate. So we obtain a Gaussian kernel of this sample, shown in \ref{fig:gaussiancurves} (b). \par
Now with the Gaussian kernel, we are able to integrate the 2D distribution of ($\delta\alpha$,$\delta\beta$) inside the contour of constant density that goes through the origin. This allows us to obtain $\mathcal{P}$, which gives us an estimation of how consistent or inconsistent two Beta distributions are.

\section{Individual Corner Plots}
The corner plots obtained from Efit5 with observable-based priors for the 21 individual objects are presented. The objects can be identified in the caption of each figure.

\figsetstart

\figsetnum{21}
\figsettitle{Corner Plots for Individual Objects} 
\figsetgrpstart
\figsetgrpnum{21.1}
\figsetgrptitle{Corner Plot for HD 984 b}
\figsetplot{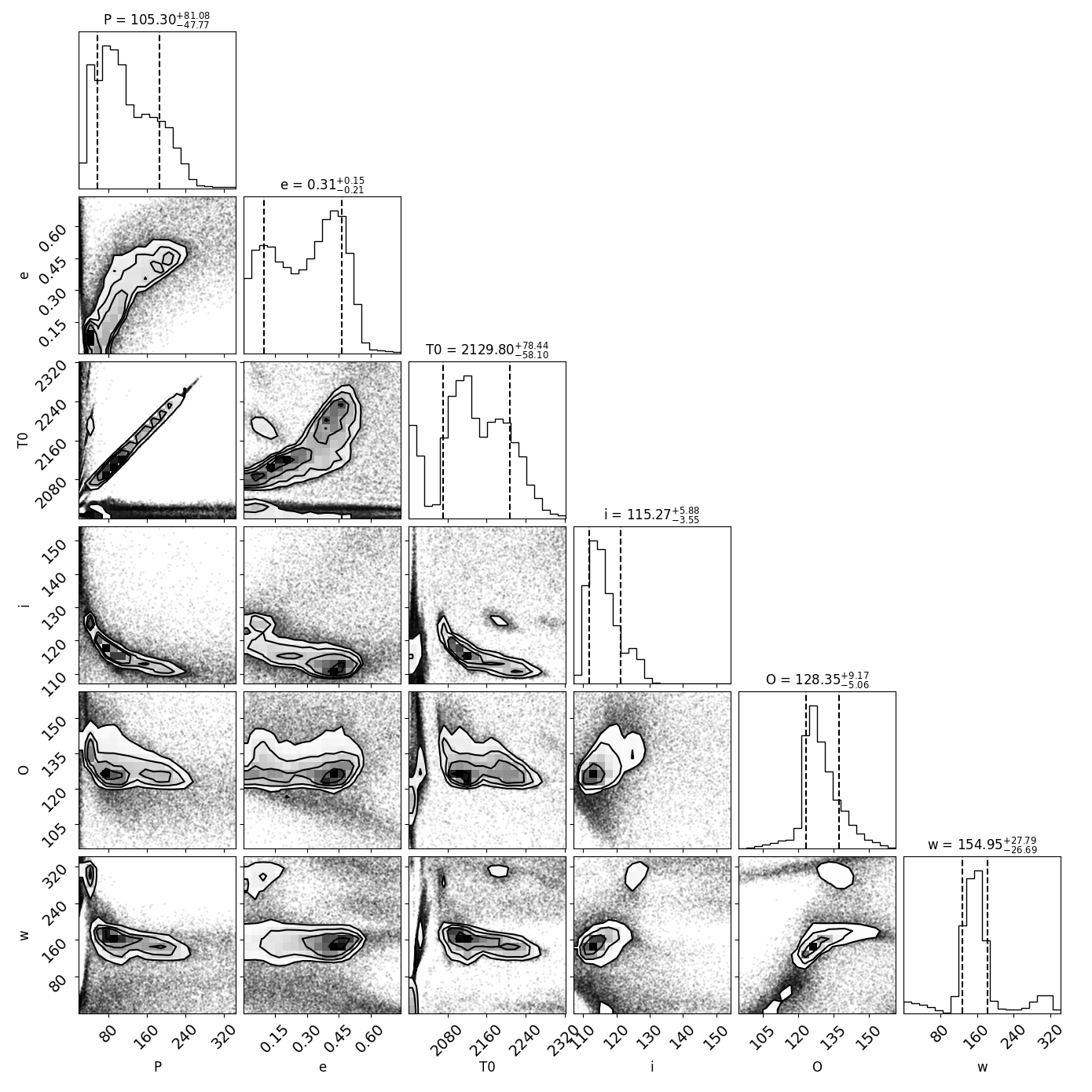}
\figsetgrpnote{}
\figsetgrpend
 
\figsetgrpstart
\figsetgrpnum{21.2}
\figsetgrptitle{Corner Plot for HD 1160 b}
\figsetplot{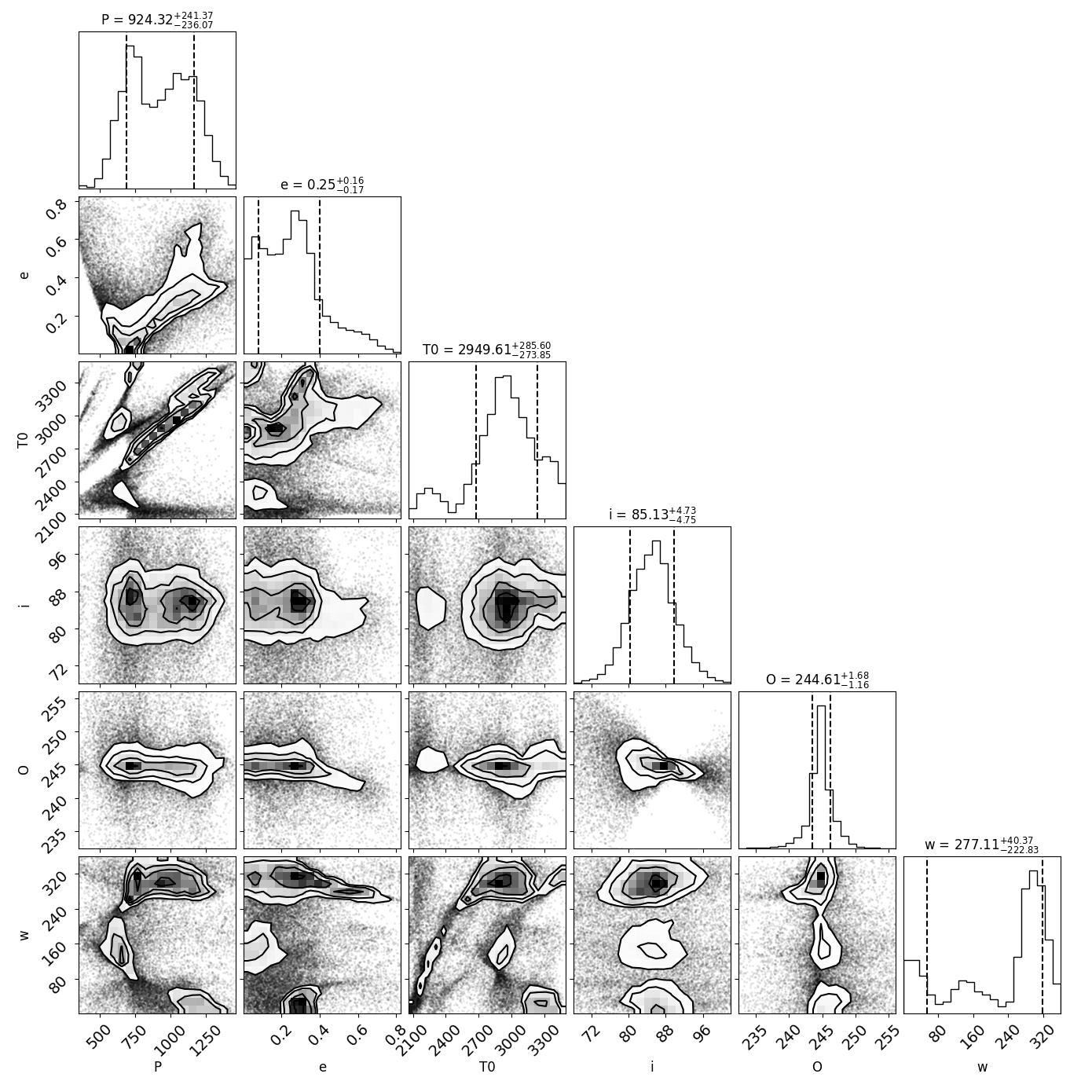}
\figsetgrpnote{}
\figsetgrpend

\figsetgrpstart
\figsetgrpnum{21.3}
\figsetgrptitle{Corner Plot for HD 19467 b}
\figsetplot{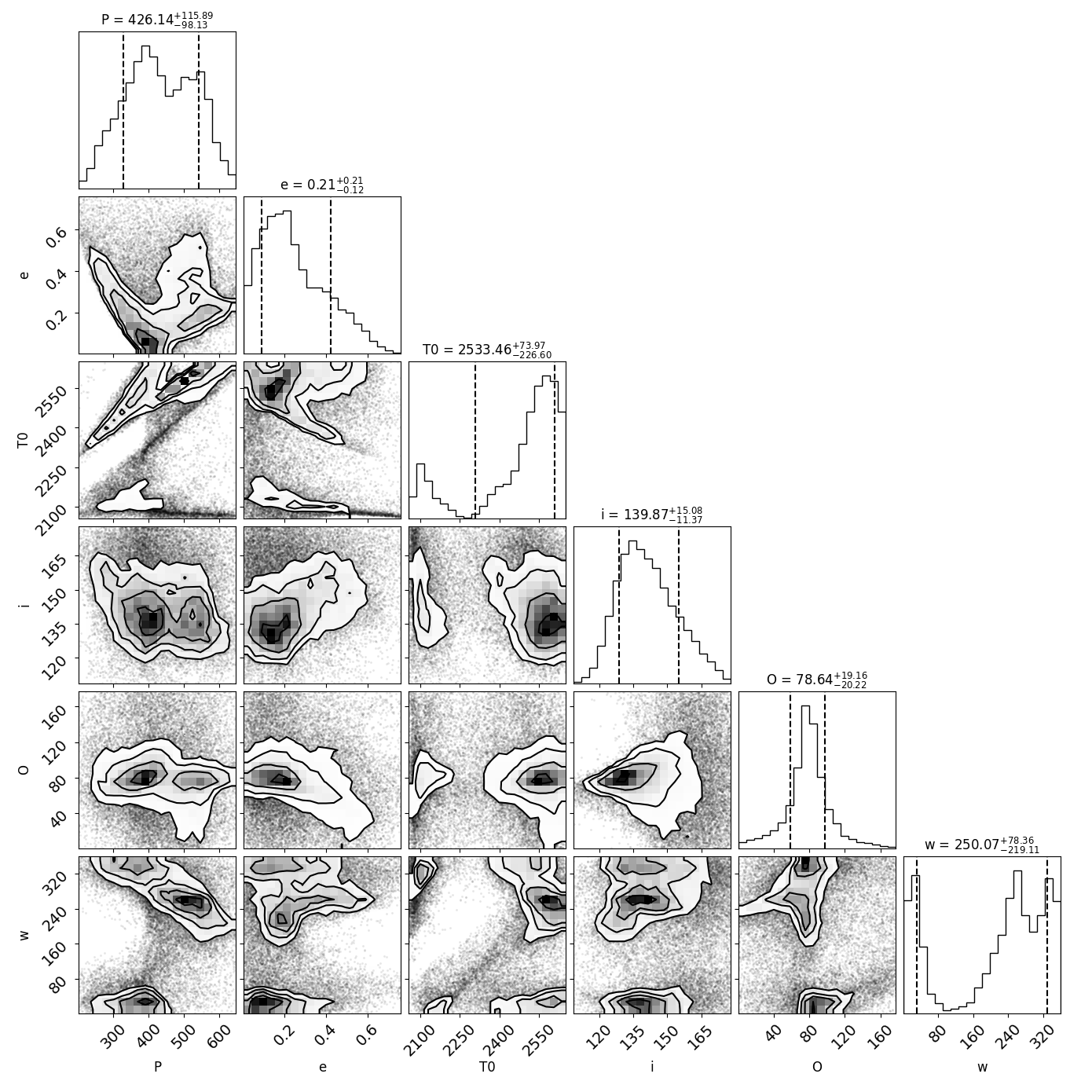}
\figsetgrpnote{}
\figsetgrpend

\figsetgrpstart
\figsetgrpnum{21.4}
\figsetgrptitle{Corner Plot for 1RXS0342+1216 b}
\figsetplot{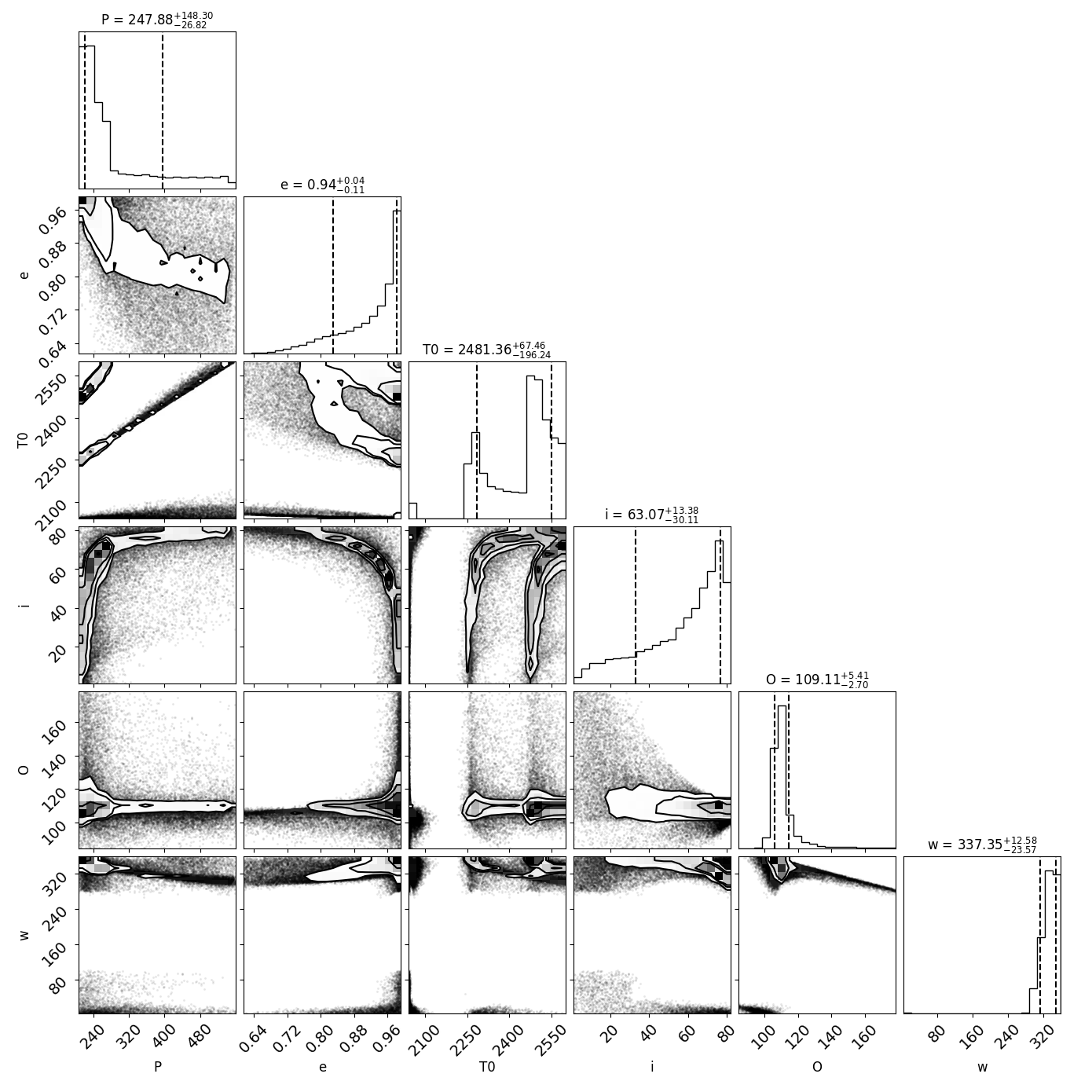}
\figsetgrpnote{}
\figsetgrpend

\figsetgrpstart
\figsetgrpnum{21.5}
\figsetgrptitle{Corner Plot for 51 Eridani b}
\figsetplot{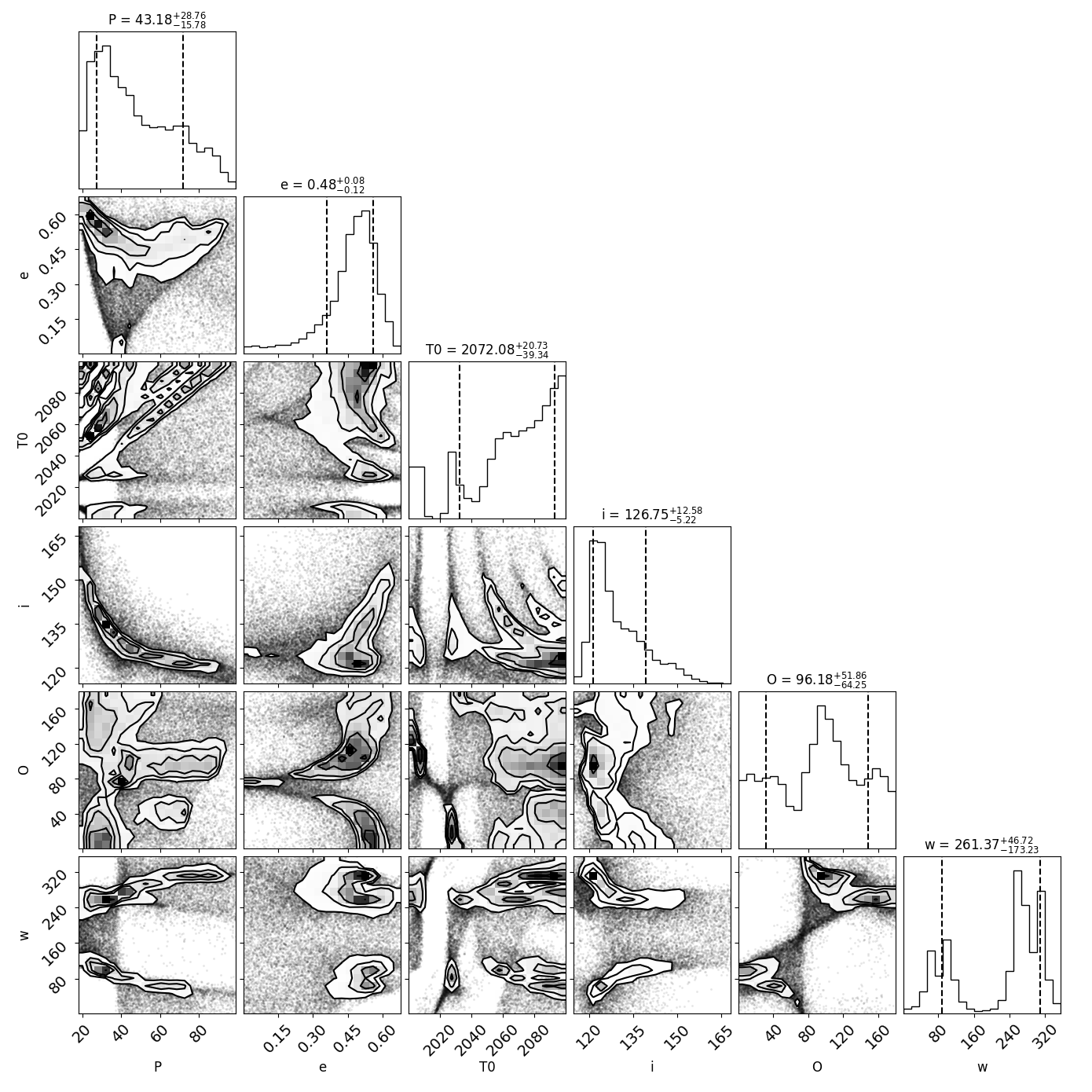}
\figsetgrpnote{}
\figsetgrpend

\figsetgrpstart
\figsetgrpnum{21.6}
\figsetgrptitle{Corner Plot for HD 49197 b}
\figsetplot{HD49197B_cornerplot_efit.png}
\figsetgrpnote{}
\figsetgrpend

\figsetgrpstart
\figsetgrpnum{21.7}
\figsetgrptitle{Corner Plot for HR 2562 b}
\figsetplot{HR2562B_cornerplot_efit.png}
\figsetgrpnote{}
\figsetgrpend

\figsetgrpstart
\figsetgrpnum{21.7}
\figsetgrptitle{Corner Plot for HR 2562 b}
\figsetplot{HR2562B_cornerplot_efit.png}
\figsetgrpnote{}
\figsetgrpend

\figsetgrpstart
\figsetgrpnum{21.8}
\figsetgrptitle{Corner Plot for HR 3549 b}
\figsetplot{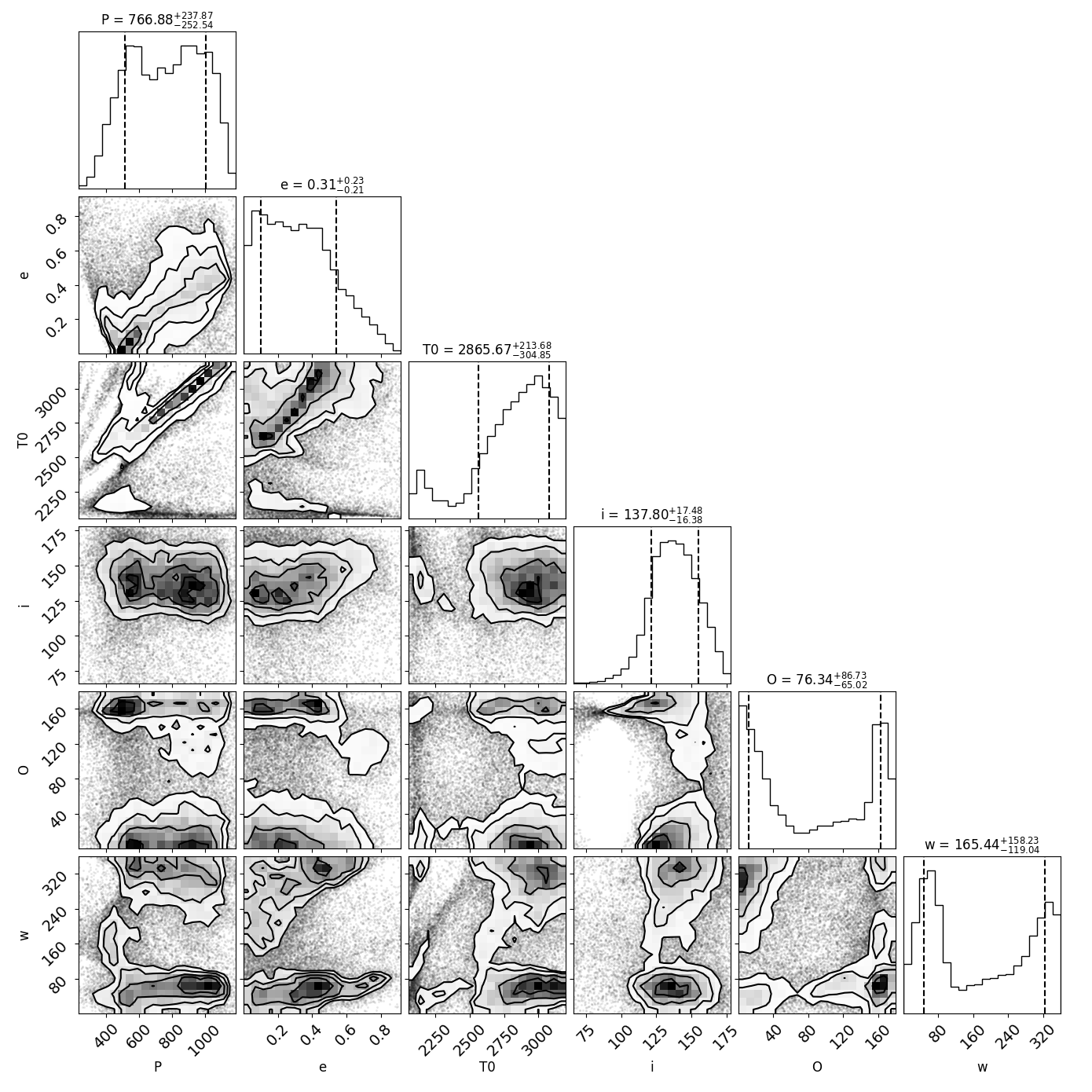}
\figsetgrpnote{}
\figsetgrpend

\figsetgrpstart
\figsetgrpnum{21.9}
\figsetgrptitle{Corner Plot for HR 95086 b}
\figsetplot{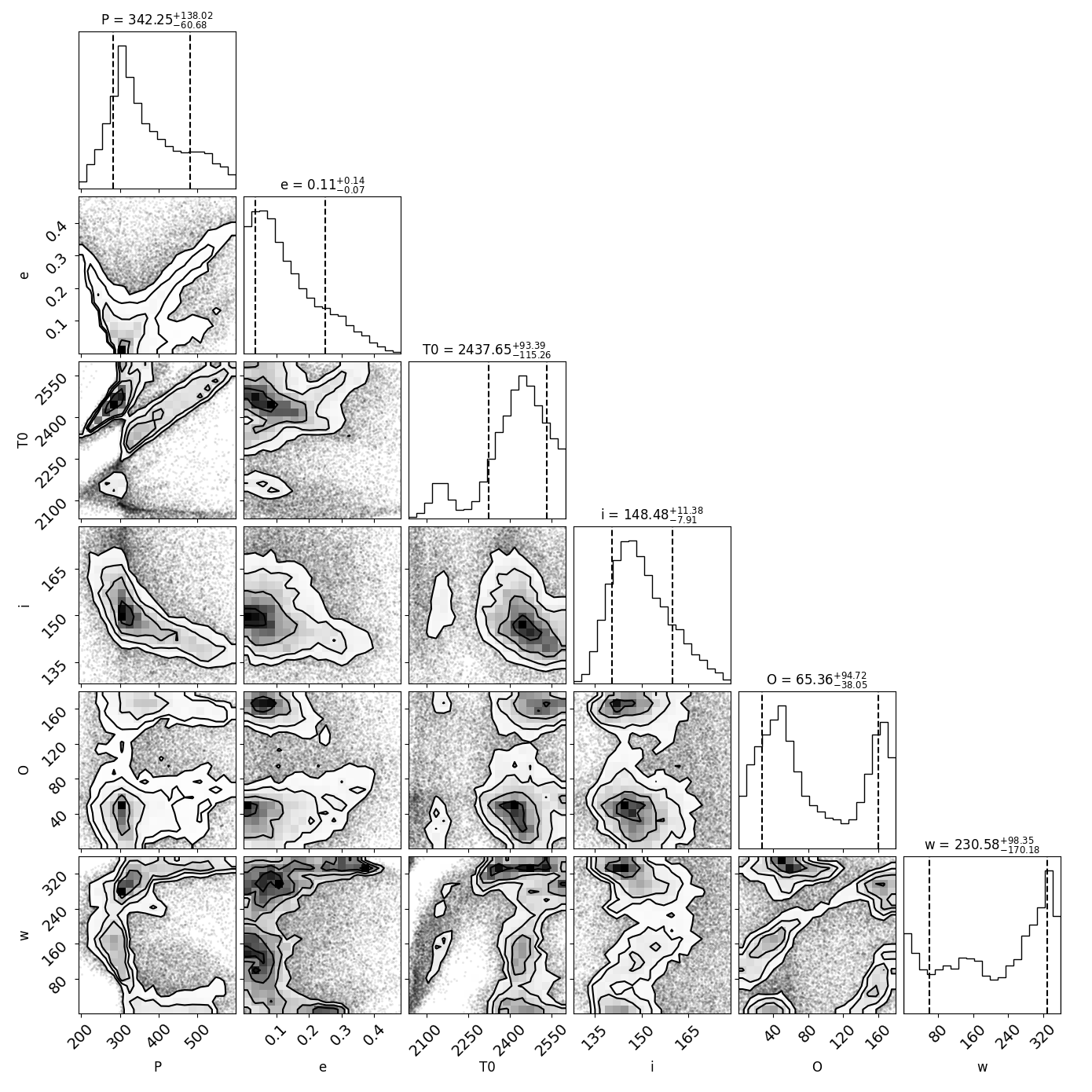}
\figsetgrpnote{}
\figsetgrpend

\figsetgrpstart
\figsetgrpnum{21.10}
\figsetgrptitle{Corner Plot for GJ 504 b}
\figsetplot{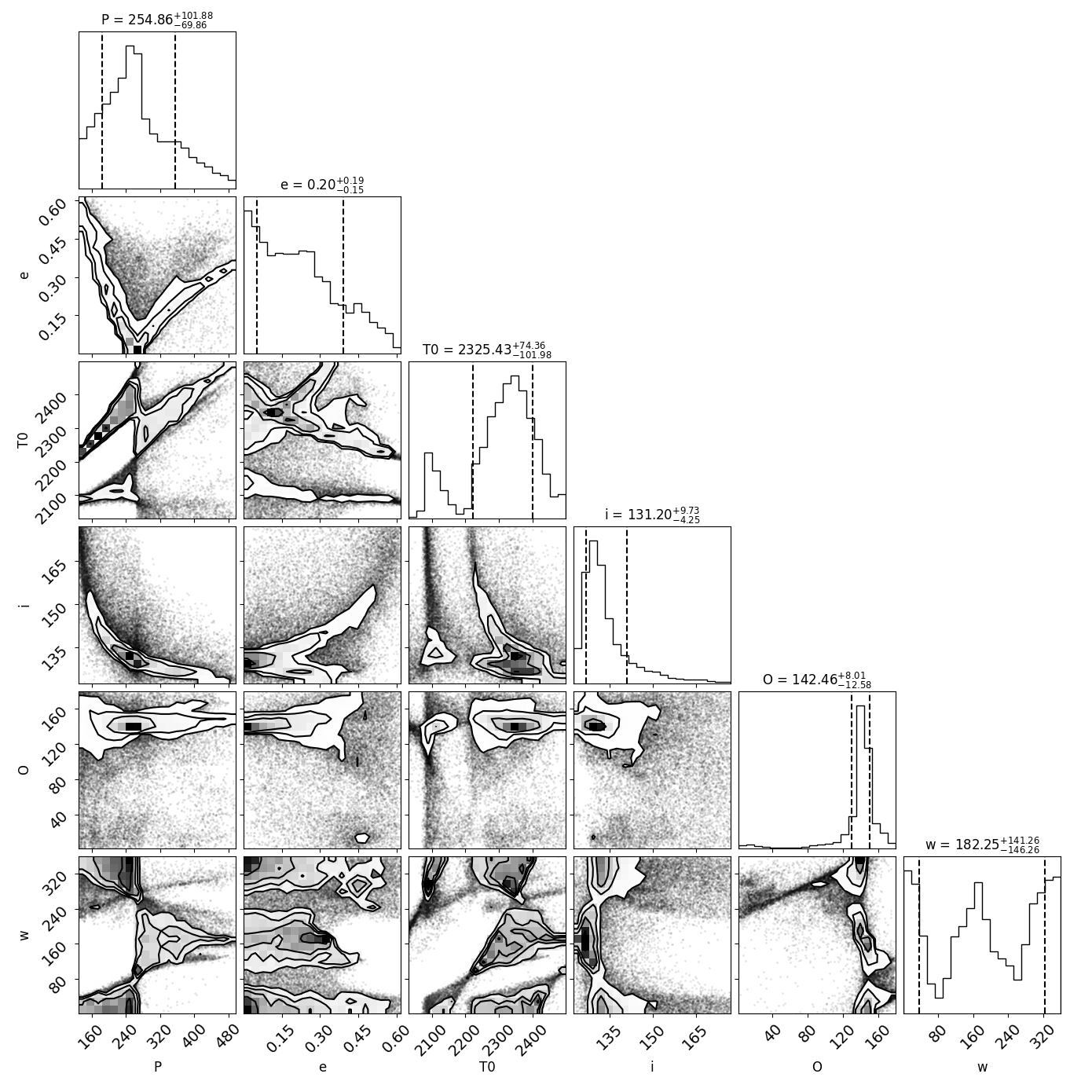}
\figsetgrpnote{}
\figsetgrpend

\figsetgrpstart
\figsetgrpnum{21.11}
\figsetgrptitle{Corner Plot for HIP 65426 b}
\figsetplot{HIP65426b_cornerplot_efit.png}
\figsetgrpnote{}
\figsetgrpend

\figsetgrpstart
\figsetgrpnum{21.12}
\figsetgrptitle{Corner Plot for PZ Tel b}
\figsetplot{PZTelB_cornerplot_efit.png}
\figsetgrpnote{}
\figsetgrpend

\figsetgrpstart
\figsetgrpnum{21.13}
\figsetgrptitle{Corner Plot for HD 206893 b}
\figsetplot{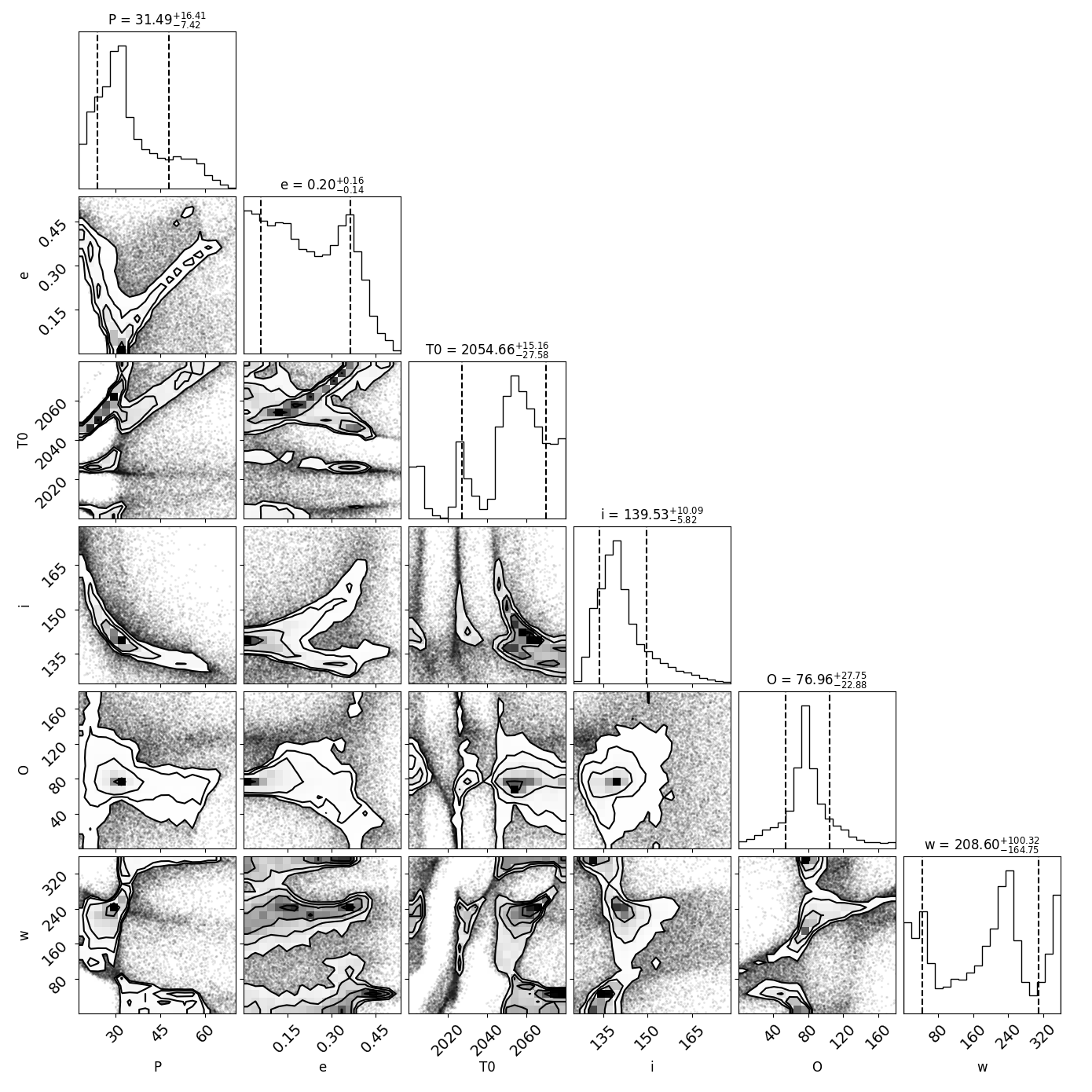}
\figsetgrpnote{}
\figsetgrpend

\figsetgrpstart
\figsetgrpnum{21.14}
\figsetgrptitle{Corner Plot for $\kappa$ And b}
\figsetplot{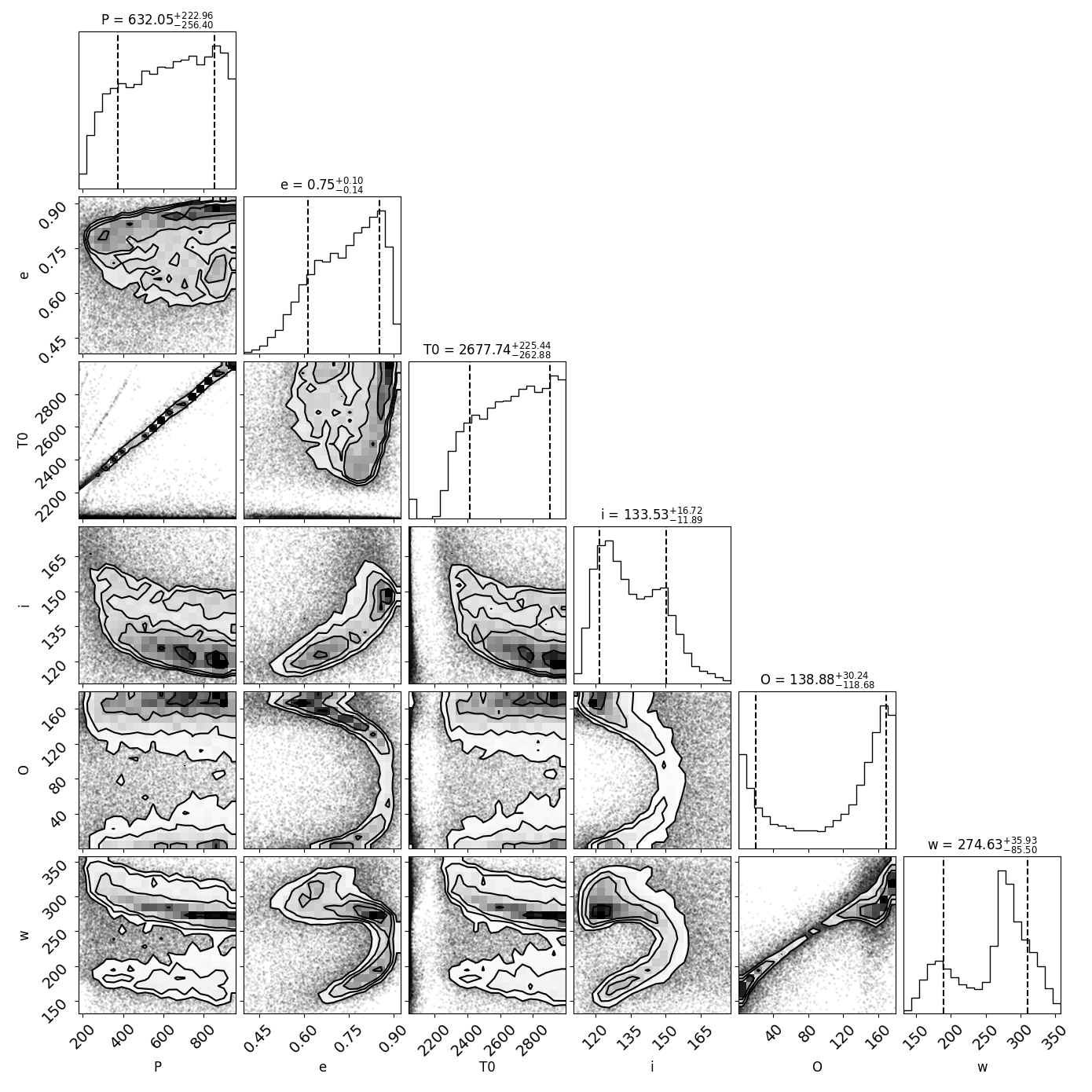}
\figsetgrpnote{}
\figsetgrpend

\figsetgrpstart
\figsetgrpnum{21.15}
\figsetgrptitle{Corner Plot for PDS 70 c}
\figsetplot{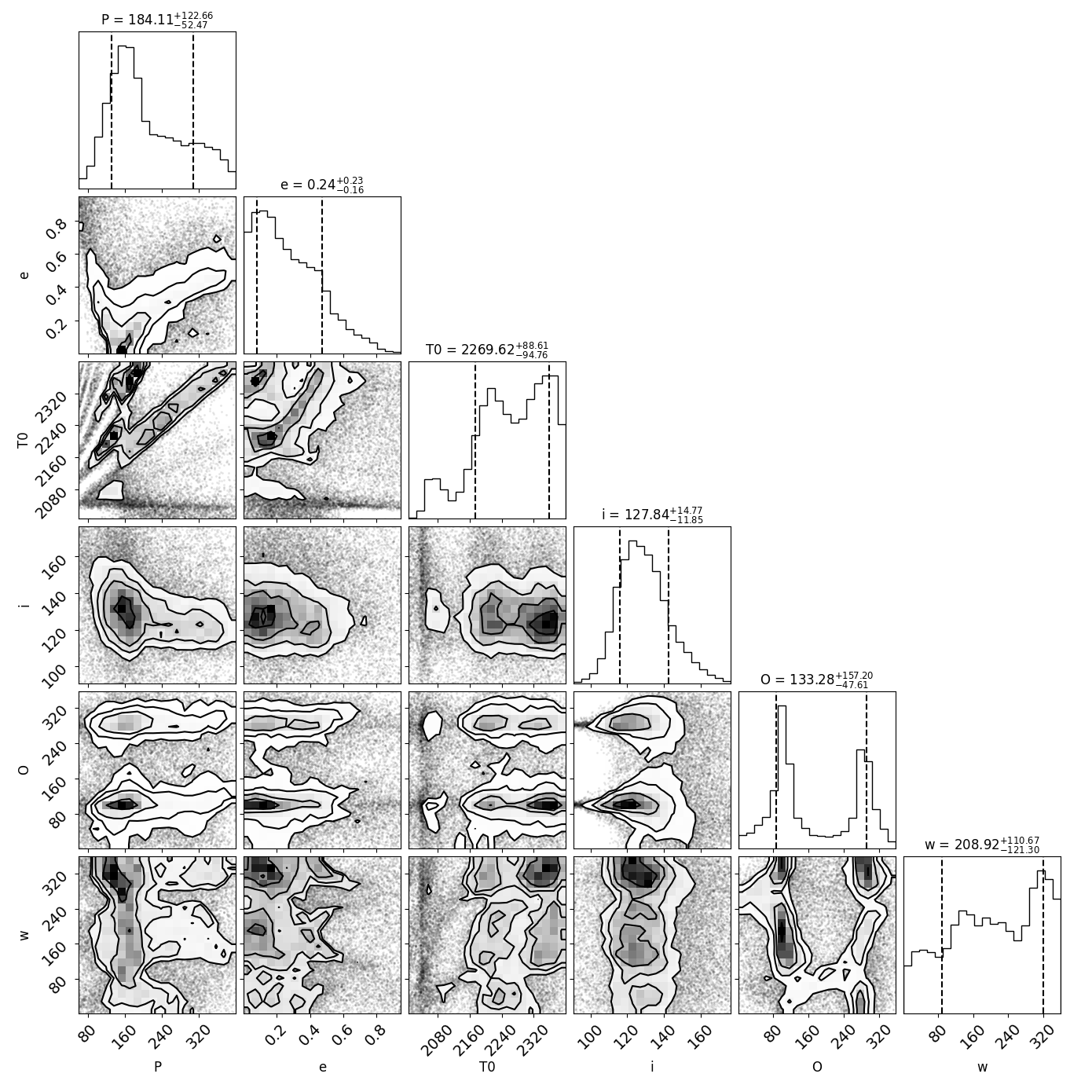}
\figsetgrpnote{}
\figsetgrpend

\figsetgrpstart
\figsetgrpnum{21.16}
\figsetgrptitle{Corner Plot for $\beta$ Pictoris b}
\figsetplot{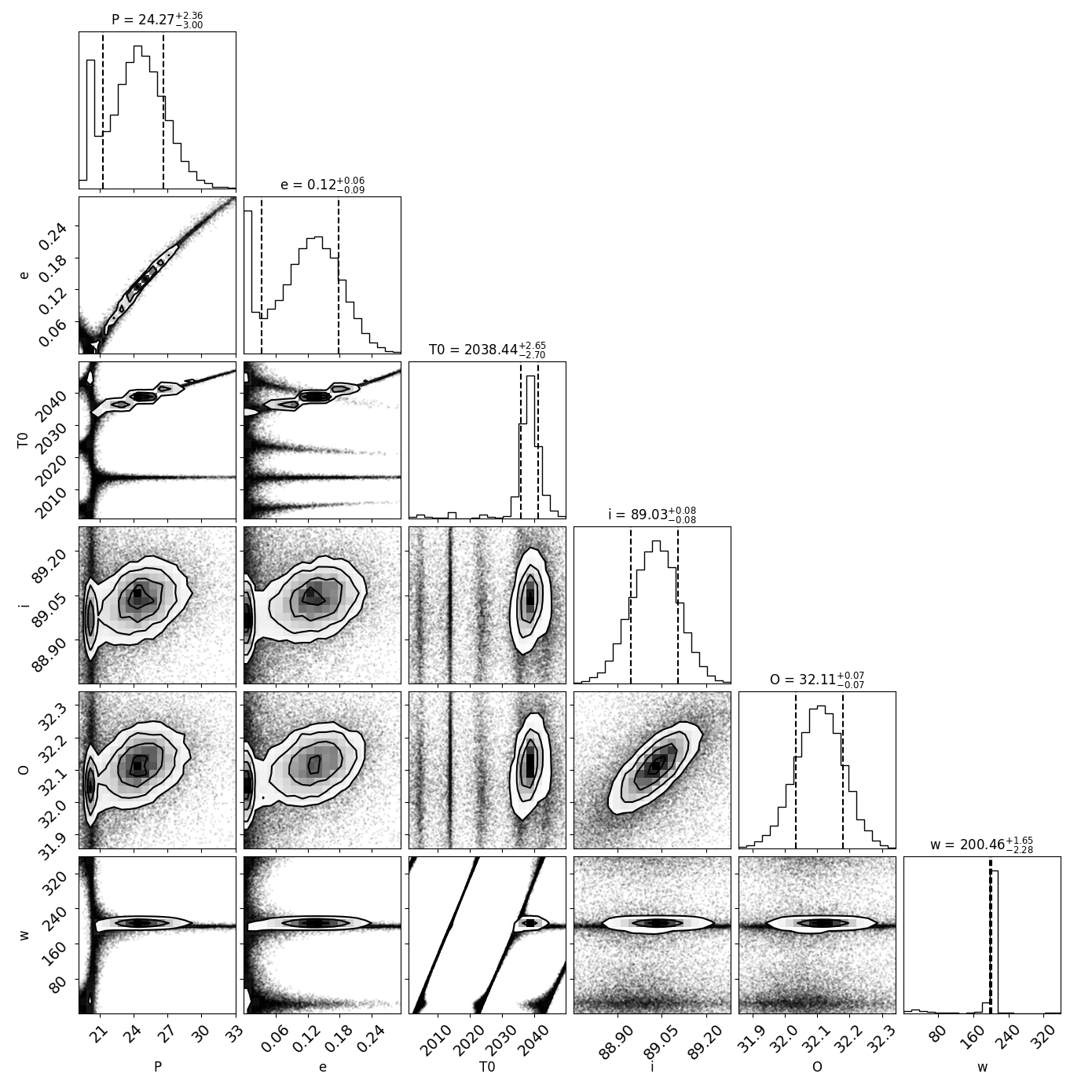}
\figsetgrpnote{}
\figsetgrpend

\figsetgrpstart
\figsetgrpnum{21.17}
\figsetgrptitle{Corner Plot for HR 8799 c}
\figsetplot{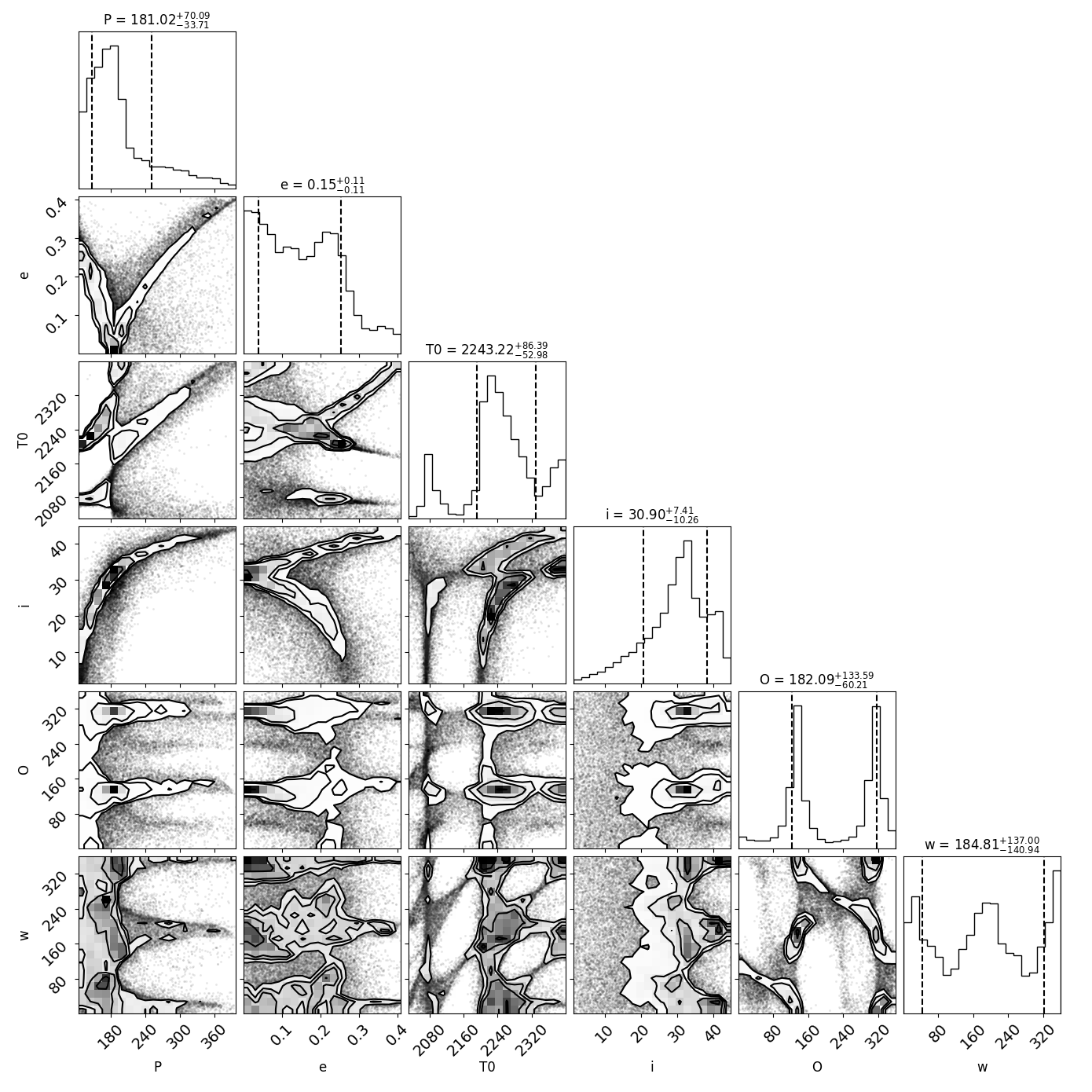}
\figsetgrpnote{}
\figsetgrpend

\figsetgrpstart
\figsetgrpnum{21.18}
\figsetgrptitle{Corner Plot for HD 4747 b}
\figsetplot{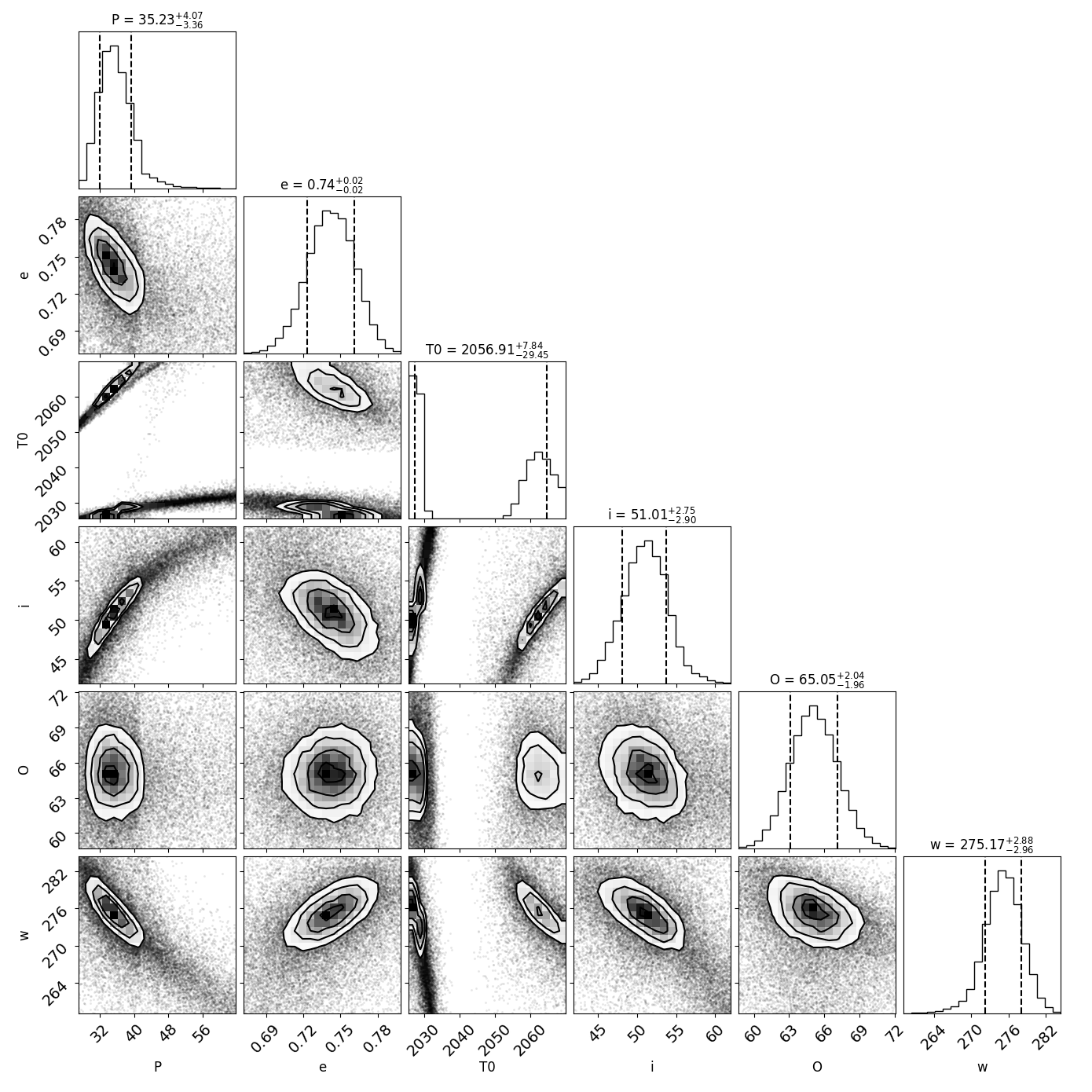}
\figsetgrpnote{}
\figsetgrpend

\figsetgrpstart
\figsetgrpnum{21.19}
\figsetgrptitle{Corner Plot for Gl 229 b}
\figsetplot{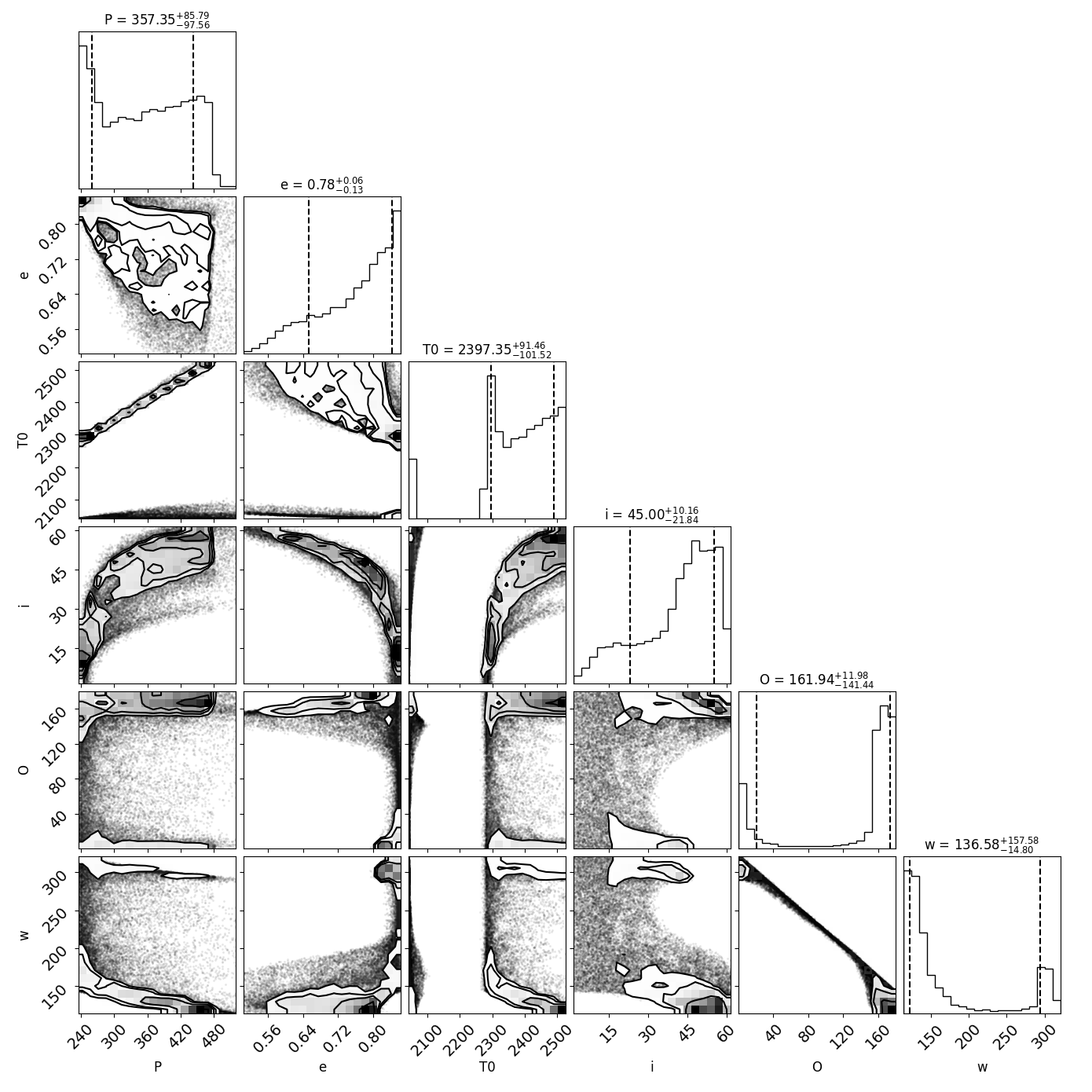}
\figsetgrpnote{}
\figsetgrpend

\figsetgrpstart
\figsetgrpnum{21.20}
\figsetgrptitle{Corner Plot for HR 7672 b}
\figsetplot{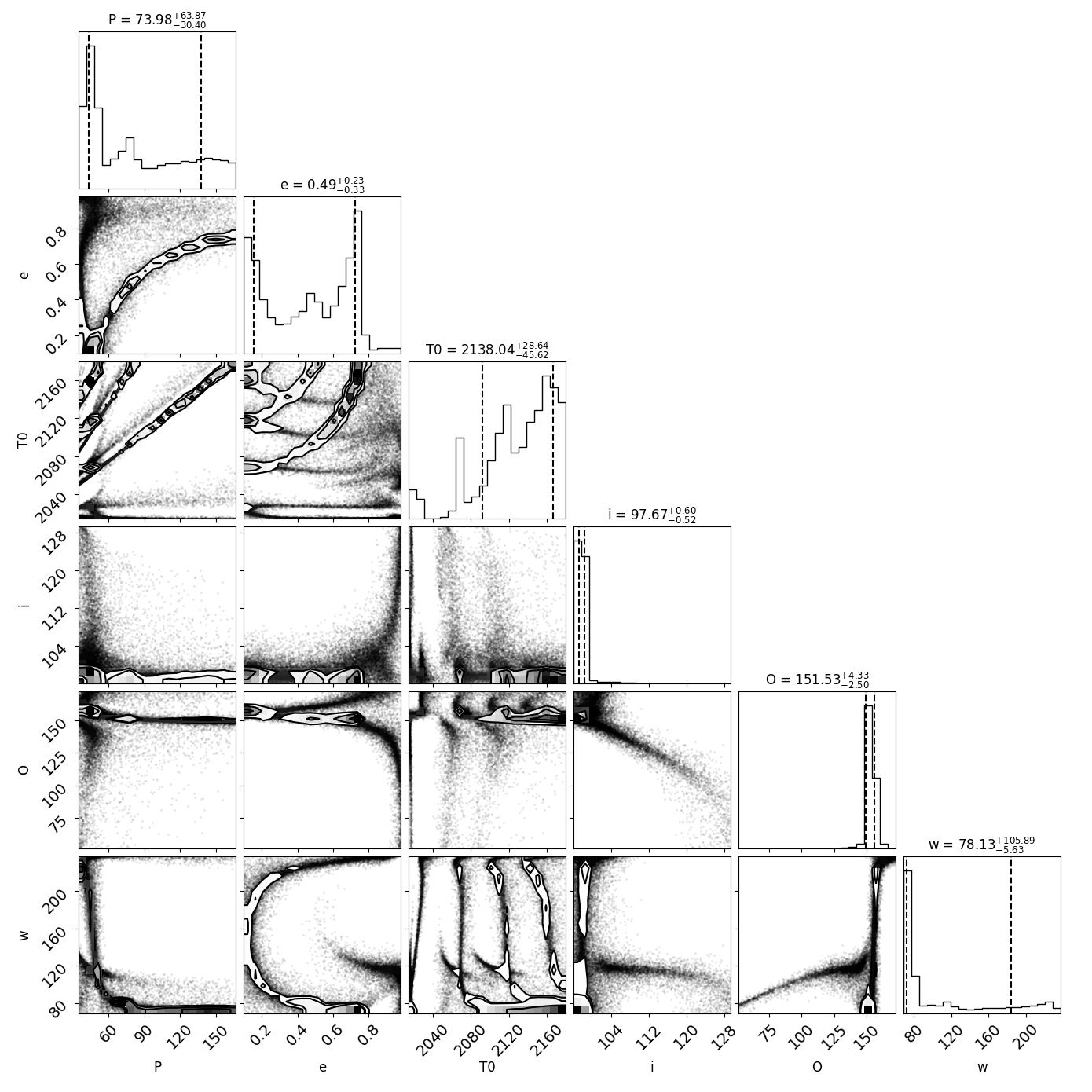}
\figsetgrpnote{}
\figsetgrpend

\figsetgrpstart
\figsetgrpnum{21.21}
\figsetgrptitle{Corner Plot for Gl 758 b}
\figsetplot{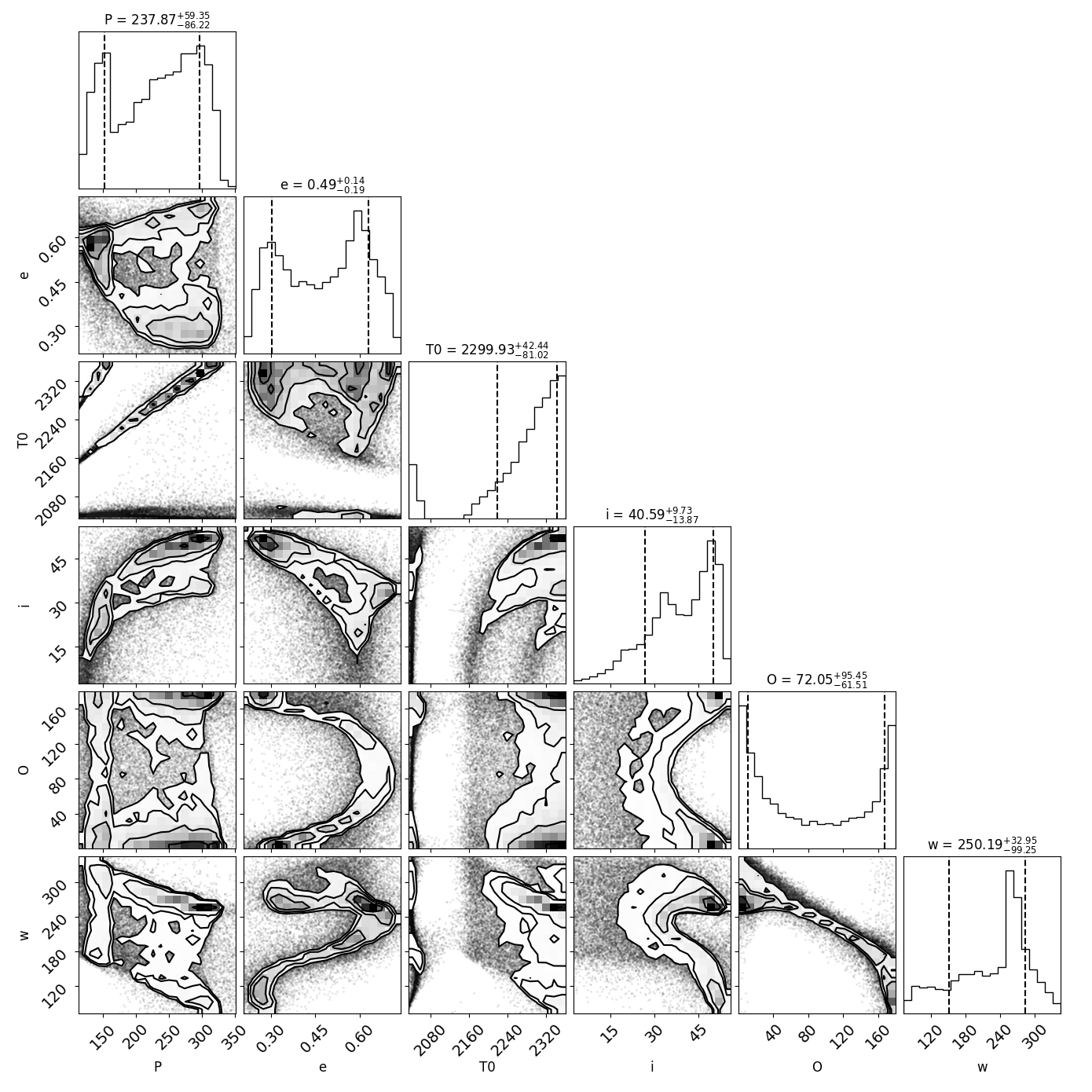}
\figsetgrpnote{}
\figsetgrpend

\figsetend

\end{document}